\documentclass[aps, twocolumn, showpacs, preprintnumbers, amsmath, amssymb, prb]{revtex4-2}
\usepackage{amsfonts}
\usepackage{mathrsfs}
\usepackage{amsmath}
\usepackage{color}
\usepackage[dvipsnames,svgnames,x11names]{xcolor}
\usepackage{natbib}
\usepackage{textcomp}
\usepackage{graphicx}
\usepackage{enumerate}
\usepackage{bm}
\usepackage{xspace}
\usepackage{epstopdf}
\usepackage{dcolumn}
\usepackage{subfigure}
\usepackage{multirow}
\usepackage{makecell}
\usepackage{lipsum}
\usepackage{array}
\usepackage{changes}
\usepackage{ulem}
\usepackage{chngcntr}
\usepackage[colorlinks=true, letterpaper=true, pdfstartview=FitV, linkcolor=blue, citecolor=blue, urlcolor=blue]{hyperref}
\usepackage{float}
\usepackage[figuresright]{rotating}
\usepackage{booktabs}
\usepackage{hyperref}
\usepackage{braket}

\makeatletter

\newcommand{\Rmnum}[1]{\expandafter\@slowromancap\romannumeral #1@}
\makeatother
\providecommand{\U}[1]{\protect\rule{.1in}{.1in}}

\newcommand{\CV}[0]{\color{black}}
\newcommand{\CIV}[0]{\color{black}}

\begin{document}

\title{Fundamental Relations as the Leading Order in Nonlinear Thermoelectric Responses with Time-Reversal Symmetry}


\author{Ying-Fei Zhang$^{1}$
}
\altaffiliation{These authors contributed equally to this work.}
\author{Zhi-Fan Zhang$^{2}$ 
}
\altaffiliation{These authors contributed equally to this work.}

\author{Hua Jiang$^{2,3}$ 
}
\email{jianghuaphy@fudan.edu.cn}

\author{Zhen-Gang Zhu$^{1,4}$
}
\email{zgzhu@ucas.ac.cn}

\author{Gang Su$^{5,6}$
}
\email{gsu@ucas.ac.cn}

\affiliation{
$^{1}$School of Physical Sciences, University of Chinese Academy of Sciences, Beijing 100049, China.\\	
$^{2}$Interdisciplinary Center for Theoretical Physics and Information Sciences, Fudan University, Shanghai 200433, China. \\
$^{3}$State Key Laboratory of Surface Physics, Fudan University, Shanghai 200433, China.\\
$^{4}$School of Electronic, Electrical and Communication Engineering, University of Chinese Academy of Sciences, Beijing 100049, China.\\
$^{5}$Kavli Institute for Theoretical Sciences, University of Chinese Academy of Sciences, Beijing 100190, China.\\
$^{6}$Institute of Theoretical Physics, Chinese Academy of Sciences, Beijing 100190, China.
}


\begin{abstract}
In recent years, nonlinear transport phenomena have garnered significant interest in both theoretical explorations and experiments. In this work, we utilize the semi-classical wave packet theory to calculate disorder-induced second-order transport coefficients: second-order electrical ($\sigma$), thermoelectric ($\alpha$), and thermal ($\kappa$) coefficients, capturing the interplay between side-jump and skew-scattering contributions in systems with time-reversal symmetry. \CV By employing a realistic model of topological insulators specifically tailored to represent $\text{Bi}_2(\text{Se}, \text{Te})_3$  materials\CIV, we quantitatively characterize the Fermi-level dependence of these second-order transport coefficients by explicitly including Coulomb impurity potentials. Furthermore, we elucidate the relationships between these coefficients, establishing the second-order Mott relation and the Wiedemann-Franz law induced by disorder. This study develops a comprehensive theoretical framework elucidating the nonlinear thermoelectric transport mechanisms in quantum material systems.
\end{abstract}
\maketitle

\section{INTRODUCTION}

Thermoelectric transport  plays an important role in condensed matter physics, such as the anomalous Hall effect and Nernst effect \cite{Nagaosa2010, Xiao2006,Bergman2010,Zhang2008,Zhang2009,Zhang2016}.
Based on the linear response theory, the Mott relation and  Wiedemann-Franz (WF) law have been observed and  proved \cite{Jonson1980,Xiao2006,Yokoyama2011,NeilAshcroft1976},
which reveals general relations between charge currents or heat currents driven by an electric field or temperature gradient in systems with broken time-reversal symmetry (TRS).
\CV However, for time-reversal symmetric systems, the corresponding linear transverse Mott and Wiedemann-Franz law are vanishing, because the linear Hall-like responses are forbidden by TRS \CIV \cite{Xiao2010,Behnia2016,Ortix2021}.

However, striking progress has been made recently: a nonlinear response may be induced in TRS materials with broken inversion symmetry ($\mathcal P$-broken) \cite{Sodemann2015}.
Importantly the nonlinear responses are dominant as leading order since the linear responses may disappear in some cases.
%
Moreover, the nonlinear responses for charge, spin, and heat are deeply rooted in the intrinsic properties of systems, for example, the Berry curvature dipole (BCD)
\cite{Sodemann2015,Wang2022,Zeng2020,Zhang2024}.
The BCD reveals that particular pattern of distribution of Berry curvature in momentum space and brings us observable effects \cite{Duan2023,Xu2018,Ma2018,Kang2019,Xiao2020a,Xiao2020b}.
For a comparison, the linear Hall effect (the intrinsic mechanism for anomalous Hall effect or quantum Hall effect \cite{Shen2017,Klitzing1980,Thouless1982,Haldane1988}) is an integral of Berry curvature over the entire momentum space. 
In some sense, nonlinear responses provide a powerful tool to study more detailed properties of systems.


These phenomena can be traced back to the second-order nonlinear Hall effect
\cite{Sodemann2015,Du2019,Du2021,Wang2021,Liu2021,Duan2022,Das2023,Adhidewata2023,Saha2023,Zhang2024,Wang2024,Isobe2020,Makushko2024}, the second-order thermoelectric effect  \cite{Gao2018,Zeng2019,Yu2021,Papaj2021,Zeng2021,Zeng2022,Zeng2022a,Qiang2023,Varshney2025,Zhang2025,Liu2025,Yu2019}, and the second-order thermal effect  \cite{Zhou2022,Zhang2016,Zhang2025,Li2024}, whose response currents are proportional to the square of the driving electric field or temperature gradient.
%
Three contributions can be identified for these effects: intrinsic (in), and two extrinsic terms, namely side-jump (sj) and skew-scattering (sk).
%

%

Identifying the mutual relationships between different dynamical quantities is a fundamental issue and may reveal deep insights into the microscopic physics of systems.
Recent progress in nonlinear transport has shown a deep and subtle interplay between charge, spin,  heat, and topological properties of Bloch bands.
%
%
In the second-order nonlinear regime induced by BCD, the anomalous Hall and Nernst coefficients are proposed to  be directly proportional to each other, $\sigma\propto \alpha$ \cite{Zeng2020,Wang2022}, and the anomalous thermal Hall coefficient satisfies  $\kappa \propto \frac{\partial \sigma }{\partial \mu}$ \cite{Zeng2020} or $\sigma \propto \frac{\partial \kappa} {\partial \mu}$ \cite{Wang2022}.
Thus, we are confronted with a key question: what are the correct fundamental relations in the nonlinear regime?
Furthermore, if disorder is inevitably present, how these relationships are modified by impurity scattering?
As far as we know, the influence of impurities on nonlinear thermoelectric effects is non-negligible.   

In this work, we perform a comprehensive analysis of the impact of impurity scattering on the second-order thermoelectric effect using the semi-classical method (Table \ref{table1}). \CV In order to better describe the actual material, we utilize the surface states of topological insulators (TIs) (Bi$_2$(Se,Te)$_3$). For this concrete model, \CIV
%
we show that skew scattering (sk) makes a dominant contribution to the second-order Hall effect where the BCD is zero due to $C_{3v}$ symmetry.
Meanwhile, we find the relationships among all the coefficients when considering Coulomb screening scattering (Table \ref{table2}).
Intriguingly, for side-jump (sj) scattering, the relation among the three second-order thermoelectric response coefficients ($\alpha_{xxy} ^{\mathrm{sj}}=- \frac{L}{3} \sigma_{xxy}^{\mathrm{sj}} $ and
$\frac{\partial \kappa_{xxy}^{\text{sj}}}{\partial \varepsilon _{F}} = \frac{L}{3e} \sigma_{xxy}^{\text{sj}}$, \CV with the Lorentz number $L =\frac{\pi ^{2} k_{B}^{2}}{3 e^{2}}$\CIV )  is independent of the scattering potential and model details.
As for sk scattering, we obtain the relations $\alpha _{xxy}^{\text{sk}} =LP\sigma _{xxy}^{\text{sk}}$ and  $\frac{\partial \kappa _{xxy}^{\text{sk}}}{\partial \varepsilon _{F}} =-\frac{L}{e} P \sigma _{xxy}^{\text{sk}}$, where $P$ is a parameter related to the Coulomb potential.

%

\begin{figure}[tb]
	\centering
	\includegraphics[width=1\linewidth]{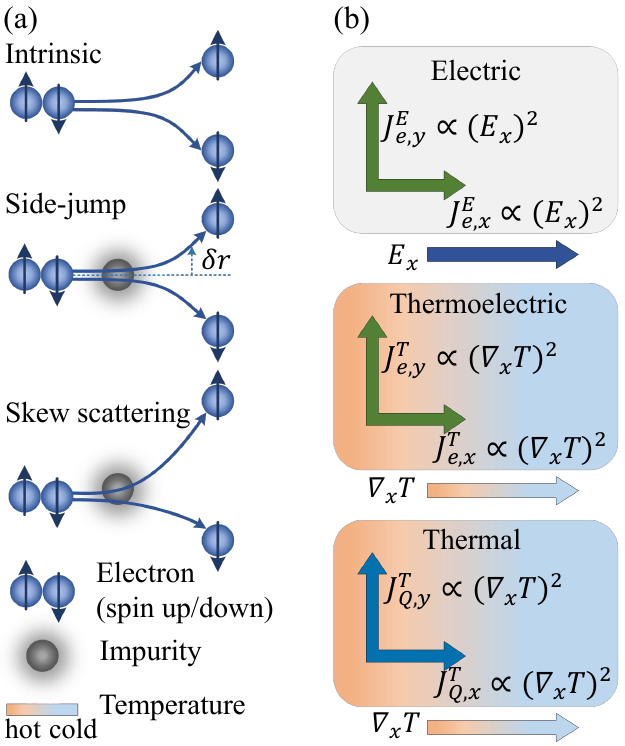}
	\caption{ (a) Illustration of the intrinsic and two other extrinsic mechanisms that can give rise to nonlinear response. (b) Diagrams of transverse and longitudinal electric, thermal and thermoelectric effects, where the electric field or temperature gradient is along the direction of $x$.}
	\label{Fig.1}
\end{figure}

\begin{table*}[t]
\renewcommand\arraystretch{2}
	\centering
	\caption{The anomalous transport coefficients with TRS. 2nd means the second-order. \CV$C_{m_1,m_2}^{\text{in/sj/sk}}$ is defined in Eq. (\ref{eq:Cm1m2}), $m_1$ labels the microscopic kernel $\mathcal{X} _{m_1}^{\text{in/sj/sk}}$ as seen in Eqs. (\ref{eq3})-(\ref{eq5}) with specific representation, and $m_2$ labels the power of $(\varepsilon -\mu)/k_BT$. 
		Physical quantities without quotation of a reference are derived in this work. \CIV}
\begin{tabular*}{\textwidth}{@{\extracolsep{\fill}}llll}
\hline\hline
2nd nonlinear effect &	intrinsic       & 	skew scattering  & side-jump \\
\hline
Electric  &	 $\sigma _{abc}^{\text{in}}=-e^{3} C_{1,0}^{\text{in}} $ \cite{Sodemann2015}  &  $\sigma _{abc}^{\text{sk}} =e^{3} \left( C_{1,0}^{\text{sk}}-C_{2,0}^{\text{sk}}\right)$\cite{Isobe2020}  &  $\sigma _{abc}^{\text{sj}} =-e^{3} \left( C_{1,0}^{\text{sj}} +C_{2,0}^{\text{sj}} - C_{3,0}^{\text{sj}}\right)$\cite{Du2019} \\
Thermoelectric \ & $\alpha _{abc}^{\text{\text{in}}} =-ek_{B}^{2} C_{1,2}^{\text{in}}$\cite{Wang2022,Zeng2020} &  $\alpha _{abc}^{\text{sk}}=-ek_B^2 \left( \frac{1}{k_{B}T }C_{3,1}^{\text{sk}} +C_{4,2}^{\text{sk}}\right)$\cite{Papaj2021} & $\alpha _{abc}^{\text{sj}} =ek_B^2 \left( \frac{1}{k_{B}T }C_{4,1}^{\text{sj}} +C_{5,2}^{\text{sj}}\right)$ \\
Thermal & $\kappa _{abc}^{\text{in}}=-k_{B}^{3} TC_{1,3}^{\text{in}}$\cite{Wang2022} & $\kappa _{abc}^{\text{sk}} =k_{B}^{2} \left( C_{3,2}^{\text{sk}} +k_{B}TC_{4,3}^{\text{sk}}\right)$ & $\kappa _{abc}^{\text{sj}} =-k_{B}^{2}\left( C_{4,2}^{\text{sj}} +k_{B}TC_{5,3}^{\text{sj}}\right)$\\
\hline\hline
	\end{tabular*}
		\label{table1}
\end{table*}

\section{Second-order nonlinear thermoelectric effects}
\label{section2}
For TRS systems driven by  an electric  field $\mathbf{E}$ or temperature gradient $\mathbf {{\nabla} T}$ , we can write the charge current  $	\mathbf{J}_{e}$   and heat current $\mathbf{J}_{Q}$ in component form as \cite{Li2017Nonlinear,Itskov2025Tensor,Ortix2021,Du2021a,Yang2020,Costi2010}


\begin{equation}
	\begin{aligned}	\label{eq1}
		{J}_{e,a} &={J}^{E}_{e,a}+{J}^{T}_{e,a}=\sigma_{abc} E_bE_c+\alpha_{abc} (-\nabla_b T)(-\nabla_c T),\\
		{J}_{Q,a} &={J}^{E}_{Q,a}+{J}^{T}_{Q,a}=\tilde{\alpha }_{abc} {E}_b{E}_c+ \kappa_{abc} (-\nabla_b T)(-\nabla_c T),
	\end{aligned}
\end{equation}
 where $a, b, c \in \{x, y, z\}$ denote the Cartesian indices, with the Einstein summation convention implied for repeated indices ($b, c$). ${J}^{E(T)}_{e,a}$ represents the $a$-th component of response electric currents driving by $\mathbf{E}$ ($\mathbf {{\nabla} T}$), and ${J}^{E(T)}_{Q,a}$ corresponds to the $a$-th component of response heat currents driving by  $\mathbf{E}$ ($\mathbf {{\nabla} T}$).
 $\sigma_{abc}$, $\alpha_{abc}$, $\tilde{\alpha}_{abc} $, and $\kappa_{abc}$  are  the second-order electric, thermoelectric, electric-thermal and thermal coefficients, respectively.  
It should be noted that Eq. (\ref{eq1}) does not take into account the mixed second-order terms of driving electric field and temperature gradient \cite{Nakai2019,Yamaguchi2024,Yang2025,Yamaguchi2024a}, which we will consider in future work.

In order to better describe the response effects of real materials, we  consider three major contributions to the response, as illustrated in Fig. \ref{Fig.1} \cite{Du2019,Papaj2021,Zhou2022,Qiang2023}.
After some steps of derivation (see  Appendix \ref{App:Derivation of Second-Order Nonlinear Thermoelectric Effects}   for more details), \CV we introduce the C-function, e.g. $C_{m_1,m_2}^{\text{in/sj/sk}}$, to unify the second-order transport coefficients \cite{Du2019,Papaj2021} in a compact form as shown in Table \ref{table1}\CIV,
\begin{equation}\label{eq:Cm1m2}
C_{m_1,m_2}^{\text{in/sj/sk}}=-\int d\varepsilon  \mathcal{X} _{m_1}^{\text{in/sj/sk}} \left( \frac{\varepsilon -\mu}{k_BT} \right) ^{m_2}\frac{\partial f_0}{\partial \varepsilon },
\end{equation}
where $ m_1 \in \{1,2,3,4,5\}$ \CV labels microscopic contributions [Eqs. (\ref{eq3})-(\ref{eq5})]\CIV, $ m_2 \in \{0,1,2,3\}$ \CV specifies the power of the thermal factor $(\varepsilon -\mu)/k_BT$\CIV, $\varepsilon$ represents energy, $k_B$ represents the Boltzmann constant, $T$ is temperature, and $f_0$ is the Fermi-Dirac  equilibrium distribution  of electrons in absence of external electric field or temperature gradient.

Unlike linear responses \cite{Qiang2023,Du2019}, the kernel functions in the second-order nonlinear response derived in this work possess distinct forms.  
The first one is an intrinsic contribution induced by BCD (we name it as intrinsic since this term is induced by the anomalous velocity due to Berry curvature of Bloch bands) 
\begin{equation}
\mathcal{X} _{1}^{\mathrm{in}} =\tau\frac{ \epsilon _{abd}}{\hbar}\int_k{\Omega _{k_d} v_{k_c}^{\text{g}}\frac{\partial f_0}{\partial \varepsilon _k}},
\label{eq3}
\end{equation}
where $\tau$ is the relaxation time, $\epsilon _{abd}$ is the Levi-Civita symbol and $a,b,c,d \in \{x,y,z\}$, $\hbar$ is the Planck constant, $\int_k =\int \frac{d^D k}{(2\pi)^D}$ with the dimension $D$, $v_{k_c}^{\text{g}}=\frac{1}{\hbar}\frac{\partial \varepsilon_k}{\partial k_c}$ is the group velocity, $\varepsilon_k$ is the energy marked by the wave vector $k$, and Berry curvature $\mathbf{\Omega}_k$, a gauge-field tensor, is crucial in all Hall-like effects.
%
It is clearly visible that this second-order response is proportional to $\tau$.   
In \CV TRS \CIV 
systems, numerous experiments investigating the second-order response have been conducted in recent years, drawing significant attention \cite{Duan2023,Xu2018,Ma2018,Kang2019,Xiao2020a,Xiao2020b}.

The side-jump scattering from the side-jump velocity and the anomalous distribution is the second mechanism we consider. And the kernel functions are
\begin{subequations}\label{eq4}
	\begin{align}
		&\mathcal{X}_{1}^{\text{sj}}  =\tau ^{2}  \int _{k} v_{k_a}^{\text{sj}}  v^{\text{g}}_{k_b} v^{\text{g}}_{k_c}\frac{\partial ^{2} f_{0} }{\partial \varepsilon _{k}^{2}},\label{eq4a}\\
		&\mathcal{X}_{2}^{\text{sj}} =\tau ^{2}\int _{k}\left( v_{k_a}^{\text{sj}}  \Gamma_{bc}  +  v_{k_b}^{\text{sj}} \Gamma_{ac}\right)\frac{\partial f_{0} }{ \partial \varepsilon _{k}},\label{eq4b}\\
		&\mathcal{X}_{3}^{\text{sj}} =\tau ^{2}  \int _{k}  v^{\text{g}}_{k_a} \int _{k^{\prime }} w_{kk^{\prime }}^{(S)} v^{\text{g}}_{k'_{c}} \delta r_{b,{k}^{\prime }{k}}\frac{\partial ^{2} f_{0} }{\partial \varepsilon _{k}^{2}},\label{eq4c}\\
		&\mathcal{X}_{p=4,5}^{\text{sj}} =\tau ^{2} \frac{2}{p-3}\int _{k} (2 v_{k_a}^{\text{g}}  v_{k_b}^{\text{sj}} -v_{k_a}^{\text{sj}}  v_{k_b}^{\text{g}} ) v^{\text{g}}_{k_c}\frac{\partial ^{p-3} f_{0} }{\partial \varepsilon _{{k}}^{p-3}},\label{eq4d}
	\end{align}
\end{subequations}
where $\Gamma_{ij}=\frac{\partial v^{\text{g}}_{k_{i}}}{\hbar \partial k_{j}} = \frac{\partial^2 \varepsilon_k}{\hbar^2 \partial k_i k_j} $ is the Hessian matrix \cite{Gao2019}, inversely proportional to the effective mass of the electron. Under the approximation of large effective mass, the contribution of this term can be neglected ($\Gamma_{ij} \rightarrow 0$).
$w_{kk'}^{(S)} $ is the symmetric scattering rate.
Overall, the sj scattering is all proportional to $\tau^2\mathbf v^{\text{sj}}_k\propto \tau$, with $\mathbf v^{\text{sj}}_k=\int_{{k}^{\prime}}w^{(S)}_{{k}{k}^{\prime}}\delta\mathbf{r}_{{k}^{\prime}{k}}\propto\tau^{-1}$ and $\delta\mathbf{r}_{{k}^{\prime}{k}}$ the coordinate shift \cite{Sinitsyn2005,Sinitsyn2006}, indicating that we might need additional scaling to distinguish sj from that of intrinsic contribution.
For electric field driving, the sj term is rather complex and can be decomposed into three parts [Eqs. (\ref{eq4a})-(\ref{eq4c})].
Apart from the term [Eq. (\ref{eq4b})] relating to the effective mass, the other two terms [Eqs. (\ref{eq4a}) and (\ref{eq4c})] can be approximately regarded as the contributions of $\mathbf v^{\text{sj}}_k$ and the square of $\mathbf v^{\text{g}}_k$, which indicates that they are highly related to the shape of the energy band in the $k$-space.
However, as for the driving of the temperature gradient, the effective mass of electrons does not directly affect the second-order thermoelectric or thermal effects based on the Eq. (\ref{eq4d}).

Lastly,  the sk scattering, which are asymmetric scattering due to the effective spin-orbit coupling of the electron or impurity, reads
\begin{subequations}\label{eq5}
\begin{align}
	&\mathcal{X}_{1}^{\text{sk}} =\tau ^{3}\int _{k}\int _{k^{\prime }} \left(\Gamma_{ac} v_{k'_{b}}^{\text{g}}  +v_{k'_{a}}^{\text{g}} \Gamma_{bc}\right) w_{kk^{\prime }}^{(A)}\frac{\partial f_{0}}{\partial \varepsilon _{k}},\label{eq5a}\\
	&\mathcal{X}_{2}^{\text{sk}} =\tau ^{3}  \int _{k}\int _{k^{\prime }} v^{\text{g}}_{k_{a}} v^{\text{g}}_{k'_{b}} v^{\text{g}}_{k'_{c}} w_{kk^{\prime }}^{(A)}\frac{\partial ^{2} f_{0} }{\partial \varepsilon _{k}^{2}},\label{eq5b}\\
	&\mathcal{X}_{q=3,4}^{\text{sk}} =\tau ^{3}\frac{4  }{q-2}\int _{k}\int _{k'}  v_{k_a}^{\text{g}}  v_{k_b}^{\text{g}}  v_{k_c}^{\text{g}}  w_{kk'}^{(A)}\frac{\partial ^{q-2} f_{0}}{\partial \varepsilon _{k}^{q-2}},\label{eq5c}
\end{align}
\end{subequations}
where $w_{k'k}^{(A)} $ is the anti-symmetric scattering rate from $k$ to $k^\prime$. 
Without loss of generality, we consider the third order skew scattering which is $w_{k'k}^{(A)}  \propto \tau^0$.
It is seen that the contribution from  sk scattering is proportional to $\tau^3$ [Eqs. (\ref{eq5a})-(\ref{eq5c})]. Since the longitudinal linear conductivity is proportional to $\tau$, the sk scattering [Eqs. (\ref{eq5a})-(\ref{eq5c})] can in principle be distinguished from the intrinsic [Eq. (\ref{eq3})] and side-jump terms [Eqs. (\ref{eq4a})-(\ref{eq4d})] (they are both proportional to $\tau$) through the scaling relation to the longitudinal linear conductivity. %

%


\CV In terms of the kernel representation (Eqs.~(\ref{eq:Cm1m2})--(\ref{eq5})), we systematically work out all the nine second-order transport coefficients shown in Table~\ref{table1}.
The intrinsic BCD-induced $\sigma^{\text{in}}_{abc}$, $\alpha^{\text{in}}_{abc}$ and $\kappa^{\text{in}}_{abc}$ in Table~\ref{table1} were established in Refs. \cite{Sodemann2015,Wang2022,Zeng2020}. 
However, these quantities are rewritten by using the kernel functions. 
$\sigma _{abc}^{\text{sj}}$ is different from that in Ref. \cite{Du2019} due to a different treatment of the shift energy induced by the external electric field.
$\alpha^{\rm sk}_{abc}$ is derived by using the same nonequilibrium distribution functions of Ref. \cite{Papaj2021}. 
%
%
$\alpha^{\text{sj}}_{abc}$, $\kappa^{\text{sk}}_{abc}$ and $\kappa^{\text{sj}}_{abc}$ are derived via our kernel representation.
These compact equations presented in Table~\ref{table1} may be used to evaluate nonlinear transport in various materials via the first-principles method or other numerical calculations, enabling a practical basis for studying.

The nine transport quantities in Table~\ref{table1} may also serve as a starting point to study the mutual relations among them, i. e.,
%
the second-order Mott relation and WF law.
Since these relations generally contain microscopic integrals that depend on the specific band structure and types of impurity,
and we therefore illustrate the formalism using a typical model for surface states of TIs, which corresponds to realistic materials and the kernel integrals can be evaluated explicitly. Thus, the second-order Mott relation and WF law induced by disorder can be exhibited analytically. 
 \CIV

\begin{figure}[tb]
	\centering
	\includegraphics[width=1\linewidth]{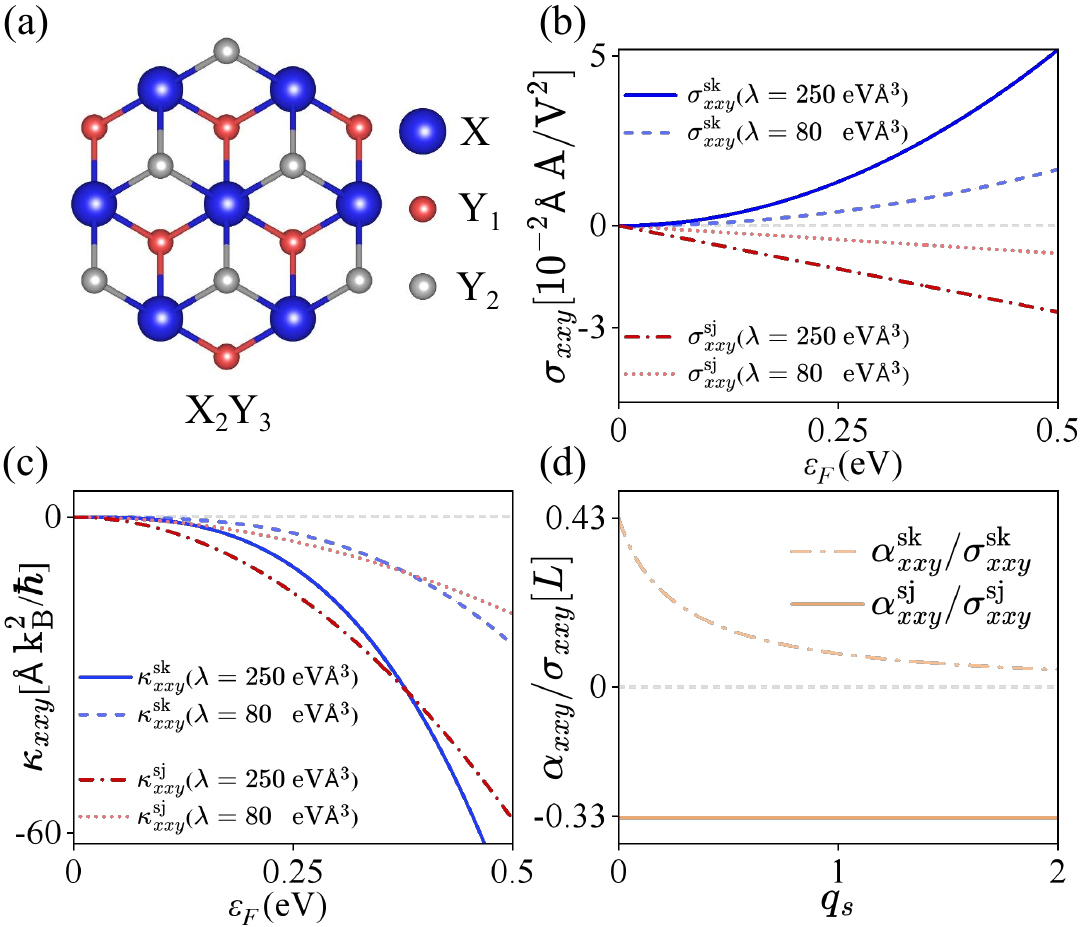}
	\caption{(a) Crystal structure of the surface of the topological insulator X$_2$Y$_3$. The hexagonal lattice is projected along the (001) direction.    (b) Variation of the second-order Hall coefficient $\sigma_{xxy}^{\text{sk/sj}}$ with Fermi energy $\varepsilon_F$ under different $\lambda$. (c) Fermi energy $\varepsilon_F$ dependence of the second-order thermal coefficient $\kappa^{\text{sk/sj}}_{xxy}$ for different $\lambda$. (d) Calculated ratio $\alpha^{\text{sk/sj}}_{xxy}/\sigma^{\text{sk/sj}}_{xxy}$ in unit of $L$ as a function of $q_s$.  The gray dot line in (b)-(d) is shown for visual guidance. Parameters used: $v=3.291$ \text{eV\AA}, $\lambda=80$ \text{eV\AA}$^3$, $250$ \text{eV\AA}$^3$, $\tau$ = $0.1$  $\text{ps}$. }
	\label{Fig.2}
\end{figure}

\section{Model and Discussions} \label{section4}
\CV The surface state of  TI X$_2$Y$_3$ (X=Bi, Y=Se, Te) is considered in the general framework developed in Sec. \ref{section2} \CIV, with its crystal structure shown in Fig. \ref{Fig.2}(a) \cite{Kuroda2010,Fu2009}. 
It is evident that the structure exhibits $C_{3v}$ symmetry and the corresponding Hamiltonian is
\begin{align}\label{eq:H(k)}
	H( k) =v( k_{x} \sigma _{y} -k_{y} \sigma _{x}) +\frac{\lambda }{2}\left( k_{+}^{3} +k_{-}^{3}\right) \sigma _{z},
\end{align}
where $v$ is the Dirac velocity, $k_{\pm}=k_x \pm i k_y= k e^{i\theta_k}$, $\sigma_{x,y,z}$ are the Pauli matrixes, and the parameter $\lambda$ serves as a quantitative descriptor for the magnitude of hexagonal warping in the electronic structure.
The energy dispersion is
$\varepsilon _{\pm,k} =\pm\sqrt{ k^{2} v^{2} + k^{6} \lambda ^{2} \cos^{2}( 3\theta_k ) }$. In addition, the Berry curvature linearly in the warping strength $\lambda$ is $\Omega _{k_z}\approx \frac{\lambda }{v}\cos(3\theta _{k}) +O\left( \lambda ^{2}\right)$.

 In principle, we need to distinguish whether the nonlinear current originates from the Berry curvature dipole, quantum metric, or extrinsic contributions (sj, sk, or other mixing terms), based on criteria such as crystal symmetry, scaling law, or other means.
Nonetheless, within the framework of our model, it should be noted that the $\tau$-independent contribution arising from quantum metric \cite{Gao2025,Liu2024a,Yu2025} is strictly forbidden due to the $\mathcal P$-broken but with TRS.
Meanwhile,
 \ due to the $C_{3v}$ symmetry, the second-order intrinsic effect induced by the BCD vanishes and only the two extrinsic contributions from impurity scattering exists \cite{Sodemann2015,Du2021}.
%
To be specific, the independent nonzero elements are only  those of the $\sigma_{xxy} = \sigma_{xyx} = \sigma_{yxx} = - \sigma_{yyy}$ \cite{Du2021}.  \CV In particular, electric frequency doubling has already been observed on Bi$_2$Se$_3$ surfaces under zero magnetic field, where an extrinsic mechanism is expected to dominate \cite{He2021}, demonstrating that the present model is directly connected to experimentally relevant real materials.\CIV

For the present Hamiltonian, the absence of momentum dependence in a $\delta$-correlated random potential results in the vanishing of $w_{k'k}^{(A)}$, implying that such impurities make no skew scattering contribution to the second-order response \cite{Isobe2020}. 
This issue can be circumvented by considering the screened  Coulomb impurities \cite{He2021,Makushko2024,Isobe2020,Lu2024} that are randomly dispersed within the sample.

Beginning with  the bare Coulomb potential in real space, $V_0(r) =\frac{e^2Q}{4\pi \varepsilon_0 \varepsilon r}$ captures the interaction between charge carriers and a charge impurity $Q$, where $r$ is the carrier-impurity separation and $\varepsilon_0(\varepsilon)$ is the vacuum (material) permittivity. Through Fourier transformation and considering Thomas-Fermi screening at long wavelength, the screened Coulomb interaction takes the form
\(V(q)=\frac{2\pi\alpha v}{q + q_{\text{TF}}}\), where  $q=|  \mathbf k-\mathbf{k'}|$ is the magnitude of the momentum transfer for an electron scattered from an initial state $\mathbf k$ to a final state  $\mathbf k'$, $\alpha$ is a dimensionless coupling constant analogous to the fine-structure constant \cite{He2021}, $q_{\text{TF}}$ represents the Thomas-Fermi wave vector,
and we define $q_s=\frac{q_{\text{TF}}}{k_{F}}$.
Experimentally, the ratio  $\frac{q_{\text{TF}}}{k_{F}}$ requires three key physical quantities, the carrier density, Fermi velocity, and effective dielectric constant, which can be extracted from Hall bar transport measurements, scanning tunnelling microscopy and angle-resolved photoemission spectroscopy(such as $\frac{q_{\text{TF}}}{k_{F}} \lesssim $ 0.4 for Bi$_2$Se$_3$ \cite{He2021}).
The scattering rate and relaxation time are both closely related to $q_s$, which indicate that the magnitude of Coulomb interaction is crucial to the second-order thermoelectric transport.
%

\begin{table}[t]
	\renewcommand\arraystretch{2}
	\centering
\caption{The second-order (2nd) Mott relation and  Wiedemann-Franz (WF) Law induced by disorder, where $P$ is a dimensionless quantity determined by the  strength of Coulomb interaction $q_s$, and $L =\frac{\pi ^{2} k_{B}^{2}}{3 e^{2}}$ is the Lorenz number.}
\begin{tabular*}{8.5cm}{@{\extracolsep{\fill}}llll}
\hline\hline
&	side-jump & skew scattering  \\
\hline
2nd Mott   & $\alpha _{xxy}^{\mathrm{sj}}=- \frac{1}{3} L\sigma _{xxy}^{\mathrm{sj}} $
& $\alpha _{xxy}^{\text{sk}} =LP\sigma _{xxy}^{\text{sk}}$
\\
2nd WF law   & $\frac{\partial \kappa _{xxy}^{\text{sj}}}{\partial \varepsilon _{F}} =\frac{L}{3e} \sigma _{xxy}^{\text{sj}}$
& $\frac{\partial \kappa _{xxy}^{\text{sk}}}{\partial \varepsilon _{F}} =-\frac{L}{e} P \sigma _{xxy}^{\text{sk}}$
\\
\hline\hline
	\end{tabular*}
	\label{table2}
\end{table}

It can be shown in this model that a nonzero warping term is essential for generating a finite second-order response due to disorder, i.e. $\lambda\neq 0$.
Following a laborious calculation, we derive the second-order electric, thermoelectric and thermal response coefficients for the $xxy$ component
(more details in Appendix \ref{App:The Surface State of the Topological Insulator}).
%
%
In Fig. \ref{Fig.2}(b), the skew scattering and side-jump contributions of the second-order electric response coefficients scale as $\sigma^{\text{sk}}_{xxy}\propto \varepsilon_F^4$ and $\sigma^{\text{sj}}_{xxy}\propto \varepsilon_F^2$, respectively.
These scaling behaviors indicate that tuning the chemical potential away from the band edge significantly enhances the second-order response, in agreement with experimental observations \cite{He2021}.
Notably, the two contributions display opposite signs, wherein the skew scattering term plays a dominant role compared to the side-jump component.
However, for  the second-order thermal response, there are different scaling behaviors.
%
$\kappa_{xxy}^{\text{sk}}\propto\varepsilon_F^5$ and $\kappa_{xxy}^{\text{sj}}\propto\varepsilon_F^3$, the skew scattering term grows more rapidly with increasing $\varepsilon_F$.  %
As seen in Fig. \ref{Fig.2}(c), at higher Fermi energies, the thermal response becomes increasingly dominated by the skew scattering mechanism, which is significantly larger than the side-jump scattering contribution.  %
Besides,  to highlight the role of hexagonal warping in the second-order disorder-induced response, we consider two representative values of the warping parameter $\lambda=80$ and 250 \text{eV\AA}$^3$ shown in Figs. \ref{Fig.2}(b) and (c), which correspond to two topological insulators Bi$_2$Se$_3$ \cite{He2021,Kuroda2010} and Bi$_2$Te$_3$ \cite{Fu2009}, respectively.  %
It is evident that an increase in the wrapping term results in a corresponding augmentation of the second-order response coefficients.

Furthermore, \CV specifically for the given model, \CIV we obtain the second-order Mott relation, $\alpha _{xxy}^{\mathrm{sj}}=-\frac{1}{3}L\sigma _{xxy}^{\mathrm{sj}} $ and $\alpha _{xxy}^{\text{sk}} =LP\sigma _{xxy}^{\text{sk}}$, and the second-order WF law, $\frac{\partial \kappa _{xxy}^{\text{sj}}}{\partial \varepsilon _{F}} =\frac{L}{3e} \sigma _{xxy}^{\text{sj}}$ and  $\frac{\partial \kappa _{xxy}^{\text{sk}}}{\partial \varepsilon _{F}} =-\frac{L}{e} P \sigma _{xxy}^{\text{sk}}$(Table \ref{table2}),
where $P$ is just a function of parameter $q_s$ (see details in Appendix \ref{App:The Surface State of the Topological Insulator}).
Overall, the coefficients that describe these relationships can be classified into two categories based on the contribution of impurities.
The skew scattering exhibits a diminishing trend with increasing the Coulomb interaction's strength $q_s$; while the side-jump contribution yields a constant ratio that is independent of $q_s$.  %
Specially for the second-order Mott relation, we simply show this trend in Fig. \ref{Fig.2}(d).
In the limit of $q_s \rightarrow 0$, we still have fundamental relations that don't depend on the details of the model ($\alpha _{xxy}^{\text{sk}} \approx 0.43L\sigma _{xxy}^{\text{sk}}$).
Similarly, the second-order WF law of the sk term can be expressed as $\frac{\partial \kappa _{xxy}^{\text{sk}}}{\partial \varepsilon _{F}} \approx - 0.43\frac{L}{e}  \sigma _{xxy}^{\text{sk}}$ when $q_s$ approaches zero  as discussed in Appendix \ref{Append:Fundamental relations in surface states of topological insulators }.
%

\section{Conclusions and Discussions}
\label{section5}

Although we try our best to obtain the general relationship of the second-order thermoelectric coefficients, there are  still some points that we don't take into consideration.  %
We only take into account the contribution of the highest-order relaxation time for sj and sk scattering.  %
We  note that there are other contributions related to the interband Berry connection that has no counterpart in linear response \cite{Gong2024,Xiao2019,Mehraeen2024}, which may also affect the relations.  %
In addition, we would like to point out that there are some other methods in the research field of topological nonlinear transport, such as  Feynman diagrams \cite{Du2021,Sinitsyn2007,Parker2019,Wang2022,Bruus2004}, density matrices \cite{Bhalla2021,Varshney2023,Varshney2025,Ma2025b} or  nonequilibrium Green's function  formalism \cite{Wei2022,Zhang2023,Li2025}.  %
\CV In particular, the effects such as self-energy and vertex corrections from electron-electron correlations \cite{Pasqua2025,Jho2017,Chen2017,Wang2025}, and Nielsen-Taylor interference \cite{Nielsen1968,Nielsen1974,Dudenhoeffer1972,Howson1988,Reizer2000} between electron-phonon and electron-impurity scattering may further alter the associated fundamental relations. \CIV
\CV Moreover, the present work is restricted to TRS systems, for which the second-order coefficients considered here constitute the leading signals. An important extension is to some magnetic systems that break $\mathcal{P}$ and $\mathcal{T}$ individually while preserving the combined $\mathcal{PT}$ symmetry \cite{Wang2021,Liu2021,Yang2025,Zhang2025}, in which the Berry connection polarizability \cite{Wang2021,Liu2021} contribution as well as zeroth-order extrinsic \cite{Huang2025,Guo2026} contributions  can qualitatively modify the associated fundamental relations. 
Extending the present framework in this direction would be an interesting topic for future study.
  \CIV

Finally, we would like to point out that the Mott relation induced by sk scattering in our theoretical work is qualitatively consistent with the recent experimental obervation \cite{Liu2025}, where the proportionality between the second-order thermoelectric \CV coefficient \CIV and the electrical \CV coefficients \CIV due to the sk is verified (see Extended Data Fig. 4 in Ref. \cite{Liu2025}). Crucially, beyond merely capturing the sk-induced Mott-like trends, our study establishes microscopic theoretical basis within semiclassical framework, as detailed in Table \ref{table1}. More importantly, aside from the experimentally verified sk-Mott term, we provide the first theoretical predictions for the remaining three fundamental relations in Table \ref{table2}. The results in our work establish a comprehensive theoretical foundation that accounts for the disorder-induced transport mechanisms.

\begin{acknowledgments}
This work is supported by the National Key R\&D Program of China (Grant No. 2024YFA1409200, No. 2022YFA1402802, and No. 2022YFA1403700), CAS Project for Young Scientists in Basic Research Grant No. YSBR-057, and the NSFC (Grant No. 12350401). G.S. is supported in part by the Innovation Program for Quantum Science and Technology (Grant No. 2024ZD0300500), NSFC (Grants No. 12534009 and No. 12447101), and the Strategic Priority Research Program of Chinese Academy of Sciences (Grant No. XDB1270000).
\end{acknowledgments}

\begin{widetext}

\appendix
\counterwithin{figure}{section}
\counterwithin{table}{section}

\renewcommand{\thefigure}{\thesection\arabic{figure}}
\renewcommand{\thetable}{\thesection\arabic{table}}
\section{Derivation of Second-Order Nonlinear Thermoelectric Effects}\label{App:Derivation of Second-Order Nonlinear Thermoelectric Effects}

\subsection{Distribution function driven by the electric field \cite{Isobe2020,Du2019} or temperature gradient \cite{Papaj2021}}\label{App:Distribution function driven by the electric fieldor temperature gradient}
We deal with spatially
homogeneous systems under steady-state conditions $ \frac{\partial f( k,\varepsilon_k )}{\partial t}=0 $  and hence the distribution function $f\left({k}, \varepsilon_k\right)$ does not depend on the spatial position, the semi-classical Boltzmann equation becomes
\begin{equation}
	\dot{\mathbf k}\cdot\frac{\partial f( k,\varepsilon_k )}{\partial \mathbf k}
	=-\mathcal C[f(k,\varepsilon_k )]
	=-\int_{{k}^{\prime}}\left[w_{{k}^{\prime}{k}}f({k},\varepsilon_k)
	-w_{{k}{k}^{\prime}}f({k}^{\prime},\varepsilon_k+\delta\varepsilon_{{k}^{\prime}{k}})\right],
\end{equation}
where $\dot{\mathbf k}=-\frac{e}{\hbar}\mathbf{E}$, the distribution function $f(k,\varepsilon_k )$ deviates from the equilibrium Fermi-Dirac distribution $f_0(k,\varepsilon_k )$ when electrons are accelerated by the
external field, the energy shift is  {$\delta\varepsilon_{ k^{\prime} k} =\varepsilon_{ k^{\prime} } -\varepsilon_k= -eE_a\delta r_{a, k^{\prime} k}$} , the collision integral $\mathcal{C}[f(k,\varepsilon_k )]$ incorporates the scattering rates $w_{ k^{\prime} k} =w_{ k' k}^{(S)}+w_{ k' k}^{(A)}$ and the explicit expression is given by \cite{Nagaosa2010},
\begin{equation}\begin{aligned}
		&w_{ k' k}^{(S)} =\frac{2\pi}{\hbar}\langle|V_{ k' k}|^2\rangle\delta(\varepsilon_{ k'}-\varepsilon_{ k}), \\
		&w_{ k k^{\prime}}^{(A)} =-\frac{(2\pi)^2}{\hbar}\int_q \text{Im}\langle V_{ k k^{\prime}}V_{ k^{\prime} q}V_{ q k}\rangle\delta(\varepsilon_{ k^{\prime}}-\varepsilon_{ k})\delta(\varepsilon_{ k^{\prime}}-\varepsilon_{ q}) ,
\end{aligned}\end{equation}
where $w_{ k' k}^{(S)}$ and $w_{ k' k}^{(A)}$ represent symmetric and antisymmetric scattering rate respectively, $\langle\cdots \rangle$ denotes the average over the impurity distribution \cite{Bruus2004}, and \( V_{ k' k} \) is the matrix element
for the single impurity scattering.

Central to nonlinear expansion is the Taylor expansion of $f({k'},\varepsilon_k+\delta\varepsilon_{kk'})$ (up to the second order)
\begin{equation}\label{eq:Taylor expansion}
	f({k} ',\varepsilon _{k'} )=f({k} ',\varepsilon _{k} +\delta \varepsilon _{{k} '{k}} )=f({k}^{\prime } ,\varepsilon _{k'} )\Bigl|_{\varepsilon _{k'} =\varepsilon _{k}} +\frac{\partial f({k} ',\varepsilon _{k'} )}{\partial \varepsilon _{k'}}\Bigl|_{\varepsilon _{k'} =\varepsilon _{k}} \delta \varepsilon _{{k} '{k}} +\frac{1}{2}\frac{\partial ^{2} f({k} ',\varepsilon _{k'} )}{\partial \varepsilon _{k'}^{2}}\Bigl|_{\varepsilon _{k'} =\varepsilon _{k}} (\delta \varepsilon _{{k} '{k}} )^{2}.
\end{equation}
In order to obtain the second-order response, we decompose $f( k,\varepsilon_k) $ as \cite{Isobe2020},
\begin{equation}
	f( k,\varepsilon_k)=f_0( k,\varepsilon_k)+f^\mathrm{scatt}( k,\varepsilon_k)+f^\mathrm{adist}( k,\varepsilon_k),
\end{equation}
where the first term $f_0 ( k,\varepsilon_k) $ is the equilibrium distribution function,
the second term $f^\text{scatt}( k,\varepsilon_k)$ represents the scattering contribution without the Berry curvature, the last term $f^\text{adist}( k,\varepsilon_k)$ denotes the anomalous distribution arising from both the Berry curvature and the energy shift $\delta\varepsilon_{ {k^{\prime}}{k}}$.
And we further expand $f^\text{scatt }( k,\varepsilon_k)$ and $f^\text{adist}( k,\varepsilon_k)$ with respect to the electric field $\mathbf E$ order ($1\leq n \leq 2$) and asymmetry scattering rate $w^{(A)}$ order ($0\leq m \leq 1$),
\begin{equation}\label{}
	\begin{aligned}
		f^{\mathrm{scatt}}( k,\varepsilon_k) &=\sum_{n\geq1}f_n^{\text{scatt}}( k,\varepsilon_k),\quad f_n^{\text{scatt}}( k,\varepsilon_k)=\sum_{m\geq0}f_n^{(m)}( k,\varepsilon_k),
		\\
		f^{\text{adist}}( k,\varepsilon_k) &=\sum_{n\geq1}f_n^{\text{adist}}( k,\varepsilon_k),\quad f_n^{\text{adist}}( k,\varepsilon_k)=\sum_{m\geq0}g_n^{(m)}( k,\varepsilon_k).
	\end{aligned}
\end{equation}
Therefore, the distribution function can be written as
\begin{equation}\label{}
	\begin{aligned}
		f( k,\varepsilon_k) =&f_0( k,\varepsilon_k)+f_1^{(0)}( k,\varepsilon_k)+f_1^{(1)}( k,\varepsilon_k)+f_2^{(0)}( k,\varepsilon_k)+f_2^{(1)}( k,\varepsilon_k)\\&+g_1^{(0)}( k,\varepsilon_k)+g_1^{(1)}( k,\varepsilon_k)+g_2^{(0)}( k,\varepsilon_k)+g_2^{(1)}( k,\varepsilon_k),
	\end{aligned}
\end{equation}
and the collision integrals with $w_{kk'}^{(S)}$ define the scattering times,
\begin{equation}
	\begin{aligned}\label{eq:scattering times}
		\int_{k'}w_{kk'}^{(S)}\left[f_n^{(m)}(k,\varepsilon_k)-f_n^{(m)}(k',\varepsilon_{k'})\right]=\frac1{\tau_n^{(m)}}f_n^{(m)}(k,\varepsilon_k),\\
		\int_{k'}w_{kk'}^{(S)}\left[g_n^{(m)}(k,\varepsilon_k)-g_n^{(m)}(k',\varepsilon_{k'})\right]=\frac1{\tilde{\tau}_n^{(m)}}g_n^{(m)}(k,\varepsilon_k),
	\end{aligned}
\end{equation}
where $\tau_n^{(m)}$ denotes the relaxation time associated with $f_n^{(m)}(k,\varepsilon_k)$, while $\tilde{\tau}_n^{(m)}$ denotes the relaxation time associated with $g_n^{(m)}(k,\varepsilon_k)$.	

Based on the results in Ref. \cite{Isobe2020}, the nonequilibrium distribution function are obtained by solving the Boltzmann equation within the relaxation time approximation. Their explicit forms are
\begin{subequations}\label{eq:f(k)}
	\begin{align}
		f_{1}^{(0)}(k,\varepsilon_k)=& \frac{eE_{b}\tau_{1}^{(0)}}{\hbar } \frac{\partial f_{0}( k,\varepsilon _{k})}{\partial k_{b}}, \\
		g_{1}^{(0)}(k,\varepsilon_k)=&{-}\tilde{\tau}_{1}^{(0)}eE_{b}\int _{k^{\prime }} w_{kk^{\prime }}^{(S)}\frac{\partial f_{0} ({k} ',\varepsilon _{k'} )}{\partial \varepsilon _{k'}}\Big|_{\varepsilon _{k'}=\varepsilon _{k}} \delta r_{b,{k}^{\prime }{k}}, \\
		f_{1}^{(1)}(k,\varepsilon_k) =&\tau_1^{(1)}\tau_{1}^{(0)}\frac{e}{\hbar}E_{b}\int _{k^{\prime }} w_{kk^{\prime }}^{(A)}\frac{\partial f_{0} (k^{\prime } ,\varepsilon _{k'} )}{\partial k'_{b}}\Bigl|_{\varepsilon _{k'} =\varepsilon _{k}}, \\
		g_{1}^{(1)}(k,\varepsilon_k)
		=&{-} \tilde{\tau} _{1}^{( 1)} \tilde{\tau} _{1}^{( 0)} eE_{b}\int _{k^{\prime }} w_{kk^{\prime }}^{(A)}\int _{k^{'' }} \left[ w_{k'k^{'' }}^{(S)}\left[ \frac{\partial f_{0} (k'',\varepsilon _{k''} )}{\partial \varepsilon _{k''}}\delta r_{b,{k}^{'' }{k} '}\right] \Bigl|_{\varepsilon _{k''} =\varepsilon _{k'}}\right] \Bigl|_{\varepsilon _{k'} =\varepsilon _{k}} \\&{-} \tilde{\tau} _{1}^{( 1)} eE_{b}\int _{k^{\prime }} w_{kk^{\prime }}^{(A)}\frac{\partial f_{0} (k',\varepsilon _{k'} )}{\partial \varepsilon _{k'}}\Bigl|_{\varepsilon _{k'} =\varepsilon _{k}} \delta r_{b,k'k}, \\
		f_{2}^{(0)}(k,\varepsilon_k)=&\tau_2^{(0)}\tau_{1}^{(0)}\frac {e^2}{\hbar^2} \frac{\partial ^{2} f_{0} (k,\varepsilon _{k} )}{\partial k_{c} \partial k_{b}} E_{c} E_{b},\\
		g_{2}^{(0)}(k,\varepsilon_k)
		=&{-}\tilde{\tau}_2^{(0)}\tilde{\tau}_{1}^{(0)}\frac{e^{2}}{\hbar } E_{b} E_{c}\frac{\partial }{\partial k_{c}}\int _{k^{\prime }} w_{kk^{\prime }}^{(S)}\left[\frac{\partial f_{0} ({k} ',\varepsilon _{k'} )}{\partial \varepsilon _{k'}}\right]\Bigl|_{\varepsilon _{k'} =\varepsilon _{k}}\delta r_{b,{k}^{\prime }{k}} \\ \nonumber&{-}\tilde{\tau}_2^{(0)}\tau_{1}^{(0)}\frac{e^{2}}{\hbar } E_{b} E_{c}\int _{k^{\prime }} w_{kk^{\prime }}^{(S)}\left[\left[\frac{\partial }{\partial \varepsilon _{k'}}\frac{\partial }{\partial k'_{c}} f_{0} (k',\varepsilon _{k'} )\right]\Bigl|_{\varepsilon _{k'} =\varepsilon _{k}} \delta r_{b,{k}^{\prime }{k}}\right],\\
		f_{2}^{(1)} (k,\varepsilon _{k} )
		=&\tau _{2}^{( 1)} \tau _{1}^{(1)} \tau _{1}^{(0)}\frac{e^{2}}{\hbar ^{2}} E_{b} E_{c}\frac{\partial }{\partial k_{c}}\left[\int _{k^{\prime }} w_{kk^{\prime }}^{(A)}\left[\frac{\partial f_{0} (k^{\prime } ,\varepsilon _{k'} )}{\partial k'_{b}} \right]\Bigl|_{\varepsilon _{k'} =\varepsilon _{k}}\right] \\ \nonumber &+\tau _{2}^{( 1)} \tau _{2}^{(0)} \tau _{1}^{(0)}\frac{e^{2}}{\hbar ^{2}} E_{c} E_{b}\int _{k^{\prime }} w_{kk^{\prime }}^{(A)}\left[\frac{\partial ^{2}f_{0} (k',\varepsilon _{k'} )}{\partial k'_{c} \partial k'_{b}} \right]\Bigl|_{\varepsilon _{k'} =\varepsilon _{k}},
	\end{align}
\end{subequations}
where we ignored the terms $\propto (\delta r_{k'k})^2$. These expressions in Eq. (\ref{eq:f(k)}) will be used in Sec. \ref{App:Second order electric Hall} for the derivation of the second-order electrical conductivity.	

On the other hand, we consider a system with in the absence of electric fields ($\mathbf E=0$) but subject to a spatially uniform temperature gradient $\nabla T=$ const. The steady-state Boltzmann equation [to distinguish the Hall effect in the previous section, write the distribution function as $F(r,k)$, and $F_0(r,k)=f_0(k,\varepsilon_k)$ denote the Fermi-Dirac  equilibrium distribution]:
\begin{equation}
	\dot{\mathbf{r}}\cdotp\frac{\partial F(r,k)}{\partial \mathbf{r}}
	=-\mathcal{C}[F(r,k)]
	=-\int_{k'}[w_{k'k}F(r,k)-w_{kk'}F({r} + \delta \mathbf{r}_{k'k}, k')],
\end{equation}
where the semiclassical equations of motion of electron wave packet is
\begin{equation}\dot{\mathbf r}=\mathbf{v}^{\text{g}}_{k}+\mathbf{v}^{\text{a}}_{k}+\mathbf{v}^{\text{sj}}_{k}=\frac{\partial\varepsilon_{{k}}}{\hbar\partial{k}}-\boldsymbol{\dot{\mathbf{k}}}\times\boldsymbol{\Omega}_{{k}}
	+\sum_{{k}^{\prime}}w_{{k}^{\prime}{k}}\delta\mathbf{r}_{{k}^{\prime}{k}},
\end{equation}
and $\delta \mathbf{r}_{k'k}$ being the real-space coordinate shift that contributes to the side-jump term
\begin{equation}
	\delta\mathbf{r}_{{k}^{\prime}{k}}=\langle u_{{k}^{\prime}}|i\partial_{\mathbf{k}^{\prime}}u_{{k}^{\prime}}\rangle-\langle u_{{k}}|i\partial_{\mathbf{k}}u_{{k}}\rangle-(\partial_{\mathbf{k}}+\partial_{\mathbf{k}^{\prime}})\mathrm{arg}\langle u_{{k}^{\prime}}|u_{{k}}\rangle.
\end{equation}

According to Ref. \cite{Papaj2021}, similar to the case of electric field driving, utilizing the idea from Eqs. (\ref{eq:Taylor expansion})-(\ref{eq:scattering times}), the nonequilibrium distribution function is given by
\begin{subequations}\label{eq:F(k)}
	\begin{align}
		F_{1}^{(0)} (r,k)=&\tau_{1}^{(0)} \dot{r}_{b} (\varepsilon _{k} -\mu )\frac{\partial F_{0}( r,k)}{\partial \varepsilon _{{k}}}\frac{\nabla _{b} T}{T},\\
		G_{1}^{(0)} (r,k)=&-\tilde\tau_{1}^{(0)} \int _{k^{\prime }} w_{k^{\prime } k}^{(S)}\left[ (\varepsilon _{k'} -\mu )\frac{\partial F_{0} (r',k')}{\partial \varepsilon _{{k} '}}\right]\Bigl|_{r'=r} \delta r_{b,k'k}\frac{\nabla _{b} T}{T},\\
		F_{1}^{(1)} (r,k)  =&-\tau_1^{(1)}\tau_{1}^{(0)}\int _{k^{\prime }} w_{k^{\prime } k}^{(A)}\left[ \dot{r}_{b} (\varepsilon _{k'} -\mu )\frac{\partial F_{0}( r',k')}{\partial \varepsilon _{{k} '}}\frac{\nabla _{b} T}{T}\right]\Bigl|_{r'=r},\\
		G_{1}^{(1)} (r,k)=&\tilde\tau_1^{(1)} \int _{k^{\prime }} w_{k^{\prime } k}^{(A)}\Big[ \tilde\tau_{1}^{(0)} \int _{k''} w_{k''k'}^{(S)}\left[\left[ (\varepsilon _{k''} -\mu )\frac{\partial F_{0} (r'',k'')}{\partial \varepsilon _{{k} ''}}\right]\Bigl|_{r''=r'} \delta r_{b,k''k'}\right]\Bigl|_{r'=r}\frac{\nabla _{b} T}{T}\\&+ (\varepsilon _{k'} -\mu )\frac{\partial F_{0}( r',k')}{\partial \varepsilon _{{k} '}}\Bigl|_{r'=r} \delta r_{b,k'k}\frac{\nabla _{b} T}{T}\Big],\\
		F_{2}^{(0)} (r,k)  =& \tau _{2}^{(0)} \tau _{1}^{(0)}\left[ (\varepsilon _{k} -\mu )^{2}\frac{\partial ^{2} F_{0}(r,k)}{\partial \varepsilon _{k}^{2}} +2(\varepsilon _{k} -\mu )\frac{\partial F_{0}(r,k)}{\partial \varepsilon _{k}}\right] \dot{r}_{b}\dot{r}_{c}\frac{\nabla _{b} T}{T}\frac{\nabla _{c} T}{T},\\
		G_{2}^{(0)} (r,k)
		=&-\tilde\tau _{2}^{( 0)}( \tilde\tau _{1}^{( 0)} +\tau _{1}^{( 0)}) \dot{r}_c\int _{k^{\prime }} w_{k^{\prime } k}^{(S)}\left[ 2(\varepsilon _{k'} -\mu )\frac{\partial F_{0}(r,k')}{\partial \varepsilon _{{k} '}} +(\varepsilon _{k'} -\mu )^{2}\frac{\partial ^{2} F_{0}(r,k')}{\partial \varepsilon _{{k} '}^{2}}\right] \delta r_{b,k'k}\frac{\nabla _{c} T}{T}\frac{\nabla _{b} T}{T},\\
		F_{2}^{(1)}(r,k)
		=&-\tau_2^{(1)}\tau_{1}^{(0)}(\tau_{1}^{(1)}+\tau_{2}^{(0)})\int_{k^{\prime}}w^{(A)}_{k^{\prime}k}{\dot{r_b}}{\dot{r_c}}\left[   2(\varepsilon _{k'}-\mu)\frac{\partial F_{0}(r,k')}{\partial \varepsilon _{k'}}+ {(\varepsilon _{k'}-\mu)^2}\frac{\partial^2 F_{0}(r,k')}{\partial \varepsilon _{k'}^2}\right] \frac{\nabla_b T}{T}\frac{\nabla_c T}{T},
	\end{align}
\end{subequations}
where $\left[ \frac{\partial G_1^{(0)}(r,k')}{\partial r}\delta r_{k',k}+\frac12\frac{\partial ^2F_0(r,k')}{\partial r^2}(\delta r_{k',k})^2 \right] \propto (\delta r_{k',k})^2$ is ignored. The analytical forms established in Eq. (\ref{eq:F(k)}) constitute a key component in the subsequent derivation of the second-order thermoelectric and thermal responses as detailed in Sec. \ref{App:Derivation of Second-order anomalous Nernst effect} and \ref{App:Derivation of Second-order  anomalous Thermal effect}.

 \subsection{Response currents driven by the electric field or temperature gradient}\label{App:Response currents driven by the electric field or temperature gradient}
For TRS systems driven by an electric  field $\mathbf{E}$ or temperature gradient $\mathbf {{\nabla} T}$, we can write the charge current  $	\mathbf{J}_{e}$   and heat current $\mathbf{J}_{Q}$  as \cite{Li2017Nonlinear,Itskov2025Tensor}
\begin{equation}\label{}
	\begin{aligned}
		\mathbf{J}_e &=\mathbf{J}^{E}_{e}+\mathbf{J}^{T}_{e}= \boldsymbol{\sigma} : \mathbf{E}\mathbf{E} \, + \, \boldsymbol{\alpha} : \left(-\nabla T\right)\left(-\nabla T\right), \\
		\mathbf{J}_Q &=\mathbf{J}^{E}_{Q}+\mathbf{J}^{T}_{Q}= \tilde{\boldsymbol{\alpha}} : \mathbf{E}\mathbf{E} \, + \, \boldsymbol{\kappa} : \left(-\nabla T\right)\left(-\nabla T\right),
	\end{aligned}
\end{equation}
where $\mathbf{J}^{E(T)}_{e}$ are response electric currents driving by $\mathbf{E}$ ($\mathbf {{\nabla} T}$), and $\mathbf{J}^{E(T)}_{Q}$ are response heat currents driving by  $\mathbf{E}$ ($\mathbf {{\nabla} T}$). The symbols $''\colon''$ denoted third-rank tensor multiplication operations.
	$\sigma$, $\alpha$, $\tilde{\alpha }$, and $\kappa$ are  the second-order electric, thermoelectric, electric-thermal and thermal coefficients, respectively.
	Furthermore, we can rewrite the above equation in matrix form considering the exchange symmetry of the external field. Take the second-order nonlinear electric $J_{e}^E$as an example,
	\begin{equation}\label{}
		\left( \begin{array}{c}
			J_{e,x}^E \\
			J_{e,y}^E \\
			J_{e,z}^E \\
		\end{array} \right)
		=\left( \begin{array}{llllll}
			\sigma _{xxx} &\sigma _{xyy} & \sigma _{xzz} &\sigma _{xyz}+\sigma _{xzy} &\sigma _{xxz}+\sigma _{xzx} &\sigma _{xxy}+\sigma _{xyx} \\
			\sigma _{yxx}&	\sigma _{yyy} &		\sigma _{yzz} &		 \sigma _{yyz}+\sigma _{yzy} &		 \sigma _{yxz}+\sigma _{yzx} &	  \sigma _{yxy}+\sigma _{yyx} \\
			\sigma _{zxx} &\sigma _{zyy} &		 \sigma _{zzz} &		 \sigma _{zyz}+\sigma _{zzy} &	      \sigma _{zxz}+\sigma _{zzx} &		\sigma _{zxy}+\sigma _{zyx} \\
		\end{array} \right)
		\left( \begin{array}{l}
			E_x E_x \\
			E_y E_y \\
			E_z E_z \\
			E_y E_z \\
			E_x E_z \\
			E_x E_y \\
		\end{array} \right)
	\end{equation}
	To make it clearer, we have provided the expression of the quantity [Eq. (1)] in the main text.

	\subsubsection{Derivation of second-order electric effect}\label{App:Second order electric Hall}
	
	In subsequent calculations, we assume that all relevant scattering times are equal, specifically, $\tau_n^{(m)}=\tilde\tau_n^{(m)}=\tau$.
	
	Firstly, the \textbf{intrinsic} second-order electric effect is given by
	\begin{equation}\label{}
		\begin{aligned}
			J_{e,a}^{E,\text{in}} &=	\sigma _{abc}^{\text{in}}E_bE_c=-e\int _{k}\left[ v^{\text{g}}_{k_a} f_{2}^{( 0)}(k,\varepsilon_k) +v^{\text{a}}_{k_a} f_{1}^{( 0)}(k,\varepsilon_k)\right] \\&=-\frac{e^{3} \tau ^{2}}{\hbar^2}\int _{k} v^{\text{g}}_{k_a}\frac{\partial ^{2} f_{0}(k,\varepsilon_k)}{\partial k_{b} \partial k_{c}} E_{b} E_{c} -\frac{e^{3} \tau}{\hbar^2}\epsilon _{abd}\int _{k} \Omega _{k_d}\frac{\partial f_{0}(k,\varepsilon_k)}{\partial k_{c}} E_{b} E_{c}.
		\end{aligned}
	\end{equation}
	
	When we consider the time-reversal symmetry system, $\int _{k} v^{\text{g}}_{k_a}\frac{\partial ^{2} f_{0}(k,\varepsilon_k)}{\partial k_{b} \partial k_{c}}$ disappears, thus
	
	\begin{equation}\label{eq:sigma_in}
		\begin{aligned}
			\sigma _{abc}^{\text{in}}  &= -\frac{e^{3} \tau }{\hbar } \epsilon _{abd}\int _{k} \Omega _{k_d} v^{\text{g}}_{k_c}\frac{\partial f_{0}(k,\varepsilon_k)}{\partial \varepsilon _{k}}.
		\end{aligned}
	\end{equation}
	
	The \textbf{side-jump} effect consists of two distinct contributions. The first originates from the side-jump velocity, while the second stems from the side-jump-induced modification of the distribution function, which can be obtained as follows:
	\begin{equation}
		\begin{aligned}
			J^{E,\text{sj}}_{e,a}=&\sigma _{abc}^{\text{sj}}E_{b} E_{c} =-e\int_kv^{\text{sj}}_{k_a}f_2^{(0)}(k,\varepsilon_k) -e\int_kv^{\text{g}}_{k_a}g_2^{(0)}(k,\varepsilon_k)\\
			=&-\frac{\tau ^{2} e^{3}}{\hbar ^{2}} E_{c} E_{b}\int _{k} v^{\text{sj}}_{k_a} \ \frac{\partial ^{2}f_{0} (k,\varepsilon _{k} )}{\partial k_{c} \partial k_{b}} {+}\frac{\tau ^{2} e^{3}}{\hbar } E_{c} E_{b}\int _{k} v^{\text{g}}_{k_a} \ \frac{\partial }{\partial k_{c}}\int _{k^{\prime }} w_{kk^{\prime }}^{(S)}\left[\frac{\partial f_{0} ({k} ',\varepsilon _{k'}) }{\partial \varepsilon _{k'}}\right]\Bigl|_{\varepsilon _{k'} =\varepsilon _{k}} \delta r_{b,{k}^{\prime }{k}} \\
			&{+}\frac{\tau ^{2} e^{3}}{\hbar } E_{c} E_{b}\int _{k} v^{\text{g}}_{k_a} \ \int _{k^{\prime }} w_{kk^{\prime }}^{(S)}
			\left(\left[\frac{\partial }{\partial \varepsilon _{k'}}\frac{\partial }{\partial k'_{c}} f_{0} (k',\varepsilon _{k'} )\right]\Bigl|_{\varepsilon _{k'} =\varepsilon _{k}}\delta r_{b,{k}^{\prime }{k}} \right),
		\end{aligned}
		\label{Jsj(2)-1}
	\end{equation}
	with
	\begin{equation}
		\begin{aligned}
			\label{eq:sigmasj}
			\sigma _{abc}^{\text{sj}}
			=&-\tau ^{2} e^{3}\int _{k} v^{\text{sj}}_{k_a}  v^{\text{g}}_{k_b} v^{\text{g}}_{k_c}\frac{\partial ^{2} f_{0} (k,\varepsilon _{k} )}{\partial \varepsilon _{k}^{2}} \\&{+} \tau ^{2} e^{3}\int _{k} v^{\text{g}}_{k_{a}}\int _{k^{\prime }} w_{kk^{\prime }}^{(S)} v^{\text{g}}_{k'_{c}} \delta r_{b,{k}^{\prime }{k}}\frac{\partial ^{2} f_{0} (k,\varepsilon _{k} )}{\partial \varepsilon _{k}^{2}}-{\tau ^{2} e^{3}}\int _{k} \left( v^{\text{sj}}_{k_a}\Gamma_{bc} {+} v^{\text{sj}}_{k_b}\Gamma_{ac} \right) \frac{\partial f_{0} (k,\varepsilon _{k} )}{\partial \varepsilon _{k}}.
	\end{aligned}\end{equation}

	Finally, the \textbf{skew scattering} by impurities can be derived as
	\begin{equation}
		\begin{aligned}\label{eq:J_sk}
			J^{E,\text{sk}}_{e,a}&=\sigma_{abc}^{\text{sk}}E_bE_c=-e\int_kv^{\text{g}}_{k_a}f_2^{(1)}(k,\varepsilon_k)\\
			&=-\frac{e^{3} \tau ^{3}}{\hbar ^{2}} E_{c} E_{b}\int _{k} v^{\text{g}}_{k_a} \partial _{k_{c}}\int _{k^{\prime }} w_{kk^{\prime }}^{(A)}\frac{\partial f_{0} (k^{\prime } ,\varepsilon _{k'} ) }{\partial k'_{b}} \Bigl|_{\varepsilon _{k'} =\varepsilon _{k}} -\frac{e^{3} \tau ^{3}}{\hbar ^{2}} E_{c} E_{b}\int _{k} v^{\text{g}}_{k_a} \int _{k^{\prime }} w_{kk^{\prime }}^{(A)}\left[\frac{\partial ^{2} f_{0} (k',\varepsilon _{k'} )}{\partial k'_{c} \partial k'_{b}} \right]\Bigl|_{\varepsilon _{k'} =\varepsilon _{k}},
		\end{aligned}
	\end{equation}
	where
	\begin{equation}
		\begin{aligned}
			\label{eq:sigmask}
			\sigma _{abc}^{\text{sk}}
			={e^{3} \tau ^{3}}\int _{k}\int _{k^{\prime }}\left( \Gamma_{ac} v^{\text{g}}_{k'_{b}}+v^{\text{g}}_{k'_{a}} \Gamma_{bc}\right)  w_{kk^{\prime }}^{(A)}\frac{\partial f_{0} (k,\varepsilon _{k} )}{\partial \varepsilon _{k}} -e^{3} \tau ^{3}\int _{k}\int _{k^{\prime }} v^{\text{g}}_{k_{a}} v^{\text{g}}_{k'_{b}} v^{\text{g}}_{k'_{c}} w_{kk^{\prime }}^{(A)}\frac{\partial ^{2} f_{0} (k,\varepsilon _{k} )}{\partial \varepsilon _{k}^{2}}.
		\end{aligned}
	\end{equation}
 \CV It is worth noting that the side-jump contribution result in Eq. (\ref{Jsj(2)-1}) and the skew-scattering contribution result in Eq. (\ref{eq:J_sk}) are presented in Ref. \cite{Du2019} and Ref. \cite{Isobe2020}, respectively. \CIV 
	
	\subsubsection{Derivation of second-order thermoelectric effect}\label{App:Derivation of Second-order anomalous Nernst effect}
	Driving by a temperature gradient, the charge current reads \cite{Xiao2006}
	\begin{equation}
		\mathbf{J}_{e}^{T} =-e\int_k \mathbf{v}^{\text{g}}_{k}F(r,k) -\frac{\nabla T}{T} \times \frac{e}{\hbar }\int_k \boldsymbol{\Omega}_k\left[( \varepsilon_k -\mu ) F( r,k)
		+k_BT\ln\left( 1+e^{ -\beta ( \varepsilon_k -\mu )}\right)\right].
	\end{equation}
	The second-order \textbf{intrinsic} charge current along $a$ direction is
	\begin{equation}
		\begin{aligned}
			J_{e,a}^{T,\text{in}}=&\alpha _{abc}^{\text{in}} (-\nabla_bT)(-\nabla_cT) =-e\int_k v^{\text{g}}_{k_a}F_{2}^{(0)}( r,k) - \frac{e}{\hbar }\int_k \epsilon _{abd}\Omega_{k_d}( \varepsilon_k -\mu ) F_{1}^{( 0)}( r,k)\frac{\nabla_b T}{T}\\
			=&-e\tau ^{2}\int _{k} v^{\text{g}}_{k_a}v^{\text{g}}_{k_b} v^{\text{g}}_{k_c}\left[( \varepsilon _{k} -\mu )^{2}\frac{\partial ^{2} F_{0}(r,k)}{\partial \varepsilon _{k}^{2}} +2( \varepsilon _{k} -\mu )\frac{\partial F_{0}(r,k)}{\partial \varepsilon _{k}}\right] \frac{\nabla _{b} T}{T}\frac{\nabla _{c} T}{T}  \\&-\frac{e\tau}{\hbar }\int _{k} \epsilon _{abd} ( \varepsilon _{k} -\mu )^2  v^{\text{g}}_{k_c}\Omega_{k_{d}}\frac{\partial F_{0}(r,k)}{\partial \varepsilon _{k}}\frac{\nabla_b T}{T}\frac{\nabla _{c} T}{T},
		\end{aligned}
		\label{2ndJ}
	\end{equation}
	when the systems with time reversal symmetry are considered, the first term of Eq. (\ref{2ndJ}) disappears and the second-order intrinsic Nernst coefficient reads
	\begin{equation}
		\begin{aligned}\label{eq:alpha_in}
			\alpha^{\text{in}}_{abc}=&-\frac{e\tau }{\hbar T^{2}} \epsilon _{abd}\int _{k} v^{\text{g}}_{k_c} \Omega_{k_{d}}( \varepsilon _{k} -\mu )^{2}\frac{\partial F_{0}(r,k)}{\partial \varepsilon _{k}}.
		\end{aligned}
	\end{equation}

	The second-order thermoelectric effect induced by the \textbf{side-jump} scattering is
	\begin{equation}	\begin{aligned}
			\mathbf{J}_{e}^{T,\text{sj}} =&-e\int_k \mathbf{v}^{\text{g}}_{k}G_{2}^{(0)}( r,k)-e\int_k \mathbf{v}^{\text{sj}}_{k}F_{2}^{(0)}( r,k) -\frac{\nabla T}{T} \times \frac{e}{\hbar }\int_k ( \varepsilon_k -\mu )\boldsymbol{\Omega}_{k} G_{1}^{( 0)}( r,k) ,
	\end{aligned}\end{equation}
	where we neglect the mixed contribution from the coordinate shift $\delta \mathbf r_{k'k}$ and the intrinsic term $\mathbf \Omega_k$. Thus
	
	\begin{equation}
		\begin{aligned}
			J_{e,a}^{T,\text{sj}} =&\alpha _{abc}^{\text{sj}}( -\nabla_bT)(-\nabla_cT)=-e\int _{k} v^{\text{sj}}_{k_a} F_{2}^{(0)} (r,k)-e\int _{k} v^{\text{g}}_{k_a} G_{2}^{(0)} (r,k)\\
			=&-e \tau ^{2}\int _{k} v^{\text{sj}}_{k_a}\dot{r}_{b}\dot{r}_{c}\left[ 2(\varepsilon _{k} -\mu )\frac{\partial F_{0} (r,k)}{\partial \varepsilon _{k}} +(\varepsilon _{k} -\mu )^{2}\frac{\partial ^{2} F_{0} (r,k)}{\partial \varepsilon _{k}^{2}}\right]\frac{\nabla _{c} T}{T}\frac{\nabla _{b} T}{T} \\&+ 2\tau ^{2} e\int _{k}\int _{k^{\prime }} v^{\text{g}}_{k_a} \dot{r}_{c} w_{k^{\prime } k}^{(S)}\left[ 2(\varepsilon _{k} -\mu )\frac{\partial F_{0}(r,k)}{\partial \varepsilon _{{k}}} +(\varepsilon _{k} -\mu )^{2}\frac{\partial ^{2} F_{0} (r,k)}{\partial \varepsilon _{{k}}^{2}}\right] \delta r_{b,k'k}\frac{\nabla _{c} T}{T}\frac{\nabla _{b} T}{T},
	\end{aligned}\end{equation}
	with
	\begin{equation}
		\begin{aligned}\label{eq:alphasj}
			\alpha _{abc}^{\text{sj}}
			=&\frac{e\tau ^{2} }{T^{2}}\int _{k} (2v^{\text{g}}_{k_a}  v^{\text{sj}}_{k_b}   -v^{\text{sj}}_{k_a}v^{\text{g}}_{k_{b}})v^{\text{g}}_{k_{c}}\left[ 2(\varepsilon _{k} -\mu )\frac{\partial F_{0} (r,k)}{\partial \varepsilon _{{k}}} +(\varepsilon _{k} -\mu )^{2}\frac{\partial ^{2} F_{0} (r,k)}{\partial \varepsilon _{{k}}^{2}}\right].
	\end{aligned}\end{equation}

	The second-order thermoelectric effect induced by the \textbf{skew} scattering is
	\begin{equation}\label{eq:J^T_sk}
		\mathbf J_{e}^{T,\text{sk}} =-e\int_k \mathbf v^{\text{g}}_{k}F_{2}^{( 1)}( r,k) -\frac{\nabla T}{T} \times \frac{e}{\hbar }\int_k \mathbf \Omega_{k}( \varepsilon_k -\mu ) F_{1}^{( 1)}( r,k),
	\end{equation}
	and the second term in Eq. (\ref{eq:J^T_sk}) disappears under time reversal symmetry,
	\begin{equation}
		\begin{aligned}
			J_{e,a}^{T,\text{sk}} =&\alpha _{abc}^{\text{sk}}( -\nabla _{b} T)( -\nabla _{c} T) =-e\int _{k} v^{\text{g}}_{k_a} F_{2}^{(1)} (r,k)\\
			=&2e\tau ^{3}\int _{k}\int _{k'} v^{\text{g}}_{k_a} v^{\text{g}}_{k_{b}}v^{\text{g}}_{k_c}w_{k'k}^{(A)} (\varepsilon _{k} -\mu )^{2}\frac{\partial ^{2} F_{0} (r,k)}{\partial \varepsilon _{k}^{2}}\frac{\nabla _{c} T}{T}\frac{\nabla _{b} T}{T} \\&+4e\tau ^{3}\int _{k}\int _{k'} v^{\text{g}}_{k_a}v^{\text{g}}_{k_{b}}v^{\text{g}}_{k_c} w_{k'k}^{(A)} (\varepsilon _{k} -\mu )\frac{\partial F_{0} (r,k)}{\partial \varepsilon _{k}}\frac{\nabla _{c} T}{T}\frac{\nabla _{b} T}{T},
		\end{aligned}
	\end{equation}
	with
	\begin{equation}
		\begin{aligned}\label{eq:alphask}
			\alpha _{abc}^{\text{sk}} =-\frac{2e\tau ^{3}}{T^{2}}\int _{k}\int _{k'} v^{\text{g}}_{k_a}v^{\text{g}}_{k_{b}}v^{\text{g}}_{k_{c}} w_{kk'}^{(A)} (\varepsilon _{k} -\mu )^{2}\frac{\partial ^{2} F_{0} (r,k)}{\partial \varepsilon _{k}^{2}} -\frac{4e\tau ^{3}}{T^{2}}\int _{k}\int _{k'} v^{\text{g}}_{k_a}v^{\text{g}}_{k_{b}}v^{\text{g}}_{k_{c}} w_{kk'}^{(A)} (\varepsilon _{k} -\mu )\frac{\partial F_{0} (r,k)}{\partial \varepsilon _{k}}.
		\end{aligned}
	\end{equation}
\CV The disorder-induced thermoelectric effect has been mentioned in Ref.\cite{Papaj2021} for some specific models. \CIV

	\subsubsection{Derivation of second-order thermal effect}\label{App:Derivation of Second-order  anomalous Thermal effect}
	The thermal magnetization at the edge can be obtained as \cite{Zhang2016}
	\begin{align}
		\mathbf M_{Q}^{\text{edge}} &=-\frac{1}{\hbar }\int _{k}\int _{\varepsilon }( \varepsilon -\mu ) F(r,k ) \Theta ( \varepsilon -\varepsilon _{k}) \mathbf \Omega_k,
	\end{align}
	where $\Theta ( \varepsilon -\varepsilon _{k})$ denotes a step function.
	The magnetization current does not contribute to the net current measured by conventional transport experiments. Therefore, the thermal transport current driven by temperature gradient $\nabla T$  is given by \cite{Zhang2016}
	\begin{equation}\begin{aligned}
			\mathbf J_{Q}^{T} &=\int _{k}( \varepsilon _{k} -\mu ) \dot{\mathbf{{r}}}F(r,k ) -\mathbf\nabla \times \mathbf M_{Q}^{\text{edge}} =\int _{k}( \varepsilon _{k} -\mu ) \dot{\mathbf r} F(r,k ) +\frac{1}{\hbar } \mathbf\nabla \times \int _{k}\int _{\varepsilon }( \varepsilon -\mu ) F(r,k ) \Theta ( \varepsilon -\varepsilon _{k}) \mathbf\Omega_k.
	\end{aligned}\end{equation}
	
	The second-order \textbf{intrinsic} thermal current is
	\begin{equation}
		\begin{aligned}
			\label{eq:j_Q,a}
			J^{T,\text{in}}_{Q,a} =&\kappa _{abc}^{\text{in}} (-\nabla _{b} T)(-\nabla _{c} T)=\int _{k}( \varepsilon _{k} -\mu ) v^{\text{g}}_{k_a}F_2^{(0)}( r,k)+\frac{1}{\hbar } \epsilon_{abd}\nabla_bT  \int _{k}\int _{\varepsilon }( \varepsilon -\mu ) \frac{\partial F_1^{(0)}(r,k )}{\partial T} \Theta ( \varepsilon -\varepsilon _{k}) \Omega_{k_d}\\ =&\tau ^{2}\int _{k} v^{\text{g}}_{k_a}v^{\text{g}}_{k_{b}} v^{\text{g}}_{k_{c}}\left[( \varepsilon _{k} -\mu )^{3}\frac{\partial ^{2} F_{0}(r,k)}{\partial \varepsilon _{k}^{2}} +2( \varepsilon _{k} -\mu )^2\frac{\partial F_{0}(r,k)}{\partial \varepsilon _{k}}\right] \frac{\nabla _{b} T}{T}\frac{\nabla _{c} T}{T}\\&+\frac{1}{\hbar } \epsilon_{abd}\nabla_bT \int_\varepsilon ( \varepsilon -\mu ) \frac{\partial F_1^{(0)}(r,k )}{\partial T}\int _{k} \Theta ( \varepsilon -\varepsilon _{k}) \Omega_{k_d},
		\end{aligned}
	\end{equation}
	when we consider the time-reversal symmetry system, the first term in Eq. (\ref{eq:j_Q,a}) disappears. Substituting Eq. (\ref{eq:F(k)}a) of $F_1^{(0)}$ into the second term of Eq. (\ref{eq:j_Q,a}), the corresponding second-order intrinsic thermal transport coefficient can be derived as \cite{Wang2022}
	
%
%
\begin{equation}
	\begin{aligned}\label{eq:kappa_in}
		{\kappa _{abc}^{\text{in}} =\frac{\tau \epsilon _{abd} }{\hbar T^{2}}\int d\varepsilon ( \varepsilon -\mu )^{3}\frac{\partial f_{0} }{\partial \varepsilon }\int _{{k}} \delta (\varepsilon -\varepsilon _{{k}} )\Omega_{k_d}v^{\text{g}}_{k_{c}}}.
	\end{aligned}
\end{equation}

The second-order thermal response arising from \textbf{side-jump} scattering is given by
\begin{equation}
	\begin{aligned}
		J_{Q,a}^{T,\text{sj}} =&\kappa _{abc}^{\text{sj}}( -\nabla _{b} T)( -\nabla _{c} T) =\int _{k}( \varepsilon _{k} -\mu ) v^{\text{sj}}_{k_a} F_{2}^{(0)} (r,k)+\int _{k}( \varepsilon _{k} -\mu ) v^{\text{g}}_{k_a} G_{2}^{(0)} (r,k)\\
		= &\tau ^{2}\int _{k} v^{\text{sj}}_{k_a}v^{\text{g}}_{k_b}v^{\text{g}}_{k_c}\left[ (\varepsilon _{k} -\mu )^{3}\frac{\partial ^{2} F_{0} (r,k)}{\partial \varepsilon _{k}^{2}} +2(\varepsilon _{k} -\mu )^{2}\frac{\partial F_{0} (r,k)}{\partial \varepsilon _{k}}\right]\frac{\nabla _{c} T}{T}\frac{\nabla _{b} T}{T} \\&- 2\tau ^{2}\int _{k}\int _{k^{\prime }} v^{\text{g}}_{k_a}v^{\text{g}}_{k_c} w_{k^{\prime } k}^{(S)} \delta r_{b,k'k}\left[ 2(\varepsilon _{k} -\mu )^{2}\frac{\partial F_{0} (r,k)}{\partial \varepsilon _{{k}}} +(\varepsilon _{k} -\mu )^{3}\frac{\partial ^{2} F_{0} (r,k)}{\partial \varepsilon _{{k}}^{2}}\right]\frac{\nabla _{c} T}{T}\frac{\nabla _{b} T}{T},
\end{aligned}\end{equation}	
thus,
\begin{equation}
	\begin{aligned}\label{eq:kappasj}
		\kappa _{abc}^{\text{sj}} = -\frac{\tau ^{2}}{T^{2}}\int _{k} (2v^{\text{g}}_{k_a} v^{\text{sj}}_{k_b}-v^{\text{sj}}_{k_a}v^{\text{g}}_{k_b})v^{\text{g}}_{k_c}\left[ 2(\varepsilon _{k} -\mu )^{2}\frac{\partial F_{0} (r,k)}{\partial \varepsilon _{{k}}} +(\varepsilon _{k} -\mu )^{3}\frac{\partial ^{2} F_{0} (r,k)}{\partial \varepsilon _{{k}}^{2}}\right].
\end{aligned}\end{equation}

The second-order thermal effect resulting from \textbf{skew} scattering can be expressed as
\begin{equation}
	\begin{aligned}
		J_{Q,a}^{T,\text{sk}} =&\kappa _{abc}^{\text{sk}}( -\nabla _{b} T)( -\nabla _{c} T) =\int _{k}( \varepsilon _{k} -\mu ) v^{\text{g}}_{k_a}F_{2}^{(1)} (r,k),
\end{aligned}\end{equation}
with
\begin{equation}
	\begin{aligned}\label{eq:kappask}
		\kappa _{abc}^{\text{sk}} =\frac{2\tau ^{3}}{T^{2}}\int _{k}\int _{k'} v^{\text{g}}_{k_a}v^{\text{g}}_{k_b}v^{\text{g}}_{k_c} w_{kk'}^{(A)} (\varepsilon _{k} -\mu )^{3}\frac{\partial ^{2} F_{0} (r,k)}{\partial \varepsilon _{k}^{2}} +\frac{4\tau ^{3}}{T^{2}}\int _{k}\int _{k'} v^{\text{g}}_{k_a}v^{\text{g}}_{k_b}v^{\text{g}}_{k_c} w_{kk'}^{(A)} (\varepsilon _{k} -\mu )^{2}\frac{\partial F_{0} (r,k)}{\partial \varepsilon _{k}}.	
\end{aligned}\end{equation}	
\CV It is worth noting that the thermal effect induced by disorder was also discussed in Ref. \cite{Zhou2022}. The formulas derived and the impurity potential used in this work are different from those in Ref. \cite{Zhou2022}. \CIV

\subsubsection{Second-order transport coefficients}
Without loss of generality, from Eqs. (\ref{eq:sigma_in}),(\ref{eq:sigmasj}),(\ref{eq:sigmask}), we can write the second-order conductivities as,
\begin{equation}\label{A38}
	\begin{aligned}
		\sigma _{abc}^{\text{in}} =&-\tau e^{3}\int d\varepsilon \frac{\epsilon _{abd}  }{\hbar }\int _{k} \Omega_{k_d} v^{\text{g}}_{k_{c}}\frac{\partial f_{0} (k,\varepsilon _{k} )}{\partial \varepsilon _{k}}\left( -\frac{\partial f_{0}}{\partial \varepsilon }\right),\\
		\sigma _{abc}^{\text{sk}} =&{\tau ^{3}e^{3}}\int d\varepsilon \int _{k}\int _{k^{\prime }} \left(\Gamma_{ac}  v^{\text{g}}_{k'_{b}} +v^{\text{g}}_{k'_{a}}\Gamma_{bc}\right) w_{kk^{\prime }}^{(A)}\frac{\partial f_{0} (k,\varepsilon _{k} )}{\partial \varepsilon _{k}}\left( -\frac{\partial f_{0}}{\partial \varepsilon }\right) \\
		& -  \tau ^{3} e^{3}\int d\varepsilon \int _{k}\int _{k^{\prime }}  v^{\text{g}}_{k_{a}} v^{\text{g}}_{k'_{b}} v^{\text{g}}_{k'_{c}} w_{kk^{\prime }}^{(A)}\frac{\partial ^{2} f_{0} (k,\varepsilon _{k} )}{\partial \varepsilon _{k}^{2}}\left( -\frac{\partial f_{0} }{\partial \varepsilon }\right),\\
		\sigma _{abc}^{\text{sj}} =&-\tau ^{2}e^{3}\int d\varepsilon \int _{k}  v^{\text{sj}}_{k_a} v^{\text{g}}_{k_{b}} v^{\text{g}}_{k_{c}}\frac{\partial ^{2} f_{0} (k,\varepsilon _{k} )}{\partial \varepsilon _{k}^{2}}\left( -\frac{\partial f_{0}}{\partial \varepsilon }\right) {+}\tau ^{2}e^{3}\int d\varepsilon \int _{k}  v^{\text{g}}_{k_{a}} \int _{k^{\prime }} w_{kk^{\prime }}^{(S)} v^{\text{g}}_{k'_{c}} \delta r_{b,{k}^{\prime }{k}}\frac{\partial ^{2} f_{0} (k,\varepsilon _{k} )}{\partial \varepsilon _{k}^{2}}\left( -\frac{\partial f_{0}}{\partial \varepsilon }\right)\\
		&-\tau ^{2}e^{3}\int d\varepsilon{}\int _{k}\left( v^{\text{sj}}_{k_a}\Gamma_{bc} {+}v^{\text{sj}}_{k_b}\Gamma_{ac}\right)\frac{\partial f_{0} (k,\varepsilon _{k} )}{\partial \varepsilon _{k}}\left( -\frac{\partial f_{0}}{\partial \varepsilon }\right),
	\end{aligned}
\end{equation}
while from Eqs. (\ref{eq:alpha_in}), (\ref{eq:alphask}), (\ref{eq:alphasj}), the second-order thermoelectric transport coefficients  are
\begin{equation}\label{A39}
	\begin{aligned}
		\alpha _{abc}^{\text{in}} =&-\tau ek_{B}^{2}\int d\varepsilon \left(\frac{\varepsilon -\mu }{k_{B} T}\right)^{2}\frac{ \epsilon _{abd}}{\hbar }\int _{k} v^{\text{g}}_{k_{c}} \Omega_{k_d}\frac{\partial F_{0} (r,k)}{\partial \varepsilon _{k}}\left( -\frac{\partial f_{0}}{\partial \varepsilon }\right),\\
		\alpha _{abc}^{\text{sj}} =&\frac{2\tau ^{2} ek_{B}}{T}\int d\varepsilon \frac{\varepsilon -\mu }{k_{B} T}\int _{k} ( 2 v^{\text{g}}_{k_{a}} v^{\text{sj}}_{k_b} -v^{\text{sj}}_{k_a} v^{\text{g}}_{k_{b}}) v^{\text{g}}_{k_{c}}\frac{\partial F_{0} (r,k)}{\partial \varepsilon _{{k}}}\left( -\frac{\partial f_{0}}{\partial \varepsilon }\right) \\&+\tau ^{2}ek_{B}^{2}\int d\varepsilon \left(\frac{\varepsilon -\mu }{k_{B} T}\right)^{2}\int _{k} ( 2 v^{\text{g}}_{k_{a}} v^{\text{sj}}_{k_b} -v^{\text{sj}}_{k_a} v^{\text{g}}_{k_{b}}) v^{\text{g}}_{k_{c}}\frac{\partial ^{2} F_{0} (r,k)}{\partial \varepsilon _{{k}}^{2}}\left( -\frac{\partial f_{0}}{\partial \varepsilon }\right),\\
		\alpha _{abc}^{\text{sk}} =&-2\tau ^{3}ek_{B}^{2}\int d\varepsilon \left(\frac{\varepsilon -\mu }{k_{B} T}\right)^{2}\int _{k}\int _{k'}  w_{kk'}^{(A)} v^{\text{g}}_{k_{a}} v^{\text{g}}_{k_{b}} v^{\text{g}}_{k_{c}}\frac{\partial ^{2} F_{0} (r,k)}{\partial \varepsilon _{k}^{2}}\left( -\frac{\partial f_{0}}{\partial \varepsilon }\right) \\&-\frac{4\tau ^{3}ek_{B}}{T}\int d\varepsilon \frac{\varepsilon -\mu }{k_{B} T}\int _{k}\int _{k'}  v^{\text{g}}_{k_{a}} v^{\text{g}}_{k_{b}} v^{\text{g}}_{k_{c}} w_{kk'}^{(A)} \frac{\partial F_{0} (r,k)}{\partial \varepsilon _{k}}\left( -\frac{\partial f_{0}}{\partial \varepsilon }\right),
	\end{aligned}
\end{equation}
and from Eqs. (\ref{eq:kappa_in}), (\ref{eq:kappasj}), (\ref{eq:kappask}), the second-order anomalous thermal conductivities
\begin{equation}\label{A40}
	\begin{aligned}
		\kappa _{abc}^{\text{in}}
		=&{-\tau k_B^3T \frac{\epsilon _{abd}}{\hbar }\int d\varepsilon \left(\frac{ \varepsilon  -\mu }{k_BT}\right)^{3} \int _{k}   v^{\text{g}}_{k_{c}}\frac{\partial  F_{0} ( r,k )}{\partial \varepsilon _k}  \Omega _{k_d} \left( -\frac{\partial f_{0}}{\partial \varepsilon }\right)},\\
		\kappa _{abc}^{\text{sj}} =&- 2\tau ^{2} k_{B}^{2}\int d\varepsilon \left(\frac{\varepsilon -\mu }{k_{B} T}\right)^{2}\int _{k} ( 2 v^{\text{g}}_{k_{a}} v^{\text{sj}}_{k_b} -v^{\text{sj}}_{k_a} v^{\text{g}}_{k_{b}}) v^{\text{g}}_{k_{c}}\frac{\partial F_{0} (r,k)}{\partial \varepsilon _{{k}}}\left( -\frac{\partial f_{0}}{\partial \varepsilon }\right) \\
		&-\tau ^{2}k_{B}^{3} T\int d\varepsilon \left(\frac{\varepsilon -\mu }{k_{B} T}\right)^{3}\int _{k} ( 2 v^{\text{g}}_{k_a} v^{\text{sj}}_{k_{b}} -v^{\text{sj}}_{k_a} v^{\text{g}}_{k_{b}}) v^{\text{g}}_{k_{c}}\frac{\partial ^{2} F_{0} (r,k)}{\partial \varepsilon _{{k}}^{2}}\left( -\frac{\partial f_{0}}{\partial \varepsilon }\right),\\
		\kappa _{abc}^{\text{sk}} =&2\tau ^{3}k_{B}^{3} T\int d\varepsilon \left(\frac{\varepsilon -\mu }{k_{B} T}\right)^{3}\int _{k}\int _{k'}  v^{\text{g}}_{k_{a}} v^{\text{g}}_{k_{b}} v^{\text{g}}_{k_{c}} w_{kk'}^{(A)}\frac{\partial ^{2} F_{0} (r,k)}{\partial \varepsilon _{k}^{2}}\left( -\frac{\partial f_{0}}{\partial \varepsilon }\right) \\
		&+4\tau ^{3}k_{B}^{2}\int d\varepsilon \left(\frac{\varepsilon -\mu }{k_{B} T}\right)^{2}\int _{k}\int _{k'}  v^{\text{g}}_{k_{a}} v^{\text{g}}_{k_{b}} v^{\text{g}}_{k_{c}} w_{kk'}^{(A)} \frac{\partial F_{0} (r,k)}{\partial \varepsilon _{k}}\left( -\frac{\partial f_{0}}{\partial \varepsilon }\right).
	\end{aligned}
\end{equation}

\CV \subsubsection{ Kernel representation of the transport coefficients}



Collecting the above results, we obtain the second-order nonlinear transport coefficients summarized in Table \ref{table1} of the main text. To show how to read off the formulas in Table \ref{table1}, we briefly show how the kernel representation is constructed and gives us the explicit expressions in Eqs.~(\ref{A38})-(\ref{A40}). Taking $\kappa^{\rm sk}_{abc}$ as an illustrative example, from Table \ref{table1}, it reads
\begin{equation}\label{A41}
	\kappa^{\rm sk}_{abc}
	=
	k_B^2\left(C^{\rm sk}_{3,2}+k_B T\, C^{\rm sk}_{4,3}\right),
\end{equation}
where the subscripts indicate $(m_1,m_2)=(3,2)$ and $(4,3)$, respectively. Using the general definition of $C^{\rm in/sj/sk}_{m_1,m_2}$ in Eq.~(\ref{eq:Cm1m2}), one finds
\begin{eqnarray}
	C^{\rm sk}_{3,2}
	&=& 
	-\int d\varepsilon
	\mathcal X^{\rm sk}_{3}
	\left(\frac{\varepsilon-\mu}{k_B T}\right)^2
	\frac{\partial f_0}{\partial \varepsilon},
	\qquad
	\mathcal X^{\rm sk}_{3}
	=
	4\tau^3
	\int_{\bm k}\int_{\bm k'}
	v^g_{k_a}v^g_{k_b}v^g_{k_c}
	w^{(A)}_{kk'}
	\frac{\partial f_0}{\partial \varepsilon_k}, 
\label{A42} \\
	C^{\rm sk}_{4,3}
	&=& 
	-\int d\varepsilon
	\mathcal X^{\rm sk}_{4}
	\left(\frac{\varepsilon-\mu}{k_B T}\right)^3
	\frac{\partial f_0}{\partial \varepsilon},
	\qquad
	\mathcal X^{\rm sk}_{4}
	=
	2\tau^3
	\int_{\bm k}\int_{\bm k'}
	v^g_{k_a}v^g_{k_b}v^g_{k_c}
	\,w^{(A)}_{kk'}
	\frac{\partial^2 f_0}{\partial \varepsilon_k^2}.
\label{A43}
\end{eqnarray}
Substituting the Eqs. (\ref{A42}) and (\ref{A43}) into Eq.~(\ref{A41}), we immediately recover the explicit form of $\kappa^{\rm sk}_{abc}$ in Eq. (\ref{A40}).
\CIV

\section{The Surface State of the Topological Insulator}\label{App:The Surface State of the Topological Insulator}
\subsection{Scattering by Coulomb impurities}\label{Append:Scattering by Coulomb}
The Hamiltonian can be written as $H( k) = \boldsymbol d_k \cdot \boldsymbol{\sigma}$, where the $d$-vector can be parameterized in spherical coordinates as
\begin{equation}\label{}
	\begin{aligned}
		\boldsymbol d_k =\left[ -vk_{y} ,vk_{x} ,\frac{\lambda }{2}\left( k_{+}^{3} +k_{-}^{3}\right)\right]  =|\boldsymbol d_k|[ \sin\phi \cos\theta_k ,\sin\phi \sin\theta_k ,\cos\phi ],
	\end{aligned}
\end{equation}
where we introduced the angle $\phi$ defined by
\begin{equation}
	\begin{aligned}
		\cos\phi =\frac{\ d_{k_z}}{|\boldsymbol d_k |} =\frac{\cos( 3\theta_k ) k^{3} \lambda }{\sqrt{ \cos^2( 3\theta_k ) k^{6} \lambda ^{2} +k^{2} v^{2}}}.
	\end{aligned}
\end{equation}
Thus, the normalized wavefunctions of the conduction band and valence band are
\begin{equation}
	\begin{aligned}\label{eq:|k_+>}
		|k_+\rangle =\left( \cos\frac{\phi }{2} ,ie^{i\theta_k } \sin\frac{\phi }{2}\right)^{T}, \ \ \ \ \ \ \ \ |k_-\rangle =\left( -\sin\frac{\phi }{2} ,i\cos\frac{\phi }{2} e^{i\theta_k }\right)^{T}.
	\end{aligned}
\end{equation}
It is noted that $\phi$ depends on $k$ which is fully and carefully taken into account in the following calculations. We may take into account the screened Coulomb interaction  $V( q) =\frac{2\pi \alpha v}{q+q_{TF}} =\frac{V_{c}}{2k_{F}\left| \sin\left(\frac{\theta _{k} -\theta _{k'}}{2}\right)\right| +q_{TF}}$, where $V_c=2\pi \alpha v$,  $q=|  \mathbf k-\mathbf{k'}|$ is the magnitude of the momentum transfer for an electron scattered from an initial state $\mathbf{k}$ to a final state  $\mathbf{k'}$,   and we can get,
\begin{equation}
	\begin{aligned}
		\langle	 k_+|k'_+\rangle
		&=\frac{1}{2}\left[ 1+e^{-i( \theta_k -\theta _{k'})}\right] +\frac{k_{F}^{2} \lambda }{4v}\left[ 1-e^{-i( \theta_k -\theta_{k '})}\right][ \cos( 3\theta_k ) +\cos( 3\theta_{k '})]+o\left( \lambda ^{2}\right).
	\end{aligned}
\end{equation}
The matrix element of the Coulomb interaction is $V_{kk^\prime} = \frac{2 \pi \alpha v}{\lvert {k} - {k}^\prime \rvert + q_{TF}} \langle {k_+} \rvert {k'_+} \rangle$, therefore, we can obtain the following expressions for $w_{kk'}^{( S)}$ and $w_{kk'}^{( A)}$ \cite{Isobe2020},
\begin{equation}
	\begin{aligned}
		w_{kk'}^{( S)} =\frac{2\pi }{\hbar }\left< | V_{kk'}| ^{2}\right> \delta ( \varepsilon _{k} -\varepsilon _{k'}) =\frac{4 \pi ^{3} n_{i} \alpha ^{2} v^{2}}{\hbar \left( 2k_{F}\left| \sin\left(\frac{\theta _{k} -\theta _{k'}}{2}\right)\right| +q_{TF}\right)^{2}}[ 1+\cos( \theta _{k} -\theta _{k'})] \delta ( \varepsilon _{k} -\varepsilon _{k'}),
	\end{aligned}
\end{equation}
and
\begin{equation}
	\begin{aligned}\label{eq:wkk'A}
		w_{kk'}^{( A)}&= -\frac{( 2\pi )^{2}}{\hbar }\int _{q}\text{Im} \langle V_{kk'} V_{k'q} V_{qk} \rangle_{\text{dis}} \delta ( \varepsilon _{k'} -\varepsilon _{k}) \delta ( \varepsilon _{k} -\varepsilon _{q}) \\
		&=-\frac{\pi ^{4} n_{i} \alpha ^{3} \lambda vC_{0}(q_s)\sin( \theta _{k} -\theta _{k'})[\cos( 3\theta _{k}) +\cos( 3\theta _{k'})]}{\hbar\left( 2\left| \sin\left(\frac{\theta _{k} -\theta _{k'}}{2}\right)\right| +q_s\right)}  \delta ( \varepsilon _{k'} -\varepsilon _{k}),	
	\end{aligned}
\end{equation}	
with
\begin{equation}
	\begin{aligned}
		C_{0}(q_s) &=\int \frac{d\theta _{k} d\theta _{k'}}{( 2\pi )^{2}}[ \cos( 3\theta _{k}) +\cos( 3\theta _{k'})] F( \theta _{k} ,\theta _{k'}),\\
		F_{1}( \theta _{k} ,\theta _{k'})=&\int \frac{d\theta _{q}}{2\pi }\frac{4}{\left( 2\left| \sin\left(\frac{\theta _{k'} -\theta _{q}}{2}\right)\right| +{q_s}\right)\left( 2\left| \sin\left(\frac{\theta _{q} -\theta _{k}}{2}\right)\right| +q_s\right)}\\
		\times &\left[ \frac{\cos( 3\theta _{k}) \sin( \theta _{q} -\theta _{k'}) + \cos( 3\theta _{k'}) \sin( \theta _{k} -\theta _{q})}{\sin( \theta _{k} -\theta _{k'})} -\cos( 3\theta _{q})\right]. 	
	\end{aligned}
\end{equation}

\subsection{Calculation of the coordinate shift and side-jump velocity}\label{Append:the Coordinate Shift and Side-Jump Velocity}
Assuming that the Fermi level lies within the conduction band, the coordinate shift along the $x$-direction due to scattering is given by the expression derived in Ref. \cite{Sinitsyn2006}:
\begin{equation}
	\begin{aligned}
		\delta \mathbf{r}_{{kk}^{'}} =\langle k_+|i\frac{\partial }{\partial \mathbf k} |k_+\rangle -\langle k'_+|i\frac{\partial }{\partial \mathbf k'} |k'_+\rangle -\left( \frac{\partial }{\partial \mathbf k} +\frac{\partial }{\partial \mathbf k'} \right) \mathrm{arg} \langle k_+|k'_+\rangle .
	\end{aligned}
\end{equation}

Expanding the upper band eigenstate in Eq. (\ref{eq:|k_+>}) to linear order in the warping parameter $\lambda$, the wavefunction takes the approximate form \cite{Makushko2024},
\begin{equation}
	\begin{aligned}
		|k_+\rangle =\begin{pmatrix}
			\cos\frac{\phi }{2}\\
			i\sin\frac{\phi }{2} e^{i\theta _{k}}
		\end{pmatrix}  \approx \frac{1}{\sqrt{2}}\begin{pmatrix}
			1+M( k)\\
			i[1-M( k)] e^{i\theta _{k}}
		\end{pmatrix},
	\end{aligned}
\end{equation}
with $M( k) =\frac{\lambda k^{3}\cos 3\theta _{k}}{2\varepsilon_ {k}} \approx \frac{\lambda k^{2}\cos 3\theta _{k}}{2v}$.


Meanwhile, we make use of the following identity from Ref. \cite{Du2019}
\begin{equation}
	\begin{aligned}
		\frac{\partial }{\partial x}\mathrm{arg} \ A=\frac{i}{|A|^{2}}\left(\frac{1}{2} \frac{\partial }{\partial x} |A|^{2} -A^{*} \frac{\partial }{\partial x} A\right),
	\end{aligned}
\end{equation}
the contributions to the coordinate shift can be evaluated as
\begin{equation}
	\begin{aligned}
		&\langle k_{+} |i\frac{\partial }{\partial k_{x}}  |k_{+} \rangle -{\frac{\partial }{\partial k_{x}}  \arg} [\langle k_{+} |k '_{+} \rangle ]\\
		&=\frac{1}{k[ 1+\cos( \theta _{k '} -\theta _{k})]}\left[ M( k')\sin \theta _{k} -\sin \theta _{k '} M( k) +\frac{3\lambda k^{2} \sin( \theta _{k '} -\theta _{k})}{2v}\cos 2\theta _{k}\right],
	\end{aligned}
\end{equation}	
and
\begin{equation}
	\begin{aligned}
		&-\langle k '_{+} |i\frac{\partial }{\partial k_{x}} |k '_{+} \rangle -{\frac{\partial }{\partial k'_{x}} \arg} [\langle k_{+} |k '_{+} \rangle ]\\
		&=\frac{1}{k'[ 1+\cos( \theta _{k '} -\theta _{k})]}\left[ M( k')\sin \theta _{k} -\sin \theta _{k '} M( k) +\frac{3\lambda k^{\prime 2} \sin( \theta _{k '} -\theta _{k})}{2v}\cos 2\theta _{k '}\right].
	\end{aligned}
\end{equation}	

Finally, the coordinate shift in the $x$-direction for the upper band is expressed compactly as,
\begin{equation}
	\begin{aligned}\label{eq:deltakk'}
		&	\delta r_{x,kk'} =\langle k_{+} |i\frac{\partial }{\partial k_{x}}|k_{+} \rangle -\langle k '_{+}|i\frac{\partial }{\partial k'_{x}}|k '_{+}\rangle -\left( \frac{\partial }{\partial k_{x}} +\frac{\partial }{\partial k'_{x}} \right) {\arg} [\langle k_{+ } |k '_{+} \rangle ]\\
		=&\frac{\lambda }{2v[ 1+\cos( \theta _{k '} -\theta _{k})]}\Big[ \frac{k^{\prime 2}}{k}\cos 3\theta _{k '}\sin \theta _{k} -\sin \theta _{k '} k\cos 3\theta _{k} +3 k \sin( \theta _{k '} -\theta _{k})\cos 2\theta _{k} +k'\cos 3\theta _{k '}\sin \theta _{k} \\&-\frac{k^{2}}{k'}\cos 3\theta _{k}\sin \theta _{k '} +3 k' \sin( \theta _{k '} -\theta _{k})\cos 2\theta _{k '}\Big],
	\end{aligned}
\end{equation}	
and we note that $\delta{r}_{x,kk'}=-\delta{r}_{x,k'k}$ \cite{Ortix2021}.

The side-jump velocity arises from the coordinate shift \( \delta \mathbf{r}_{k'k} \) occurring during scattering. It can be expressed as $\mathbf{v}^{\mathrm{sj}}_k = \int _{k'} w_{kk'}^{(S)} \delta \mathbf{r}_{k'k}$, and the corresponding side-jump velocity in the $x$ component reads as follows
\begin{equation}
	\begin{aligned}
		v^{\text{sj}}_{k_x} =\int_{k'} w_{kk'}^{( S)} \delta r_{x,k'k}
		=&-\frac{2n_{i} \pi ^{3} \alpha ^{2} \lambda }{ \hbar }\int \frac{k^{2} d\theta _{k'}}{( 2\pi )^{2}}\frac{1}{\left( 2k_{F}\left| \sin\left(\frac{\theta _{k} -\theta _{k'}}{2}\right)\right| +q_{TF}\right)^{2}}\\
		\times &[ 2\cos 3\theta _{k '}\sin \theta _{k} -2\sin \theta _{k '}\cos 3\theta _{k} +3\sin( \theta _{k '} -\theta _{k})\cos 2\theta _{k} +3\sin( \theta _{k '} -\theta _{k})\cos 2\theta _{k '}],
	\end{aligned}
	\label{vsj}
\end{equation}
in which we simplify the $\delta( \varepsilon _{k} -\varepsilon _{k'})$ function as follows, neglecting the higher order terms of $\lambda$,
\begin{equation}
	\delta ( \varepsilon _{k} -\varepsilon _{k'})  \approx \frac{\delta ( k-k')}{v }\left[ 1-\frac{3( \cos( 3\theta_k ))^{2} k^{5} \lambda ^{2}}{k v^{2}}\right]	\approx  \frac{\delta ( k-k')}{v }.
\end{equation}

\subsection{Second-order transport coefficients of surface states of topological insulators}\label{Append:Second-order transport coefficients}
\subsubsection{Second-order \CV electrical coefficients \CIV contributed by side-jump and skew scattering}

For the sj contribution from Eq. (\ref{eq:sigmasj}), the second-order \CV electrical coefficients \CIV in the $xxy$ component can be expressed as follows,

\begin{equation}\begin{aligned}
		\sigma _{xxy}^{\text{sj}} =&-\tau ^{2} e^{3}\int _{k} v^{\text{sj}}_{k_{x}} v^{\text{g}}_{k_{x}} v^{\text{g}}_{k_y}\frac{\partial ^{2} f_{0} (k,\varepsilon _{k} )}{\partial \varepsilon _{k}^{2}}+\tau ^{2} e^{3}\int _{k} v^{\text{g}}_{k_{x}}\int _{k^{\prime }} w_{kk^{\prime }}^{(S)} v^{\text{g}}_{k'_{y}} \delta r_{x,{k}^{\prime }{k}}\frac{\partial ^{2} f_{0} (k,\varepsilon _{k} )}{\partial \varepsilon _{k}^{2}}-\frac{2\tau ^{2} e^{3}}{\hbar }\int _{k} v^{\text{sj}}_{k_{x}} \frac{\partial v^{\text{g}}_{k_{x}}}{\partial k_{y}}\frac{\partial f_{0} (k,\varepsilon _{k} )}{\partial \varepsilon _{k}}\label{sigmaxxysj}
\end{aligned}\end{equation}

using the Sommerfeld expansion and substituting Eq. (\ref{vsj}) into Eq. (\ref{sigmaxxysj}), we derive
\begin{equation}
	\begin{aligned}
		\sigma _{xxy}^{\text{sj}}& =-\frac{e^{3} \tau ^{2} n_{i} \pi \alpha ^{2} \lambda }{2\hbar ^{3}} [{3}Q_{1}(q_s){+} Q_{1}'(q_s)]
		\approx -\frac{{3}e^{3} \tau ^{2}\lambda n_{i}\alpha ^{2} \pi   }{2\hbar ^{3}} Q_{1}(q_s),
	\end{aligned}
\end{equation}
where
\begin{equation}
	\begin{aligned}
		Q_{1}(q_s) =&\int \frac{d\theta_k d\theta_{k'}}{( 2\pi )^{2}}\frac{-\cos\theta_k \sin\theta_k }{\left( 2\left| \sin\left(\frac{\theta _{k} -\theta _{k'}}{2}\right)\right| +q_s\right)^{2}}\\
		\times &[ 2\cos 3\theta _{k '}\sin \theta _{k} -2\sin \theta _{k '}\cos 3\theta _{k} +3\sin( \theta _{k '} -\theta _{k})\cos 2\theta _{k} +3\sin( \theta _{k '} -\theta _{k})\cos 2\theta _{k '}],
	\end{aligned}
\end{equation}
and
\begin{equation}
	\begin{aligned}
		Q_{1}'( q_{s}) =&\int \frac{d\theta _{k} d\theta _{k'}}{( 2\pi )^{2}} \frac{\cos\theta _{k} \sin\theta _{k'}}{\left( 2\left| \sin\left(\frac{\theta _{k} -\theta _{k'}}{2}\right)\right| +q_{s}\right)^{2}}		\\
		\times &[ 2\cos 3\theta _{k'}\sin \theta _{k} -2\sin \theta _{k'}\cos 3\theta _{k} +3\sin( \theta _{k'} -\theta _{k})\cos 2\theta _{k} +3\sin( \theta _{k'} -\theta _{k})\cos 2\theta _{k'}],
	\end{aligned}
\end{equation}
and we note that, when $q_s\to 0$, $Q_1=0.1251$ which can be seen in Fig. \ref{Fig-A} (b), $Q_1' = 2.49*10^{-4}$ (not shown in Fig. (\ref{Fig-A})).

From Eq. (\ref{eq:sigmask}), we obtain the expression of the second-order Hall coefficient using the Sommerfeld expansion $\sigma_{xxy}^{\text{sk}}$,
\begin{equation}
	\begin{aligned}
		\sigma _{xxy}^{\text{sk}}
		&=\frac{2e^{3} \tau ^{3}}{\hbar }\int _{k}\int _{k^{\prime }}\frac{\partial v^{\text{g}}_{k_x}}{\partial k_{y}} v^{\text{g}}_{k'_{x}} w_{kk^{\prime }}^{(A)}\frac{\partial f_{0} (k,\varepsilon _{k} )}{\partial \varepsilon _{k}} -e^{3} \tau ^{3}\int _{k}\int _{k^{\prime }} v^{\text{g}}_{k_x} v^{\text{g}}_{k'_{x}} v^{\text{g}}_{k'_{y}} w_{kk^{\prime }}^{(A)}\frac{\partial ^{2} f_{0} (k,\varepsilon _{k} )}{\partial \varepsilon _{k}^{2}}=\frac{e^{3} \tau ^{3} \lambda n{_{i}} \alpha ^{3} \pi ^{2}  \varepsilon _{F}}{\hbar ^{4}} Q_{2}( q_{s}),
	\end{aligned}
\end{equation}
where we have $\frac{\partial v^{\text{g}}_{k_{x}}}{\partial k_{y}} =\ \left( \sin\theta _{k}\frac{\partial }{\partial k} +\frac{\cos\theta _{k}}{k}\frac{\partial }{\partial \theta _{k}}\right)\frac{v \cos\theta _{k}}{\hbar } =-\frac{\cos\theta _{k}}{k}\frac{v\sin\theta _{k}}{\hbar }$, and
\begin{equation}
	Q_{2}( q_{s}) =C_{0}(q_s)\int \frac{d\theta _{k} d \theta _{k'}}{( 2\pi )^{2}}\frac{\sin( \theta _{k} -\theta _{k'})}{2\left| \sin\left(\frac{\theta -\theta_{k'} }{2}\right)\right| +q_{s}} [ \cos( 3\theta _{k}) +\cos( 3\theta _{k'})] \cos\theta _{k} \cos \theta _{k'} \sin \theta _{k'},
\end{equation}
we note that when $q_s\to 0$, the integrating factor $Q_2=0.037$ as shown in Fig. \ref{Fig-A} (c).

\begin{figure}[tb]
	\centering
	\includegraphics[width=1\linewidth]{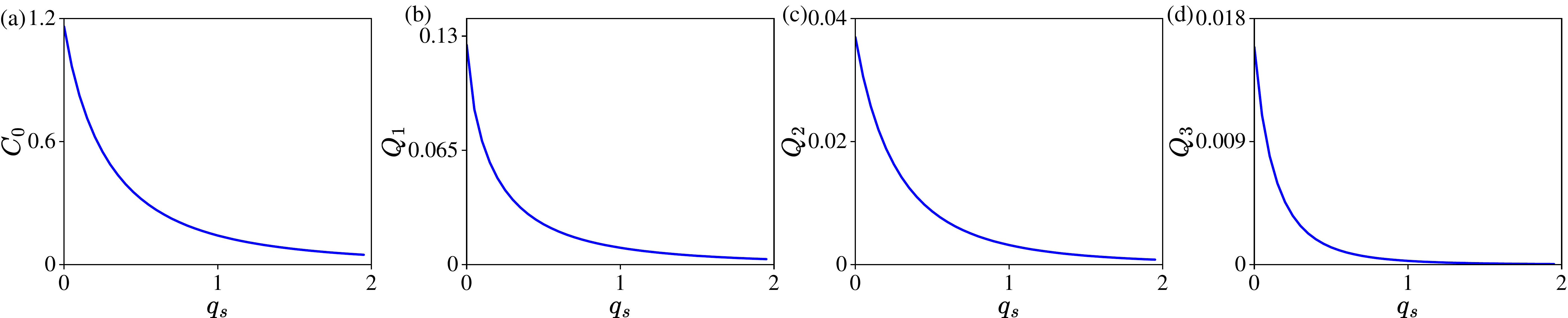}
	\caption{Dependence of the numerical coefficients on the Coulomb screening strength $q_s$. (a)-(d) show the variation of the coefficients  $Q_1$, $Q_2$, $Q_3$ and $P$, respectively, as functions of $q_s$.}
	\label{Fig-A}
\end{figure}

\subsubsection{Second-order thermoelectric coefficients contributed by side-jump and skew scattering}

From Eq. (\ref{eq:alphasj}), we derive the second-order thermoelectric coefficient associated with the $xxy$ component arising from the contribution of \text{sj} as follows,
\begin{equation}	\begin{aligned}
		\alpha _{xxy}^{\text{sj}} &=\frac{\tau ^{2} e}{T^{2}}\int _{k} v^{\text{g}}_{k_x} v^{\text{sj}}_{k_x}v^{\text{g}}_{k_y}\left[ 2(\varepsilon _{k} -\mu )\frac{\partial F_{0}(r,k)}{\partial \varepsilon _{{k}}} +(\varepsilon _{k} -\mu )^{2}\frac{\partial ^{2} F_{0}(r,k)}{\partial \varepsilon _{{k}}^{2}}\right] =\frac{ek_{B}^{2} \tau ^{2} \lambda n_{i} \alpha ^{2}  \pi ^{3} }{6\hbar ^{3}} Q_{1}(q_s),
\end{aligned}\end{equation}
as seen in Fig. \ref{Fig-A} (b), when $q_s\to 0$, $Q_1=0.1251$.

Similarly, from Eq. (\ref{eq:alphask}), we can express the second-order \CV thermoelectric coefficients \CIV in the $xxy$ component contributed by \text{sk},
\begin{equation}
	\begin{aligned}
		\alpha _{xxy}^{\text{sk}} &=-\frac{2e\tau ^{3}}{T^{2}}\int _{k}\int _{k'} w_{kk'}^{(A)} v^{\text{g}}_{k_x} v^{\text{g}}_{k_x} v^{\text{g}}_{k_y} (\varepsilon _{k} -\mu )^{2}\frac{\partial ^{2} F_{0}( r,k)}{\partial \varepsilon _{k}^{2}} -\frac{4e\tau ^{3}}{T^{2}}\int _{k}\int _{k'} w_{kk'}^{(A)} v^{\text{g}}_{k_x} v^{\text{g}}_{k_x} v^{\text{g}}_{k_y} (\varepsilon _{k} -\mu )\frac{\partial F_{0}( r,k)}{\partial \varepsilon _{k}} \\& =\frac{e k_{B}^{2} \tau ^{3} \lambda n_{i} \alpha ^{3} \pi ^{4} \varepsilon _{F}}{3\hbar ^{4}}Q_{3}(q_s),
	\end{aligned}
\end{equation}
with
\begin{equation}
	\begin{aligned}
		Q_{3}(q_s) =C_0(q_s)\int \frac{d\theta_k d\theta_{k'}}{( 2\pi )^{2}}\frac{\sin( \theta _{k} -\theta_{k'})[ \cos( 3\theta _{k}) +\cos( 3\theta _{k'})] \ }{\left( 2\left| \sin\left(\frac{\theta _{k} -\theta _{k'}}{2}\right)\right| +q_{s}\right)} \cos\theta _{k} \cos\theta _{k} \sin\theta _{k},
	\end{aligned}
\end{equation}
as seen in Fig. \ref{Fig-A} (d), when $q_s\to 0$, $Q_3=0.0159$.

\subsubsection{Second-order thermal \CV coefficients\CIV contributed by side-jump and skew scattering}
From Eq. (\ref{eq:kappasj}), the second-order thermal \CV coefficients \CIV contributed by side jumping in $xxy$ direction is given by
\begin{equation}
	\begin{aligned}
		\kappa _{xxy}^{\text{sj}} &=-\frac{\tau ^{2}}{T^{2}}\int _{k} v^{\text{g}}_{k_x} v^{\text{sj}}_{k_x}v^{\text{g}}_{k_y}\left[ 2(\varepsilon _{k} -\mu )^{2}\frac{\partial F_{0}( r,k)}{\partial \varepsilon _{{k}}} +(\varepsilon _{k} -\mu )^{3}\frac{\partial ^{2} F_{0}( r,k)}{\partial \varepsilon _{{k}}^{2}}\right]\\
		&=-  \frac{k_{B}^{2}\tau ^{2}\lambda n_{i} \alpha ^{2} \pi ^{3}  \varepsilon _{F}}{6\hbar ^{3}} Q_{1}(q_s).
	\end{aligned}
\end{equation}

From Eq. (\ref{eq:kappask}), the second-order thermal \CV coefficients \CIV contributed by skew scattering in $xxy$ direction is given by
\begin{equation}
	\begin{aligned}
		\kappa _{xxy}^{\text{sk}} &=\frac{4\tau ^{3}}{T^{2}}\int _{k}\int _{k^{\prime }} v^{\text{g}}_{k_x} v^{\text{g}}_{k_x} v^{\text{g}}_{k_y} w_{k k'}^{(A)} (\varepsilon _{k} -\mu )^{2}\frac{\partial F_{0}( r,k)}{\partial \varepsilon _{{k}}}+\frac{2\tau ^{3}}{T^{2}}\int _{k}\int _{k^{\prime }} v^{\text{g}}_{k_x} v^{\text{g}}_{k_x} v^{\text{g}}_{k_y} w_{k k'}^{(A)} (\varepsilon _{k} -\mu )^{3}\frac{\partial ^{2} F_{0}( r,k)}{\partial \varepsilon _{{k}}^{2}}\\
		&=-\frac{k_{B}^{2}\tau ^{3} \lambda n_{i} \alpha ^{3} \pi ^{4} \varepsilon _{F}^{2}}{6\hbar ^{4}} Q_{3}(q_s).
	\end{aligned}
\end{equation}
\begin{table*}[t]
		\renewcommand\arraystretch{2}
		\centering
		\caption{The anomalous transport coefficients with TRS, where $\sigma_{xxy}=\frac{e^3}{2\hbar^4 }\tilde{\sigma}_{xxy}$, $\alpha_{xxy}=\frac{e k_B^2 }{6\hbar^4}\tilde{\alpha}_{xxy}$, $\kappa_{xxy}=\frac{k_B^2 }{6\hbar^4} \tilde{\kappa}_{xxy}$, and $n_i$ is the impurity concentration, $Q_{1,2,3}$  are numerical factors.}
		\begin{ruledtabular}
			\begin{tabular}{ccc}
				&	side-jump & skew scattering  \\
				\hline
				$\tilde{\sigma}_{xxy}$ & $-{3} {\tau^2\lambda n_{i}  \alpha ^{2}\pi \hbar }Q_{1}$
				& $ {2\tau^3\lambda n_{i}  \alpha ^{3}\pi^2 \varepsilon_F}Q_2$\\
				$\tilde{\alpha}_{xxy}$  &$ {\tau^2\lambda n_{i}  \alpha ^{2}\pi^3 \hbar} Q_{1}$ &$ { 2\tau^3\lambda n_{i}  \alpha ^{3}\pi^4 \varepsilon_F}Q_3$
				\\
				$\tilde{\kappa}_{xxy}$ &$-{ \tau^2\lambda n_{i}  \alpha ^{2}\pi^3 \hbar\varepsilon_F} Q_{1}$
				& $-\tau^3\lambda n_{i}  \alpha ^{3}\pi^4 \varepsilon_F^2 Q_3$ \\
			\end{tabular}
		\end{ruledtabular}
		\label{tableA1}
	\end{table*}

	\subsection{Fundamental relations in surface states of topological insulators}\label{Append:Fundamental relations in surface states of topological insulators }
\CV	\subsubsection{Generalized Mott relation and Wiedemann Franz law forms}

In this subsection, we restrict ourselves to the disorder-induced contributions. From Eq.~(\ref{eq:Cm1m2}) and Table~\ref{table1}, the second-order thermoelectric and thermal coefficients can be rewritten in the generalized Mott relation and WF law forms
\begin{align}
	\alpha^{\eta}_{abc} &= L\mathcal{M}^{\eta}_{abc}\sigma^{\eta}_{abc}, \label{eq:general_Mott}\\
	\frac{\partial \kappa^{\eta}_{abc}}{\partial \varepsilon_F} &= \frac{L}{e}\mathcal{W}^{\eta}_{abc}\sigma^{\eta}_{abc}, \label{eq:general_WF}
\end{align}
where $\eta\in\{\mathrm{sj},\mathrm{sk}\}$ and $L=\pi^2 k_B^2/(3e^2)$ is the Lorentz number. The coefficients $\mathcal{M}^{\eta}_{abc}$ and $\mathcal{W}^{\eta}_{abc}$ are in general, but are determined by the band structure and impurity model through
\begin{align}
	\mathcal{M}^{\mathrm{sj}}_{abc}
	&=
	-\frac{3}{\pi^2}
	\frac{\dfrac{1}{k_B T}C^{\mathrm{sj}}_{4,1}+C^{\mathrm{sj}}_{5,2}}
	{C^{\mathrm{sj}}_{1,0}+C^{\mathrm{sj}}_{2,0}-C^{\mathrm{sj}}_{3,0}}, \label{eq:general_M_sj}\\
	\mathcal{M}^{\mathrm{sk}}_{abc}
	&=
	-\frac{3}{\pi^2}
	\frac{\dfrac{1}{k_B T}C^{\mathrm{sk}}_{3,1}+C^{\mathrm{sk}}_{4,2}}
	{C^{\mathrm{sk}}_{1,0}-C^{\mathrm{sk}}_{2,0}}, \label{eq:general_M_sk}\\
	\mathcal{W}^{\mathrm{sj}}_{abc}
	&=
	\frac{3}{\pi^2}
	\frac{\partial_{\varepsilon_F}\!\left(C^{\mathrm{sj}}_{4,2}+k_B T\,C^{\mathrm{sj}}_{5,3}\right)}
	{C^{\mathrm{sj}}_{1,0}+C^{\mathrm{sj}}_{2,0}-C^{\mathrm{sj}}_{3,0}}, \label{eq:general_W_sj}\\
	\mathcal{W}^{\mathrm{sk}}_{abc}
	&=
	\frac{3}{\pi^2}
	\frac{\partial_{\varepsilon_F}\!\left(C^{\mathrm{sk}}_{3,2}+k_B T\,C^{\mathrm{sk}}_{4,3}\right)}
	{C^{\mathrm{sk}}_{1,0}-C^{\mathrm{sk}}_{2,0}}. \label{eq:general_W_sk}
\end{align}
Therefore, within the general semiclassical framework, the disorder-induced second-order Mott relation and WF law can be written in a unified form, while their coefficients remain model-dependent in general. For the surface states of TIs with screened Coulomb impurities considered below, these coefficients reduce to the simple results presented in the following subsection.
\CIV

	\subsubsection{Fundamental relations in surface states of topological insulators}
	Based on the combined results above, the second-order electrical, thermoelectric, and thermal transport coefficients arising from the contributions of sk
	and sj are compiled in Table \ref{tableA1}. Furthermore, these coefficients provide the basis for obtaining the second-order Mott relation and WF law results induced by disorder detailed in Table II of the main text.
	\begin{figure}[h]
		\centering
		\includegraphics[width=0.5\linewidth]{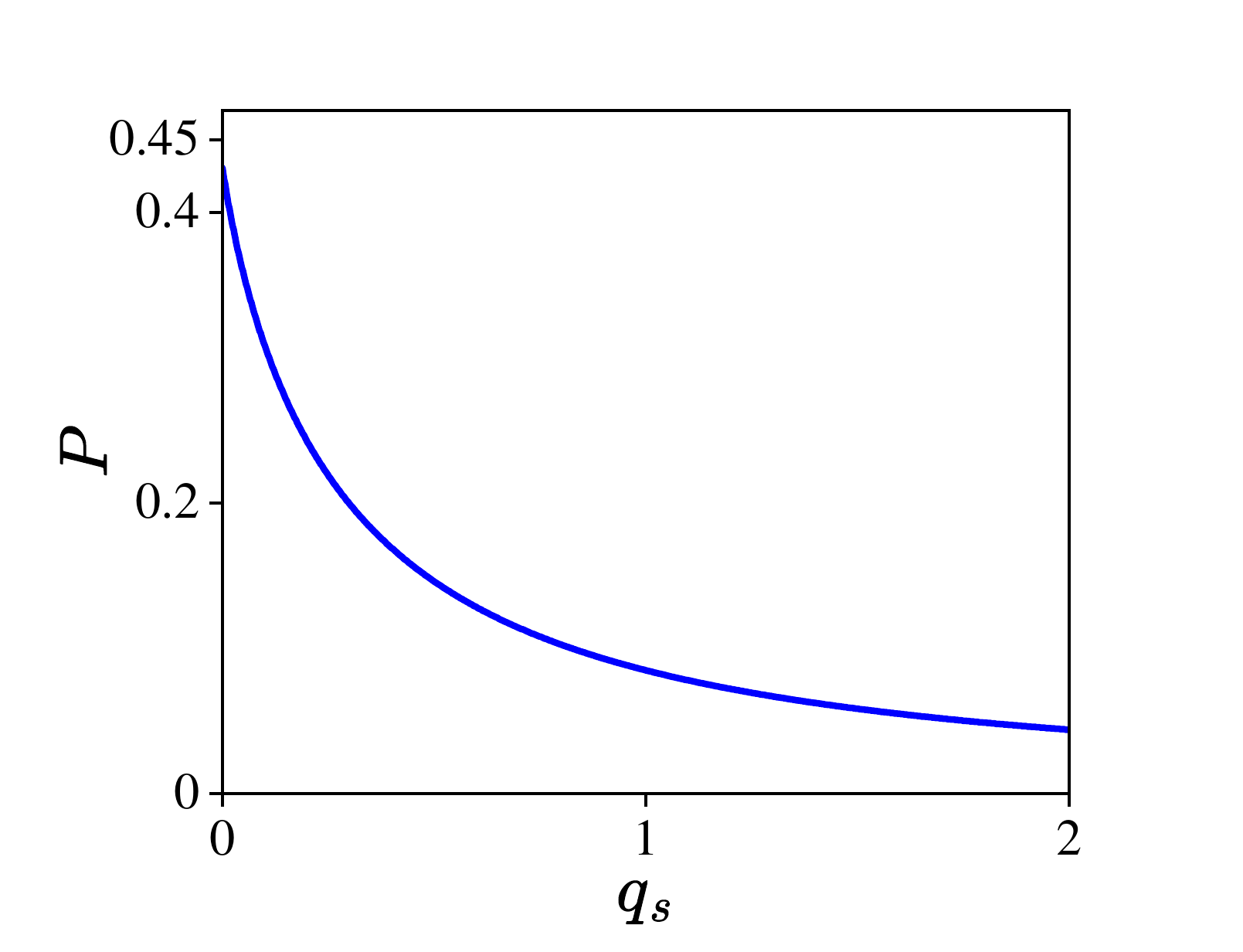}
		\caption{ Dependence of the dimensionless quantity $P$ on the Coulomb screening strength $q_s$.}
		\label{Fig-A2}
	\end{figure}
	
	For skew scattering contribution, we can have  the Mott relation and WF law
	\begin{equation}
		\begin{aligned}
			\alpha _{xxy}^{\text{sk}} =LP\sigma _{xxy}^{\text{sk}},
			\quad   \frac{\partial \kappa _{xxy}^{\text{sk}}}{\partial \varepsilon _{F}} =-\frac{L}{e} P \sigma _{xxy}^{\text{sk}},
		\end{aligned}
	\end{equation}
	with $P={Q_3}/{Q_2}$ is dimensionless quantity determined by the  strength of Coulomb interaction $q_s$, as seen in Fig. \ref{Fig-A2}, reflecting the skew scattering contribution $\alpha_{xxy}^{\text{sk}}/\sigma_{xxy}^{\text{sk}}$ exhibits a diminishing trend with increasing $q_s$. 
	
	For side-jumping term, we have
	\begin{equation}
		\begin{aligned}
			\alpha _{xxy}^{\mathrm{sj}}=- \frac{1}{3} L\sigma _{xxy}^{\mathrm{sj}}, \quad \frac{\partial \kappa _{xxy}^{\text{sj}}}{\partial \varepsilon _{F}} =\frac{L}{3e} \sigma _{xxy}^{\text{sj}},
		\end{aligned}
	\end{equation}
	which are independent of Coulomb interaction.
\end{widetext}


\begin{thebibliography}{71}%
	\makeatletter
	\providecommand \@ifxundefined [1]{%
		\@ifx{#1\undefined}
	}%
	\providecommand \@ifnum [1]{%
		\ifnum #1\expandafter \@firstoftwo
		\else \expandafter \@secondoftwo
		\fi
	}%
	\providecommand \@ifx [1]{%
		\ifx #1\expandafter \@firstoftwo
		\else \expandafter \@secondoftwo
		\fi
	}%
	\providecommand \natexlab [1]{#1}%
	\providecommand \enquote  [1]{``#1''}%
	\providecommand \bibnamefont  [1]{#1}%
	\providecommand \bibfnamefont [1]{#1}%
	\providecommand \citenamefont [1]{#1}%
	\providecommand \href@noop [0]{\@secondoftwo}%
	\providecommand \href [0]{\begingroup \@sanitize@url \@href}%
	\providecommand \@href[1]{\@@startlink{#1}\@@href}%
	\providecommand \@@href[1]{\endgroup#1\@@endlink}%
	\providecommand \@sanitize@url [0]{\catcode `\\12\catcode `\$12\catcode
		`\&12\catcode `\#12\catcode `\^12\catcode `\_12\catcode `\%12\relax}%
	\providecommand \@@startlink[1]{}%
	\providecommand \@@endlink[0]{}%
	\providecommand \url  [0]{\begingroup\@sanitize@url \@url }%
	\providecommand \@url [1]{\endgroup\@href {#1}{\urlprefix }}%
	\providecommand \urlprefix  [0]{URL }%
	\providecommand \Eprint [0]{\href }%
	\providecommand \doibase [0]{https://doi.org/}%
	\providecommand \selectlanguage [0]{\@gobble}%
	\providecommand \bibinfo  [0]{\@secondoftwo}%
	\providecommand \bibfield  [0]{\@secondoftwo}%
	\providecommand \translation [1]{[#1]}%
	\providecommand \BibitemOpen [0]{}%
	\providecommand \bibitemStop [0]{}%
	\providecommand \bibitemNoStop [0]{.\EOS\space}%
	\providecommand \EOS [0]{\spacefactor3000\relax}%
	\providecommand \BibitemShut  [1]{\csname bibitem#1\endcsname}%
	\let\auto@bib@innerbib\@empty
	\bibitem [{\citenamefont {Nagaosa}\ \emph {et~al.}(2010)\citenamefont
		{Nagaosa}, \citenamefont {Sinova}, \citenamefont {Onoda}, \citenamefont
		{MacDonald},\ and\ \citenamefont {Ong}}]{Nagaosa2010}%
	\BibitemOpen
	\bibfield  {author} {\bibinfo {author} {\bibfnamefont {N.}~\bibnamefont
			{Nagaosa}}, \bibinfo {author} {\bibfnamefont {J.}~\bibnamefont {Sinova}},
		\bibinfo {author} {\bibfnamefont {S.}~\bibnamefont {Onoda}}, \bibinfo
		{author} {\bibfnamefont {A.~H.}\ \bibnamefont {MacDonald}},\ and\ \bibinfo
		{author} {\bibfnamefont {N.~P.}\ \bibnamefont {Ong}},\ }\bibfield  {title}
	{\bibinfo {title} {Anomalous {H}all effect},\ }\href
	{https://doi.org/10.1103/revmodphys.82.1539} {\bibfield  {journal} {\bibinfo
			{journal} {Rev. Modern Phys.}\ }\textbf {\bibinfo {volume} {82}},\ \bibinfo
		{pages} {1539} (\bibinfo {year} {2010})}\BibitemShut {NoStop}%
	\bibitem [{\citenamefont {Xiao}\ \emph {et~al.}(2006)\citenamefont {Xiao},
		\citenamefont {Yao}, \citenamefont {Fang},\ and\ \citenamefont
		{Niu}}]{Xiao2006}%
	\BibitemOpen
	\bibfield  {author} {\bibinfo {author} {\bibfnamefont {D.}~\bibnamefont
			{Xiao}}, \bibinfo {author} {\bibfnamefont {Y.}~\bibnamefont {Yao}}, \bibinfo
		{author} {\bibfnamefont {Z.}~\bibnamefont {Fang}},\ and\ \bibinfo {author}
		{\bibfnamefont {Q.}~\bibnamefont {Niu}},\ }\bibfield  {title} {\bibinfo
		{title} {Berry-phase effect in anomalous thermoelectric transport},\ }\href
	{https://doi.org/10.1103/physrevlett.97.026603} {\bibfield  {journal}
		{\bibinfo  {journal} {Phys. Rev. Lett.}\ }\textbf {\bibinfo {volume} {97}},\
		\bibinfo {pages} {026603} (\bibinfo {year} {2006})}\BibitemShut {NoStop}%
	\bibitem [{\citenamefont {Bergman}\ and\ \citenamefont
		{Oganesyan}(2010)}]{Bergman2010}%
	\BibitemOpen
	\bibfield  {author} {\bibinfo {author} {\bibfnamefont {D.~L.}\ \bibnamefont
			{Bergman}}\ and\ \bibinfo {author} {\bibfnamefont {V.}~\bibnamefont
			{Oganesyan}},\ }\bibfield  {title} {\bibinfo {title} {Theory of
			dissipationless {N}ernst effects},\ }\href
	{https://doi.org/10.1103/physrevlett.104.066601} {\bibfield  {journal}
		{\bibinfo  {journal} {Phys. Rev. Lett.}\ }\textbf {\bibinfo {volume} {104}},\
		\bibinfo {pages} {066601} (\bibinfo {year} {2010})}\BibitemShut {NoStop}%
	\bibitem [{\citenamefont {Zhang}\ \emph {et~al.}(2008)\citenamefont {Zhang},
		\citenamefont {Tewari}, \citenamefont {Yakovenko},\ and\ \citenamefont
		{Das~Sarma}}]{Zhang2008}%
	\BibitemOpen
	\bibfield  {author} {\bibinfo {author} {\bibfnamefont {C.}~\bibnamefont
			{Zhang}}, \bibinfo {author} {\bibfnamefont {S.}~\bibnamefont {Tewari}},
		\bibinfo {author} {\bibfnamefont {V.~M.}\ \bibnamefont {Yakovenko}},\ and\
		\bibinfo {author} {\bibfnamefont {S.}~\bibnamefont {Das~Sarma}},\ }\bibfield
	{title} {\bibinfo {title} {Anomalous {N}ernst effect from a
			chirald-density-wave state in underdoped cuprate superconductors},\ }\href
	{https://doi.org/10.1103/physrevb.78.174508} {\bibfield  {journal} {\bibinfo
			{journal} {Phys. Rev. B}\ }\textbf {\bibinfo {volume} {78}},\ \bibinfo
		{pages} {174508} (\bibinfo {year} {2008})}\BibitemShut {NoStop}%
	\bibitem [{\citenamefont {Zhang}\ \emph {et~al.}(2009)\citenamefont {Zhang},
		\citenamefont {Tewari},\ and\ \citenamefont {Das~Sarma}}]{Zhang2009}%
	\BibitemOpen
	\bibfield  {author} {\bibinfo {author} {\bibfnamefont {C.}~\bibnamefont
			{Zhang}}, \bibinfo {author} {\bibfnamefont {S.}~\bibnamefont {Tewari}},\ and\
		\bibinfo {author} {\bibfnamefont {S.}~\bibnamefont {Das~Sarma}},\ }\bibfield
	{title} {\bibinfo {title} {Berry-phase-mediated topological thermoelectric
			transport in gapped single and bilayer graphene},\ }\href
	{https://doi.org/10.1103/physrevb.79.245424} {\bibfield  {journal} {\bibinfo
			{journal} {Phys. Rev. B}\ }\textbf {\bibinfo {volume} {79}},\ \bibinfo
		{pages} {245424} (\bibinfo {year} {2009})}\BibitemShut {NoStop}%
	\bibitem [{\citenamefont {Zhang}(2016)}]{Zhang2016}%
	\BibitemOpen
	\bibfield  {author} {\bibinfo {author} {\bibfnamefont {L.}~\bibnamefont
			{Zhang}},\ }\bibfield  {title} {\bibinfo {title} {Berry curvature and various
			thermal {H}all effects},\ }\href
	{https://doi.org/10.1088/1367-2630/18/10/103039} {\bibfield  {journal}
		{\bibinfo  {journal} {New J. Phys.}\ }\textbf {\bibinfo {volume} {18}},\
		\bibinfo {pages} {103039} (\bibinfo {year} {2016})}\BibitemShut {NoStop}%
	\bibitem [{\citenamefont {Jonson}\ and\ \citenamefont
		{Mahan}(1980)}]{Jonson1980}%
	\BibitemOpen
	\bibfield  {author} {\bibinfo {author} {\bibfnamefont {M.}~\bibnamefont
			{Jonson}}\ and\ \bibinfo {author} {\bibfnamefont {G.~D.}\ \bibnamefont
			{Mahan}},\ }\bibfield  {title} {\bibinfo {title} {Mott's formula for the
			thermopower and the {W}iedemann-{F}ranz law},\ }\href
	{https://doi.org/10.1103/physrevb.21.4223} {\bibfield  {journal} {\bibinfo
			{journal} {Phys. Rev. B}\ }\textbf {\bibinfo {volume} {21}},\ \bibinfo
		{pages} {4223} (\bibinfo {year} {1980})}\BibitemShut {NoStop}%
	\bibitem [{\citenamefont {Yokoyama}\ and\ \citenamefont
		{Murakami}(2011)}]{Yokoyama2011}%
	\BibitemOpen
	\bibfield  {author} {\bibinfo {author} {\bibfnamefont {T.}~\bibnamefont
			{Yokoyama}}\ and\ \bibinfo {author} {\bibfnamefont {S.}~\bibnamefont
			{Murakami}},\ }\bibfield  {title} {\bibinfo {title} {Transverse magnetic heat
			transport on the surface of a topological insulator},\ }\href
	{https://doi.org/10.1103/physrevb.83.161407} {\bibfield  {journal} {\bibinfo
			{journal} {Phys. Rev. B}\ }\textbf {\bibinfo {volume} {83}},\ \bibinfo
		{pages} {161407} (\bibinfo {year} {2011})}\BibitemShut {NoStop}%
	\bibitem [{\citenamefont {Neil~Ashcroft}(1976)}]{NeilAshcroft1976}%
	\BibitemOpen
	\bibfield  {author} {\bibinfo {author} {\bibfnamefont {N.~M.}\ \bibnamefont
			{Neil~Ashcroft}},\ }\bibinfo {title} {Solid state physics}\ (\bibinfo
	{publisher} {Brooks Cole; New edition (2 Jan. 1976)},\ \bibinfo {year}
	{1976})\BibitemShut {NoStop}%
	\bibitem [{\citenamefont {Xiao}\ \emph {et~al.}(2010)\citenamefont {Xiao},
		\citenamefont {Chang},\ and\ \citenamefont {Niu}}]{Xiao2010}%
	\BibitemOpen
	\bibfield  {author} {\bibinfo {author} {\bibfnamefont {D.}~\bibnamefont
			{Xiao}}, \bibinfo {author} {\bibfnamefont {M.-C.}\ \bibnamefont {Chang}},\
		and\ \bibinfo {author} {\bibfnamefont {Q.}~\bibnamefont {Niu}},\ }\bibfield
	{title} {\bibinfo {title} {Berry phase effects on electronic properties},\
	}\href {https://doi.org/10.1103/revmodphys.82.1959} {\bibfield  {journal}
		{\bibinfo  {journal} {Rev. Modern Phys.}\ }\textbf {\bibinfo {volume} {82}},\
		\bibinfo {pages} {1959} (\bibinfo {year} {2010})}\BibitemShut {NoStop}%
	\bibitem [{\citenamefont {Behnia}\ and\ \citenamefont
		{Aubin}(2016)}]{Behnia2016}%
	\BibitemOpen
	\bibfield  {author} {\bibinfo {author} {\bibfnamefont {K.}~\bibnamefont
			{Behnia}}\ and\ \bibinfo {author} {\bibfnamefont {H.}~\bibnamefont {Aubin}},\
	}\bibfield  {title} {\bibinfo {title} {Nernst effect in metals and
			superconductors: a review of concepts and experiments},\ }\href
	{https://doi.org/10.1088/0034-4885/79/4/046502} {\bibfield  {journal}
		{\bibinfo  {journal} {Rep. Prog. Phys.}\ }\textbf {\bibinfo {volume} {79}},\
		\bibinfo {pages} {046502} (\bibinfo {year} {2016})}\BibitemShut {NoStop}%
	\bibitem [{\citenamefont {Ortix}(2021)}]{Ortix2021}%
	\BibitemOpen
	\bibfield  {author} {\bibinfo {author} {\bibfnamefont {C.}~\bibnamefont
			{Ortix}},\ }\bibfield  {title} {\bibinfo {title} {Nonlinear {H}all effect
			with time-reversal symmetry: Theory and material realizations},\ }\href
	{https://doi.org/10.1002/qute.202100056} {\bibfield  {journal} {\bibinfo
			{journal} {Adv. Quantum Technol.}\ }\textbf {\bibinfo {volume} {4}},\
		\bibinfo {pages} {2100056} (\bibinfo {year} {2021})}\BibitemShut {NoStop}%
	\bibitem [{\citenamefont {Sodemann}\ and\ \citenamefont
		{Fu}(2015)}]{Sodemann2015}%
	\BibitemOpen
	\bibfield  {author} {\bibinfo {author} {\bibfnamefont {I.}~\bibnamefont
			{Sodemann}}\ and\ \bibinfo {author} {\bibfnamefont {L.}~\bibnamefont {Fu}},\
	}\bibfield  {title} {\bibinfo {title} {Quantum {N}onlinear {H}all {E}ffect
			{I}nduced by {B}erry {C}urvature {D}ipole in {T}ime-{R}eversal {I}nvariant
			{M}aterials},\ }\href {https://doi.org/10.1103/physrevlett.115.216806}
	{\bibfield  {journal} {\bibinfo  {journal} {Phys. Rev. Lett.}\ }\textbf
		{\bibinfo {volume} {115}},\ \bibinfo {pages} {216806} (\bibinfo {year}
		{2015})}\BibitemShut {NoStop}%
	\bibitem [{\citenamefont {Wang}\ \emph {et~al.}(2022)\citenamefont {Wang},
		\citenamefont {Zhu},\ and\ \citenamefont {Su}}]{Wang2022}%
	\BibitemOpen
	\bibfield  {author} {\bibinfo {author} {\bibfnamefont {Y.}~\bibnamefont
			{Wang}}, \bibinfo {author} {\bibfnamefont {Z.-G.}\ \bibnamefont {Zhu}},\ and\
		\bibinfo {author} {\bibfnamefont {G.}~\bibnamefont {Su}},\ }\bibfield
	{title} {\bibinfo {title} {Quantum theory of nonlinear thermal response},\
	}\href {https://doi.org/10.1103/physrevb.106.035148} {\bibfield  {journal}
		{\bibinfo  {journal} {Phys. Rev. B}\ }\textbf {\bibinfo {volume} {106}},\
		\bibinfo {pages} {035148} (\bibinfo {year} {2022})}\BibitemShut {NoStop}%
	\bibitem [{\citenamefont {Zeng}\ \emph {et~al.}(2020)\citenamefont {Zeng},
		\citenamefont {Nandy},\ and\ \citenamefont {Tewari}}]{Zeng2020}%
	\BibitemOpen
	\bibfield  {author} {\bibinfo {author} {\bibfnamefont {C.}~\bibnamefont
			{Zeng}}, \bibinfo {author} {\bibfnamefont {S.}~\bibnamefont {Nandy}},\ and\
		\bibinfo {author} {\bibfnamefont {S.}~\bibnamefont {Tewari}},\ }\bibfield
	{title} {\bibinfo {title} {Fundamental relations for anomalous thermoelectric
			transport coefficients in the nonlinear regime},\ }\href
	{https://doi.org/10.1103/physrevresearch.2.032066} {\bibfield  {journal}
		{\bibinfo  {journal} {Phys. Rev. Research}\ }\textbf {\bibinfo {volume}
			{2}},\ \bibinfo {pages} {032066} (\bibinfo {year} {2020})}\BibitemShut
	{NoStop}%
	\bibitem [{\citenamefont {Zhang}\ \emph {et~al.}(2024)\citenamefont {Zhang},
		\citenamefont {Zhu},\ and\ \citenamefont {Su}}]{Zhang2024}%
	\BibitemOpen
	\bibfield  {author} {\bibinfo {author} {\bibfnamefont {Z.-F.}\ \bibnamefont
			{Zhang}}, \bibinfo {author} {\bibfnamefont {Z.-G.}\ \bibnamefont {Zhu}},\
		and\ \bibinfo {author} {\bibfnamefont {G.}~\bibnamefont {Su}},\ }\bibfield
	{title} {\bibinfo {title} {Intrinsic second-order spin current},\ }\href
	{https://doi.org/10.1103/physrevb.110.174434} {\bibfield  {journal} {\bibinfo
			{journal} {Phys. Rev. B}\ }\textbf {\bibinfo {volume} {110}},\ \bibinfo
		{pages} {174434} (\bibinfo {year} {2024})}\BibitemShut {NoStop}%
	\bibitem [{\citenamefont {Duan}\ \emph {et~al.}(2023)\citenamefont {Duan},
		\citenamefont {Qin}, \citenamefont {Chen}, \citenamefont {Yang},
		\citenamefont {Qiu}, \citenamefont {Huang}, \citenamefont {Liu},
		\citenamefont {Li}, \citenamefont {Bi}, \citenamefont {Meng}, \citenamefont
		{Xi}, \citenamefont {Yao}, \citenamefont {Ideue}, \citenamefont {Lian},
		\citenamefont {Iwasa},\ and\ \citenamefont {Yuan}}]{Duan2023}%
	\BibitemOpen
	\bibfield  {author} {\bibinfo {author} {\bibfnamefont {S.}~\bibnamefont
			{Duan}}, \bibinfo {author} {\bibfnamefont {F.}~\bibnamefont {Qin}}, \bibinfo
		{author} {\bibfnamefont {P.}~\bibnamefont {Chen}}, \bibinfo {author}
		{\bibfnamefont {X.}~\bibnamefont {Yang}}, \bibinfo {author} {\bibfnamefont
			{C.}~\bibnamefont {Qiu}}, \bibinfo {author} {\bibfnamefont {J.}~\bibnamefont
			{Huang}}, \bibinfo {author} {\bibfnamefont {G.}~\bibnamefont {Liu}}, \bibinfo
		{author} {\bibfnamefont {Z.}~\bibnamefont {Li}}, \bibinfo {author}
		{\bibfnamefont {X.}~\bibnamefont {Bi}}, \bibinfo {author} {\bibfnamefont
			{F.}~\bibnamefont {Meng}}, \bibinfo {author} {\bibfnamefont {X.}~\bibnamefont
			{Xi}}, \bibinfo {author} {\bibfnamefont {J.}~\bibnamefont {Yao}}, \bibinfo
		{author} {\bibfnamefont {T.}~\bibnamefont {Ideue}}, \bibinfo {author}
		{\bibfnamefont {B.}~\bibnamefont {Lian}}, \bibinfo {author} {\bibfnamefont
			{Y.}~\bibnamefont {Iwasa}},\ and\ \bibinfo {author} {\bibfnamefont
			{H.}~\bibnamefont {Yuan}},\ }\bibfield  {title} {\bibinfo {title} {Berry
			curvature dipole generation and helicity-to-spin conversion at
			symmetry-mismatched heterointerfaces},\ }\href
	{https://doi.org/10.1038/s41565-023-01417-z} {\bibfield  {journal} {\bibinfo
			{journal} {Nat. Nanotechnol.}\ }\textbf {\bibinfo {volume} {18}},\ \bibinfo
		{pages} {867} (\bibinfo {year} {2023})}\BibitemShut {NoStop}%
	\bibitem [{\citenamefont {Xu}\ \emph {et~al.}(2018)\citenamefont {Xu},
		\citenamefont {Ma}, \citenamefont {Shen}, \citenamefont {Fatemi},
		\citenamefont {Wu}, \citenamefont {Chang}, \citenamefont {Chang},
		\citenamefont {Valdivia}, \citenamefont {Chan}, \citenamefont {Gibson},
		\citenamefont {Zhou}, \citenamefont {Liu}, \citenamefont {Watanabe},
		\citenamefont {Taniguchi}, \citenamefont {Lin}, \citenamefont {Cava},
		\citenamefont {Fu}, \citenamefont {Gedik},\ and\ \citenamefont
		{Jarillo-Herrero}}]{Xu2018}%
	\BibitemOpen
	\bibfield  {author} {\bibinfo {author} {\bibfnamefont {S.-Y.}\ \bibnamefont
			{Xu}}, \bibinfo {author} {\bibfnamefont {Q.}~\bibnamefont {Ma}}, \bibinfo
		{author} {\bibfnamefont {H.}~\bibnamefont {Shen}}, \bibinfo {author}
		{\bibfnamefont {V.}~\bibnamefont {Fatemi}}, \bibinfo {author} {\bibfnamefont
			{S.}~\bibnamefont {Wu}}, \bibinfo {author} {\bibfnamefont {T.-R.}\
			\bibnamefont {Chang}}, \bibinfo {author} {\bibfnamefont {G.}~\bibnamefont
			{Chang}}, \bibinfo {author} {\bibfnamefont {A.~M.~M.}\ \bibnamefont
			{Valdivia}}, \bibinfo {author} {\bibfnamefont {C.-K.}\ \bibnamefont {Chan}},
		\bibinfo {author} {\bibfnamefont {Q.~D.}\ \bibnamefont {Gibson}}, \bibinfo
		{author} {\bibfnamefont {J.}~\bibnamefont {Zhou}}, \bibinfo {author}
		{\bibfnamefont {Z.}~\bibnamefont {Liu}}, \bibinfo {author} {\bibfnamefont
			{K.}~\bibnamefont {Watanabe}}, \bibinfo {author} {\bibfnamefont
			{T.}~\bibnamefont {Taniguchi}}, \bibinfo {author} {\bibfnamefont
			{H.}~\bibnamefont {Lin}}, \bibinfo {author} {\bibfnamefont {R.~J.}\
			\bibnamefont {Cava}}, \bibinfo {author} {\bibfnamefont {L.}~\bibnamefont
			{Fu}}, \bibinfo {author} {\bibfnamefont {N.}~\bibnamefont {Gedik}},\ and\
		\bibinfo {author} {\bibfnamefont {P.}~\bibnamefont {Jarillo-Herrero}},\
	}\bibfield  {title} {\bibinfo {title} {Electrically switchable {B}erry
			curvature dipole in the monolayer topological insulator {WT}e$_2$},\ }\href
	{https://doi.org/10.1038/s41567-018-0189-6} {\bibfield  {journal} {\bibinfo
			{journal} {Nat. Phys.}\ }\textbf {\bibinfo {volume} {14}},\ \bibinfo {pages}
		{900} (\bibinfo {year} {2018})}\BibitemShut {NoStop}%
	\bibitem [{\citenamefont {Ma}\ \emph {et~al.}(2018)\citenamefont {Ma},
		\citenamefont {Xu}, \citenamefont {Shen}, \citenamefont {MacNeill},
		\citenamefont {Fatemi}, \citenamefont {Chang}, \citenamefont {Mier~Valdivia},
		\citenamefont {Wu}, \citenamefont {Du}, \citenamefont {Hsu}, \citenamefont
		{Fang}, \citenamefont {Gibson}, \citenamefont {Watanabe}, \citenamefont
		{Taniguchi}, \citenamefont {Cava}, \citenamefont {Kaxiras}, \citenamefont
		{Lu}, \citenamefont {Lin}, \citenamefont {Fu}, \citenamefont {Gedik},\ and\
		\citenamefont {Jarillo-Herrero}}]{Ma2018}%
	\BibitemOpen
	\bibfield  {author} {\bibinfo {author} {\bibfnamefont {Q.}~\bibnamefont
			{Ma}}, \bibinfo {author} {\bibfnamefont {S.-Y.}\ \bibnamefont {Xu}}, \bibinfo
		{author} {\bibfnamefont {H.}~\bibnamefont {Shen}}, \bibinfo {author}
		{\bibfnamefont {D.}~\bibnamefont {MacNeill}}, \bibinfo {author}
		{\bibfnamefont {V.}~\bibnamefont {Fatemi}}, \bibinfo {author} {\bibfnamefont
			{T.-R.}\ \bibnamefont {Chang}}, \bibinfo {author} {\bibfnamefont {A.~M.}\
			\bibnamefont {Mier~Valdivia}}, \bibinfo {author} {\bibfnamefont
			{S.}~\bibnamefont {Wu}}, \bibinfo {author} {\bibfnamefont {Z.}~\bibnamefont
			{Du}}, \bibinfo {author} {\bibfnamefont {C.-H.}\ \bibnamefont {Hsu}},
		\bibinfo {author} {\bibfnamefont {S.}~\bibnamefont {Fang}}, \bibinfo {author}
		{\bibfnamefont {Q.~D.}\ \bibnamefont {Gibson}}, \bibinfo {author}
		{\bibfnamefont {K.}~\bibnamefont {Watanabe}}, \bibinfo {author}
		{\bibfnamefont {T.}~\bibnamefont {Taniguchi}}, \bibinfo {author}
		{\bibfnamefont {R.~J.}\ \bibnamefont {Cava}}, \bibinfo {author}
		{\bibfnamefont {E.}~\bibnamefont {Kaxiras}}, \bibinfo {author} {\bibfnamefont
			{H.-Z.}\ \bibnamefont {Lu}}, \bibinfo {author} {\bibfnamefont
			{H.}~\bibnamefont {Lin}}, \bibinfo {author} {\bibfnamefont {L.}~\bibnamefont
			{Fu}}, \bibinfo {author} {\bibfnamefont {N.}~\bibnamefont {Gedik}},\ and\
		\bibinfo {author} {\bibfnamefont {P.}~\bibnamefont {Jarillo-Herrero}},\
	}\bibfield  {title} {\bibinfo {title} {Observation of the nonlinear {H}all
			effect under time-reversal-symmetric conditions},\ }\href
	{https://doi.org/10.1038/s41586-018-0807-6} {\bibfield  {journal} {\bibinfo
			{journal} {Nature}\ }\textbf {\bibinfo {volume} {565}},\ \bibinfo {pages}
		{337} (\bibinfo {year} {2018})}\BibitemShut {NoStop}%
	\bibitem [{\citenamefont {Kang}\ \emph {et~al.}(2019)\citenamefont {Kang},
		\citenamefont {Li}, \citenamefont {Sohn}, \citenamefont {Shan},\ and\
		\citenamefont {Mak}}]{Kang2019}%
	\BibitemOpen
	\bibfield  {author} {\bibinfo {author} {\bibfnamefont {K.}~\bibnamefont
			{Kang}}, \bibinfo {author} {\bibfnamefont {T.}~\bibnamefont {Li}}, \bibinfo
		{author} {\bibfnamefont {E.}~\bibnamefont {Sohn}}, \bibinfo {author}
		{\bibfnamefont {J.}~\bibnamefont {Shan}},\ and\ \bibinfo {author}
		{\bibfnamefont {K.~F.}\ \bibnamefont {Mak}},\ }\bibfield  {title} {\bibinfo
		{title} {Nonlinear anomalous {H}all effect in few-layer {WT}e$_2$},\ }\href
	{https://doi.org/10.1038/s41563-019-0294-7} {\bibfield  {journal} {\bibinfo
			{journal} {Nat. Mater.}\ }\textbf {\bibinfo {volume} {18}},\ \bibinfo {pages}
		{324} (\bibinfo {year} {2019})}\BibitemShut {NoStop}%
	\bibitem [{\citenamefont {Xiao}\ \emph
		{et~al.}(2020{\natexlab{a}})\citenamefont {Xiao}, \citenamefont {Wang},
		\citenamefont {Wang}, \citenamefont {Pemmaraju}, \citenamefont {Wang},
		\citenamefont {Muscher}, \citenamefont {Sie}, \citenamefont {Nyby},
		\citenamefont {Devereaux}, \citenamefont {Qian}, \citenamefont {Zhang},\ and\
		\citenamefont {Lindenberg}}]{Xiao2020a}%
	\BibitemOpen
	\bibfield  {author} {\bibinfo {author} {\bibfnamefont {J.}~\bibnamefont
			{Xiao}}, \bibinfo {author} {\bibfnamefont {Y.}~\bibnamefont {Wang}}, \bibinfo
		{author} {\bibfnamefont {H.}~\bibnamefont {Wang}}, \bibinfo {author}
		{\bibfnamefont {C.~D.}\ \bibnamefont {Pemmaraju}}, \bibinfo {author}
		{\bibfnamefont {S.}~\bibnamefont {Wang}}, \bibinfo {author} {\bibfnamefont
			{P.}~\bibnamefont {Muscher}}, \bibinfo {author} {\bibfnamefont {E.~J.}\
			\bibnamefont {Sie}}, \bibinfo {author} {\bibfnamefont {C.~M.}\ \bibnamefont
			{Nyby}}, \bibinfo {author} {\bibfnamefont {T.~P.}\ \bibnamefont {Devereaux}},
		\bibinfo {author} {\bibfnamefont {X.}~\bibnamefont {Qian}}, \bibinfo {author}
		{\bibfnamefont {X.}~\bibnamefont {Zhang}},\ and\ \bibinfo {author}
		{\bibfnamefont {A.~M.}\ \bibnamefont {Lindenberg}},\ }\bibfield  {title}
	{\bibinfo {title} {Berry curvature memory through electrically driven
			stacking transitions},\ }\href {https://doi.org/10.1038/s41567-020-0947-0}
	{\bibfield  {journal} {\bibinfo  {journal} {Nat. Phys.}\ }\textbf {\bibinfo
			{volume} {16}},\ \bibinfo {pages} {1028} (\bibinfo {year}
		{2020}{\natexlab{a}})}\BibitemShut {NoStop}%
	\bibitem [{\citenamefont {Xiao}\ \emph
		{et~al.}(2020{\natexlab{b}})\citenamefont {Xiao}, \citenamefont {Shao},
		\citenamefont {Zhang},\ and\ \citenamefont {Jiang}}]{Xiao2020b}%
	\BibitemOpen
	\bibfield  {author} {\bibinfo {author} {\bibfnamefont {R.-C.}\ \bibnamefont
			{Xiao}}, \bibinfo {author} {\bibfnamefont {D.-F.}\ \bibnamefont {Shao}},
		\bibinfo {author} {\bibfnamefont {Z.-Q.}\ \bibnamefont {Zhang}},\ and\
		\bibinfo {author} {\bibfnamefont {H.}~\bibnamefont {Jiang}},\ }\bibfield
	{title} {\bibinfo {title} {Two-dimensional metals for piezoelectriclike
			devices based on {B}erry-curvature dipole},\ }\href
	{https://doi.org/10.1103/physrevapplied.13.044014} {\bibfield  {journal}
		{\bibinfo  {journal} {Phys. Rev. Applied}\ }\textbf {\bibinfo {volume}
			{13}},\ \bibinfo {pages} {044014} (\bibinfo {year}
		{2020}{\natexlab{b}})}\BibitemShut {NoStop}%
	\bibitem [{\citenamefont {Shen}(2017)}]{Shen2017}%
	\BibitemOpen
	\bibfield  {author} {\bibinfo {author} {\bibfnamefont {S.-Q.}\ \bibnamefont
			{Shen}},\ }\bibfield  {title} {\bibinfo {title} {Topological insulators:
			{D}irac equation in condensed matter},\ }\href
	{https://doi.org/10.1007/978-981-10-4606-3} {\bibfield  {journal} {\bibinfo
			{journal} {Springer Ser. Solid-State Sci.}\ }\bibinfo {series} {0171-1873},\
		\textbf {\bibinfo {volume} {187}},\ \bibinfo {pages} {266} (\bibinfo {year}
		{2017})}\BibitemShut {NoStop}%
	\bibitem [{\citenamefont {Klitzing}\ \emph {et~al.}(1980)\citenamefont
		{Klitzing}, \citenamefont {Dorda},\ and\ \citenamefont
		{Pepper}}]{Klitzing1980}%
	\BibitemOpen
	\bibfield  {author} {\bibinfo {author} {\bibfnamefont {K.~v.}\ \bibnamefont
			{Klitzing}}, \bibinfo {author} {\bibfnamefont {G.}~\bibnamefont {Dorda}},\
		and\ \bibinfo {author} {\bibfnamefont {M.}~\bibnamefont {Pepper}},\
	}\bibfield  {title} {\bibinfo {title} {New method for high-accuracy
			determination of the fine-structure constant based on quantized {H}all
			resistance},\ }\href {https://doi.org/10.1103/physrevlett.45.494} {\bibfield
		{journal} {\bibinfo  {journal} {Phys. Rev. Lett.}\ }\textbf {\bibinfo
			{volume} {45}},\ \bibinfo {pages} {494} (\bibinfo {year} {1980})}\BibitemShut
	{NoStop}%
	\bibitem [{\citenamefont {Thouless}\ \emph {et~al.}(1982)\citenamefont
		{Thouless}, \citenamefont {Kohmoto}, \citenamefont {Nightingale},\ and\
		\citenamefont {den Nijs}}]{Thouless1982}%
	\BibitemOpen
	\bibfield  {author} {\bibinfo {author} {\bibfnamefont {D.~J.}\ \bibnamefont
			{Thouless}}, \bibinfo {author} {\bibfnamefont {M.}~\bibnamefont {Kohmoto}},
		\bibinfo {author} {\bibfnamefont {M.~P.}\ \bibnamefont {Nightingale}},\ and\
		\bibinfo {author} {\bibfnamefont {M.}~\bibnamefont {den Nijs}},\ }\bibfield
	{title} {\bibinfo {title} {Quantized {H}all conductance in a two-dimensional
			periodic potential},\ }\href {https://doi.org/10.1103/physrevlett.49.405}
	{\bibfield  {journal} {\bibinfo  {journal} {Phys. Rev. Lett.}\ }\textbf
		{\bibinfo {volume} {49}},\ \bibinfo {pages} {405} (\bibinfo {year}
		{1982})}\BibitemShut {NoStop}%
	\bibitem [{\citenamefont {Haldane}(1988)}]{Haldane1988}%
	\BibitemOpen
	\bibfield  {author} {\bibinfo {author} {\bibfnamefont {F.~D.~M.}\
			\bibnamefont {Haldane}},\ }\bibfield  {title} {\bibinfo {title} {Model for a
			quantum {H}all effect without landau levels: Condensed-matter realization of
			the ''parity anomaly''},\ }\href
	{https://doi.org/10.1103/physrevlett.61.2015} {\bibfield  {journal} {\bibinfo
			{journal} {Phys. Rev. Lett.}\ }\textbf {\bibinfo {volume} {61}},\ \bibinfo
		{pages} {2015} (\bibinfo {year} {1988})}\BibitemShut {NoStop}%
		\bibitem [{\citenamefont {Du}\ \emph {et~al.}(2021)\citenamefont {Du},
			\citenamefont {Wang}, \citenamefont {Sun}, \citenamefont {Lu},\ and\
			\citenamefont {Xie}}]{Du2021}%
		\BibitemOpen
		\bibfield  {author} {\bibinfo {author} {\bibfnamefont {Z.~Z.}\ \bibnamefont
				{Du}}, \bibinfo {author} {\bibfnamefont {C.~M.}\ \bibnamefont {Wang}},
			\bibinfo {author} {\bibfnamefont {H. -P.}\ \bibnamefont {Sun}}, \bibinfo
			{author} {\bibfnamefont {H. -Z.}\ \bibnamefont {Lu}},\ and\ \bibinfo {author}
			{\bibfnamefont {X.~C.}\ \bibnamefont {Xie}},\ }\bibfield  {title} {\bibinfo
			{title} {Quantum theory of the nonlinear {H}all effect},\ }\href
		{https://doi.org/10.1038/s41467-021-25273-4} {\bibfield  {journal} {\bibinfo
				{journal} {Nat. Commun.}\ }\textbf {\bibinfo {volume} {12}},\ \bibinfo
			{pages} {5038} (\bibinfo {year} {2021})}\BibitemShut {NoStop}%
	\bibitem [{\citenamefont {Du}\ \emph {et~al.}(2019)\citenamefont {Du},
		\citenamefont {Wang}, \citenamefont {Li}, \citenamefont {Lu},\ and\
		\citenamefont {Xie}}]{Du2019}%
	\BibitemOpen
	\bibfield  {author} {\bibinfo {author} {\bibfnamefont {Z.~Z.}\ \bibnamefont
			{Du}}, \bibinfo {author} {\bibfnamefont {C.~M.}\ \bibnamefont {Wang}},
		\bibinfo {author} {\bibfnamefont {S.}~\bibnamefont {Li}}, \bibinfo {author}
		{\bibfnamefont {H.-Z.}\ \bibnamefont {Lu}},\ and\ \bibinfo {author}
		{\bibfnamefont {X.~C.}\ \bibnamefont {Xie}},\ }\bibfield  {title} {\bibinfo
		{title} {Disorder-induced nonlinear {H}all effect with time-reversal
			symmetry},\ }\href {https://doi.org/10.1038/s41467-019-10941-3} {\bibfield
		{journal} {\bibinfo  {journal} {Nat. Commun.}\ }\textbf {\bibinfo {volume}
			{10}},\ \bibinfo {pages} {3047} (\bibinfo {year} {2019})}\BibitemShut
	{NoStop}%
	\bibitem [{\citenamefont {Wang}\ \emph {et~al.}(2021)\citenamefont {Wang},
		\citenamefont {Gao},\ and\ \citenamefont {Xiao}}]{Wang2021}%
	\BibitemOpen
	\bibfield  {author} {\bibinfo {author} {\bibfnamefont {C.}~\bibnamefont
			{Wang}}, \bibinfo {author} {\bibfnamefont {Y.}~\bibnamefont {Gao}},\ and\
		\bibinfo {author} {\bibfnamefont {D.}~\bibnamefont {Xiao}},\ }\bibfield
	{title} {\bibinfo {title} {Intrinsic nonlinear {H}all effect in
			antiferromagnetic tetragonal {C}u{M}n{A}s},\ }\href
	{https://doi.org/10.1103/physrevlett.127.277201} {\bibfield  {journal}
		{\bibinfo  {journal} {Phys. Rev. Lett.}\ }\textbf {\bibinfo {volume} {127}},\
		\bibinfo {pages} {277201} (\bibinfo {year} {2021})}\BibitemShut {NoStop}%
	\bibitem [{\citenamefont {Liu}\ \emph {et~al.}(2021)\citenamefont {Liu},
		\citenamefont {Zhao}, \citenamefont {Huang}, \citenamefont {Wu},
		\citenamefont {Sheng}, \citenamefont {Xiao},\ and\ \citenamefont
		{Yang}}]{Liu2021}%
	\BibitemOpen
	\bibfield  {author} {\bibinfo {author} {\bibfnamefont {H.}~\bibnamefont
			{Liu}}, \bibinfo {author} {\bibfnamefont {J.}~\bibnamefont {Zhao}}, \bibinfo
		{author} {\bibfnamefont {Y.-X.}\ \bibnamefont {Huang}}, \bibinfo {author}
		{\bibfnamefont {W.}~\bibnamefont {Wu}}, \bibinfo {author} {\bibfnamefont
			{X.-L.}\ \bibnamefont {Sheng}}, \bibinfo {author} {\bibfnamefont
			{C.}~\bibnamefont {Xiao}},\ and\ \bibinfo {author} {\bibfnamefont {S.~A.}\
			\bibnamefont {Yang}},\ }\bibfield  {title} {\bibinfo {title} {Intrinsic
			second-order anomalous {H}all effect and its application in compensated
			antiferromagnets},\ }\href {https://doi.org/10.1103/physrevlett.127.277202}
	{\bibfield  {journal} {\bibinfo  {journal} {Phys. Rev. Lett.}\ }\textbf
		{\bibinfo {volume} {127}},\ \bibinfo {pages} {277202} (\bibinfo {year}
		{2021})}\BibitemShut {NoStop}%
	\bibitem [{\citenamefont {Duan}\ \emph {et~al.}(2022)\citenamefont {Duan},
		\citenamefont {Jian}, \citenamefont {Gao}, \citenamefont {Peng},
		\citenamefont {Zhong}, \citenamefont {Feng}, \citenamefont {Mao},\ and\
		\citenamefont {Yao}}]{Duan2022}%
	\BibitemOpen
	\bibfield  {author} {\bibinfo {author} {\bibfnamefont {J.}~\bibnamefont
			{Duan}}, \bibinfo {author} {\bibfnamefont {Y.}~\bibnamefont {Jian}}, \bibinfo
		{author} {\bibfnamefont {Y.}~\bibnamefont {Gao}}, \bibinfo {author}
		{\bibfnamefont {H.}~\bibnamefont {Peng}}, \bibinfo {author} {\bibfnamefont
			{J.}~\bibnamefont {Zhong}}, \bibinfo {author} {\bibfnamefont
			{Q.}~\bibnamefont {Feng}}, \bibinfo {author} {\bibfnamefont {J.}~\bibnamefont
			{Mao}},\ and\ \bibinfo {author} {\bibfnamefont {Y.}~\bibnamefont {Yao}},\
	}\bibfield  {title} {\bibinfo {title} {Giant second-order nonlinear {H}all
			effect in twisted bilayer graphene},\ }\href
	{https://doi.org/10.1103/physrevlett.129.186801} {\bibfield  {journal}
		{\bibinfo  {journal} {Phys. Rev. Lett.}\ }\textbf {\bibinfo {volume} {129}},\
		\bibinfo {pages} {186801} (\bibinfo {year} {2022})}\BibitemShut {NoStop}%
	\bibitem [{\citenamefont {Das}\ \emph {et~al.}(2023)\citenamefont {Das},
		\citenamefont {Lahiri}, \citenamefont {Atencia}, \citenamefont {Culcer},\
		and\ \citenamefont {Agarwal}}]{Das2023}%
	\BibitemOpen
	\bibfield  {author} {\bibinfo {author} {\bibfnamefont {K.}~\bibnamefont
			{Das}}, \bibinfo {author} {\bibfnamefont {S.}~\bibnamefont {Lahiri}},
		\bibinfo {author} {\bibfnamefont {R.~B.}\ \bibnamefont {Atencia}}, \bibinfo
		{author} {\bibfnamefont {D.}~\bibnamefont {Culcer}},\ and\ \bibinfo {author}
		{\bibfnamefont {A.}~\bibnamefont {Agarwal}},\ }\bibfield  {title} {\bibinfo
		{title} {Intrinsic nonlinear conductivities induced by the quantum metric},\
	}\href {https://doi.org/10.1103/physrevb.108.l201405} {\bibfield  {journal}
		{\bibinfo  {journal} {Phys. Rev. B}\ }\textbf {\bibinfo {volume} {108}},\
		\bibinfo {pages} {l201405} (\bibinfo {year} {2023})}\BibitemShut {NoStop}%
	\bibitem [{\citenamefont {Adhidewata}\ \emph {et~al.}(2023)\citenamefont
		{Adhidewata}, \citenamefont {Komalig}, \citenamefont {Ukhtary}, \citenamefont
		{Nugraha}, \citenamefont {Gunara},\ and\ \citenamefont
		{Hasdeo}}]{Adhidewata2023}%
	\BibitemOpen
	\bibfield  {author} {\bibinfo {author} {\bibfnamefont {J.~M.}\ \bibnamefont
			{Adhidewata}}, \bibinfo {author} {\bibfnamefont {R.~W.~M.}\ \bibnamefont
			{Komalig}}, \bibinfo {author} {\bibfnamefont {M.~S.}\ \bibnamefont
			{Ukhtary}}, \bibinfo {author} {\bibfnamefont {A.~R.~T.}\ \bibnamefont
			{Nugraha}}, \bibinfo {author} {\bibfnamefont {B.~E.}\ \bibnamefont
			{Gunara}},\ and\ \bibinfo {author} {\bibfnamefont {E.~H.}\ \bibnamefont
			{Hasdeo}},\ }\bibfield  {title} {\bibinfo {title} {Trigonal warping effects
			on optical properties of anomalous {H}all materials},\ }\href
	{https://doi.org/10.1103/physrevb.107.155415} {\bibfield  {journal} {\bibinfo
			{journal} {Phys. Rev. B}\ }\textbf {\bibinfo {volume} {107}},\ \bibinfo
		{pages} {155415} (\bibinfo {year} {2023})}\BibitemShut {NoStop}%
	\bibitem [{\citenamefont {Saha}\ and\ \citenamefont
		{Narayan}(2023)}]{Saha2023}%
	\BibitemOpen
	\bibfield  {author} {\bibinfo {author} {\bibfnamefont {S.}~\bibnamefont
			{Saha}}\ and\ \bibinfo {author} {\bibfnamefont {A.}~\bibnamefont {Narayan}},\
	}\bibfield  {title} {\bibinfo {title} {Nonlinear {H}all effect in {R}ashba
			systems with hexagonal warping},\ }\href
	{https://doi.org/10.1088/1361-648x/acf1eb} {\bibfield  {journal} {\bibinfo
			{journal} {J. Phys. Condens. Matter}\ }\textbf {\bibinfo {volume} {35}},\
		\bibinfo {pages} {485301} (\bibinfo {year} {2023})}\BibitemShut {NoStop}%
	\bibitem [{\citenamefont {Wang}\ \emph {et~al.}(2024)\citenamefont {Wang},
		\citenamefont {Zhang}, \citenamefont {Zhu},\ and\ \citenamefont
		{Su}}]{Wang2024}%
	\BibitemOpen
	\bibfield  {author} {\bibinfo {author} {\bibfnamefont {Y.}~\bibnamefont
			{Wang}}, \bibinfo {author} {\bibfnamefont {Z.}~\bibnamefont {Zhang}},
		\bibinfo {author} {\bibfnamefont {Z.-G.}\ \bibnamefont {Zhu}},\ and\ \bibinfo
		{author} {\bibfnamefont {G.}~\bibnamefont {Su}},\ }\bibfield  {title}
	{\bibinfo {title} {Intrinsic nonlinear {O}hmic current},\ }\href
	{https://doi.org/10.1103/physrevb.109.085419} {\bibfield  {journal} {\bibinfo
			{journal} {Phys. Rev. B}\ }\textbf {\bibinfo {volume} {109}},\ \bibinfo
		{pages} {085419} (\bibinfo {year} {2024})}\BibitemShut {NoStop}%
	\bibitem [{\citenamefont {Isobe}\ \emph {et~al.}(2020)\citenamefont {Isobe},
		\citenamefont {Xu},\ and\ \citenamefont {Fu}}]{Isobe2020}%
	\BibitemOpen
	\bibfield  {author} {\bibinfo {author} {\bibfnamefont {H.}~\bibnamefont
			{Isobe}}, \bibinfo {author} {\bibfnamefont {S.-Y.}\ \bibnamefont {Xu}},\ and\
		\bibinfo {author} {\bibfnamefont {L.}~\bibnamefont {Fu}},\ }\bibfield
	{title} {\bibinfo {title} {High-frequency rectification via chiral {B}loch
			electrons},\ }\href {https://doi.org/10.1126/sciadv.aay2497} {\bibfield
		{journal} {\bibinfo  {journal} {Sci. Adv.}\ }\textbf {\bibinfo {volume}
			{6}},\ \bibinfo {pages} {eaay2497} (\bibinfo {year} {2020})}\BibitemShut
	{NoStop}%
	\bibitem [{\citenamefont {Makushko}\ \emph {et~al.}(2024)\citenamefont
		{Makushko}, \citenamefont {Kovalev}, \citenamefont {Zabila}, \citenamefont
		{Ilyakov}, \citenamefont {Ponomaryov}, \citenamefont {Arshad}, \citenamefont
		{Prajapati}, \citenamefont {de~Oliveira}, \citenamefont {Deinert},
		\citenamefont {Chekhonin}, \citenamefont {Veremchuk}, \citenamefont {Kosub},
		\citenamefont {Skourski}, \citenamefont {Ganss}, \citenamefont {Makarov},\
		and\ \citenamefont {Ortix}}]{Makushko2024}%
	\BibitemOpen
	\bibfield  {author} {\bibinfo {author} {\bibfnamefont {P.}~\bibnamefont
			{Makushko}}, \bibinfo {author} {\bibfnamefont {S.}~\bibnamefont {Kovalev}},
		\bibinfo {author} {\bibfnamefont {Y.}~\bibnamefont {Zabila}}, \bibinfo
		{author} {\bibfnamefont {I.}~\bibnamefont {Ilyakov}}, \bibinfo {author}
		{\bibfnamefont {A.}~\bibnamefont {Ponomaryov}}, \bibinfo {author}
		{\bibfnamefont {A.}~\bibnamefont {Arshad}}, \bibinfo {author} {\bibfnamefont
			{G.~L.}\ \bibnamefont {Prajapati}}, \bibinfo {author} {\bibfnamefont {T.~V.
				A.~G.}\ \bibnamefont {de~Oliveira}}, \bibinfo {author} {\bibfnamefont
			{J.-C.}\ \bibnamefont {Deinert}}, \bibinfo {author} {\bibfnamefont
			{P.}~\bibnamefont {Chekhonin}}, \bibinfo {author} {\bibfnamefont
			{I.}~\bibnamefont {Veremchuk}}, \bibinfo {author} {\bibfnamefont
			{T.}~\bibnamefont {Kosub}}, \bibinfo {author} {\bibfnamefont
			{Y.}~\bibnamefont {Skourski}}, \bibinfo {author} {\bibfnamefont
			{F.}~\bibnamefont {Ganss}}, \bibinfo {author} {\bibfnamefont
			{D.}~\bibnamefont {Makarov}},\ and\ \bibinfo {author} {\bibfnamefont
			{C.}~\bibnamefont {Ortix}},\ }\bibfield  {title} {\bibinfo {title} {A tunable
			room-temperature nonlinear {H}all effect in elemental bismuth thin films},\
	}\href {https://doi.org/10.1038/s41928-024-01118-y} {\bibfield  {journal}
		{\bibinfo  {journal} {Nat. Electron.}\ }\textbf {\bibinfo {volume} {7}},\
		\bibinfo {pages} {207} (\bibinfo {year} {2024})}\BibitemShut {NoStop}%
	\bibitem [{\citenamefont {Gao}\ and\ \citenamefont {Xiao}(2018)}]{Gao2018}%
	\BibitemOpen
	\bibfield  {author} {\bibinfo {author} {\bibfnamefont {Y.}~\bibnamefont
			{Gao}}\ and\ \bibinfo {author} {\bibfnamefont {D.}~\bibnamefont {Xiao}},\
	}\bibfield  {title} {\bibinfo {title} {Orbital magnetic quadrupole moment and
			nonlinear anomalous thermoelectric transport},\ }\href
	{https://doi.org/10.1103/physrevb.98.060402} {\bibfield  {journal} {\bibinfo
			{journal} {Phys. Rev. B}\ }\textbf {\bibinfo {volume} {98}},\ \bibinfo
		{pages} {060402} (\bibinfo {year} {2018})}\BibitemShut {NoStop}%
	\bibitem [{\citenamefont {Zeng}\ \emph {et~al.}(2019)\citenamefont {Zeng},
		\citenamefont {Nandy}, \citenamefont {Taraphder},\ and\ \citenamefont
		{Tewari}}]{Zeng2019}%
	\BibitemOpen
	\bibfield  {author} {\bibinfo {author} {\bibfnamefont {C.}~\bibnamefont
			{Zeng}}, \bibinfo {author} {\bibfnamefont {S.}~\bibnamefont {Nandy}},
		\bibinfo {author} {\bibfnamefont {A.}~\bibnamefont {Taraphder}},\ and\
		\bibinfo {author} {\bibfnamefont {S.}~\bibnamefont {Tewari}},\ }\bibfield
	{title} {\bibinfo {title} {Nonlinear {N}ernst effect in bilayer {WT}e$_2$},\
	}\href {https://doi.org/10.1103/physrevb.100.245102} {\bibfield  {journal}
		{\bibinfo  {journal} {Phys. Rev. B}\ }\textbf {\bibinfo {volume} {100}},\
		\bibinfo {pages} {245102} (\bibinfo {year} {2019})}\BibitemShut {NoStop}%
	\bibitem [{\citenamefont {Yu}\ \emph {et~al.}(2021)\citenamefont {Yu},
		\citenamefont {Zhu},\ and\ \citenamefont {Su}}]{Yu2021}%
	\BibitemOpen
	\bibfield  {author} {\bibinfo {author} {\bibfnamefont {X.-Q.}\ \bibnamefont
			{Yu}}, \bibinfo {author} {\bibfnamefont {Z.-G.}\ \bibnamefont {Zhu}},\ and\
		\bibinfo {author} {\bibfnamefont {G.}~\bibnamefont {Su}},\ }\bibfield
	{title} {\bibinfo {title} {Hexagonal warping induced nonlinear planar
			{N}ernst effect in nonmagnetic topological insulators},\ }\href
	{https://doi.org/10.1103/physrevb.103.035410} {\bibfield  {journal} {\bibinfo
			{journal} {Phys. Rev. B}\ }\textbf {\bibinfo {volume} {103}},\ \bibinfo
		{pages} {035410} (\bibinfo {year} {2021})}\BibitemShut {NoStop}%
	\bibitem [{\citenamefont {Papaj}\ and\ \citenamefont {Fu}(2021)}]{Papaj2021}%
	\BibitemOpen
	\bibfield  {author} {\bibinfo {author} {\bibfnamefont {M.}~\bibnamefont
			{Papaj}}\ and\ \bibinfo {author} {\bibfnamefont {L.}~\bibnamefont {Fu}},\
	}\bibfield  {title} {\bibinfo {title} {Enhanced anomalous {N}ernst effect in
			disordered {D}irac and {W}eyl materials},\ }\href
	{https://doi.org/10.1103/physrevb.103.075424} {\bibfield  {journal} {\bibinfo
			{journal} {Phys. Rev. B}\ }\textbf {\bibinfo {volume} {103}},\ \bibinfo
		{pages} {075424} (\bibinfo {year} {2021})}\BibitemShut {NoStop}%
	\bibitem [{\citenamefont {Zeng}\ \emph {et~al.}(2021)\citenamefont {Zeng},
		\citenamefont {Nandy},\ and\ \citenamefont {Tewari}}]{Zeng2021}%
	\BibitemOpen
	\bibfield  {author} {\bibinfo {author} {\bibfnamefont {C.}~\bibnamefont
			{Zeng}}, \bibinfo {author} {\bibfnamefont {S.}~\bibnamefont {Nandy}},\ and\
		\bibinfo {author} {\bibfnamefont {S.}~\bibnamefont {Tewari}},\ }\bibfield
	{title} {\bibinfo {title} {Nonlinear transport in {W}eyl semimetals induced
			by {B}erry curvature dipole},\ }\href
	{https://doi.org/10.1103/physrevb.103.245119} {\bibfield  {journal} {\bibinfo
			{journal} {Phys. Rev. B}\ }\textbf {\bibinfo {volume} {103}},\ \bibinfo
		{pages} {245119} (\bibinfo {year} {2021})}\BibitemShut {NoStop}%
	\bibitem [{\citenamefont {Zeng}\ \emph
		{et~al.}(2022{\natexlab{a}})\citenamefont {Zeng}, \citenamefont {Nandy},\
		and\ \citenamefont {Tewari}}]{Zeng2022}%
	\BibitemOpen
	\bibfield  {author} {\bibinfo {author} {\bibfnamefont {C.}~\bibnamefont
			{Zeng}}, \bibinfo {author} {\bibfnamefont {S.}~\bibnamefont {Nandy}},\ and\
		\bibinfo {author} {\bibfnamefont {S.}~\bibnamefont {Tewari}},\ }\bibfield
	{title} {\bibinfo {title} {Chiral anomaly induced nonlinear {N}ernst and
			thermal {H}all effects in {W}eyl semimetals},\ }\href
	{https://doi.org/10.1103/physrevb.105.125131} {\bibfield  {journal} {\bibinfo
			{journal} {Phys. Rev. B}\ }\textbf {\bibinfo {volume} {105}},\ \bibinfo
		{pages} {125131} (\bibinfo {year} {2022}{\natexlab{a}})}\BibitemShut
	{NoStop}%
	\bibitem [{\citenamefont {Zeng}\ \emph
		{et~al.}(2022{\natexlab{b}})\citenamefont {Zeng}, \citenamefont {Yu},
		\citenamefont {Yu},\ and\ \citenamefont {Yao}}]{Zeng2022a}%
	\BibitemOpen
	\bibfield  {author} {\bibinfo {author} {\bibfnamefont {C.}~\bibnamefont
			{Zeng}}, \bibinfo {author} {\bibfnamefont {X.-Q.}\ \bibnamefont {Yu}},
		\bibinfo {author} {\bibfnamefont {Z.-M.}\ \bibnamefont {Yu}},\ and\ \bibinfo
		{author} {\bibfnamefont {Y.}~\bibnamefont {Yao}},\ }\bibfield  {title}
	{\bibinfo {title} {Band tilt induced nonlinear {N}ernst effect in topological
			insulators: An efficient generation of high-performance spin polarization},\
	}\href {https://doi.org/10.1103/physrevb.106.l081121} {\bibfield  {journal}
		{\bibinfo  {journal} {Phys. Rev. B}\ }\textbf {\bibinfo {volume} {106}},\
		\bibinfo {pages} {l081121} (\bibinfo {year}
		{2022}{\natexlab{b}})}\BibitemShut {NoStop}%
	\bibitem [{\citenamefont {Qiang}\ \emph {et~al.}(2023)\citenamefont {Qiang},
		\citenamefont {Du}, \citenamefont {Lu},\ and\ \citenamefont
		{Xie}}]{Qiang2023}%
	\BibitemOpen
	\bibfield  {author} {\bibinfo {author} {\bibfnamefont {X.-B.}\ \bibnamefont
			{Qiang}}, \bibinfo {author} {\bibfnamefont {Z.~Z.}\ \bibnamefont {Du}},
		\bibinfo {author} {\bibfnamefont {H.-Z.}\ \bibnamefont {Lu}},\ and\ \bibinfo
		{author} {\bibfnamefont {X.~C.}\ \bibnamefont {Xie}},\ }\bibfield  {title}
	{\bibinfo {title} {Topological and disorder corrections to the transverse
			{W}iedemann-{F}ranz law and {M}ott relation in kagome magnets and {D}irac
			materials},\ }\href {https://doi.org/10.1103/physrevb.107.l161302} {\bibfield
		{journal} {\bibinfo  {journal} {Phys. Rev. B}\ }\textbf {\bibinfo {volume}
			{107}},\ \bibinfo {pages} {161302} (\bibinfo {year} {2023})}\BibitemShut
	{NoStop}%
\bibitem [{\citenamefont {Varshney}\ and\ \citenamefont
	{Agarwal}(2025)}]{Varshney2025}%
\BibitemOpen
\bibfield  {author} {\bibinfo {author} {\bibfnamefont {H.}~\bibnamefont
		{Varshney}}\ and\ \bibinfo {author} {\bibfnamefont {A.}~\bibnamefont
		{Agarwal}},\ }\bibfield  {title} {\bibinfo {title} {Intrinsic nonlinear
		{N}ernst and {S}eebeck effect},\ }\href
{https://doi.org/10.1088/1367-2630/adf5dd} {\bibfield  {journal} {\bibinfo
		{journal} {New J. Phys.}\ }\textbf {\bibinfo {volume} {27}},\ \bibinfo
	{pages} {083506} (\bibinfo {year} {2025})}\BibitemShut {NoStop}%
	\bibitem [{\citenamefont {Zhang}\ \emph {et~al.}(2025)\citenamefont {Zhang},
		\citenamefont {Zhang}, \citenamefont {Zhu},\ and\ \citenamefont
		{Su}}]{Zhang2025}%
	\BibitemOpen
	\bibfield  {author} {\bibinfo {author} {\bibfnamefont {Y.-F.}\ \bibnamefont
			{Zhang}}, \bibinfo {author} {\bibfnamefont {Z.-F.}\ \bibnamefont {Zhang}},
		\bibinfo {author} {\bibfnamefont {Z.-G.}\ \bibnamefont {Zhu}},\ and\ \bibinfo
		{author} {\bibfnamefont {G.}~\bibnamefont {Su}},\ }\bibfield  {title}
	{\bibinfo {title} {Second-order intrinsic {W}iedemann-{F}ranz law},\ }\href
	{https://doi.org/10.1103/physrevb.111.165424} {\bibfield  {journal} {\bibinfo
			{journal} {Phys. Rev. B}\ }\textbf {\bibinfo {volume} {111}},\ \bibinfo
		{pages} {165424} (\bibinfo {year} {2025})}\BibitemShut {NoStop}%
\bibitem [{\citenamefont {Liu}\ \emph {et~al.}(2025)\citenamefont {Liu},
	\citenamefont {Li}, \citenamefont {Zhang}\ \emph {et~al.}}]{Liu2025}%
\BibitemOpen
\bibfield  {author} {\bibinfo {author} {\bibfnamefont {H.}~\bibnamefont
		{Liu}}, \bibinfo {author} {\bibfnamefont {J.}~\bibnamefont {Li}}, \bibinfo
	{author} {\bibfnamefont {Z.}~\bibnamefont {Zhang}}, \emph {et~al.},\
}\bibfield  {title} {\bibinfo {title} {Nonlinear {N}ernst effect in trilayer
		graphene at zero magnetic field},\ }\href
{https://doi.org/10.1038/s41565-025-01963-8} {\bibfield  {journal} {\bibinfo
		{journal} {Nat. Nanotechnol.}\ }\textbf {\bibinfo {volume} {20}},\ \bibinfo
	{pages} {1221--1227} (\bibinfo {year} {2025})}\BibitemShut {NoStop}%
			\bibitem [{\citenamefont {Yu}\ \emph{et~al.}(2019)}]{Yu2019}
	\BibitemOpen
	\bibfield {author} {\bibinfo {author} {\bibfnamefont {X.-Q.}\ \bibnamefont {Yu}},
		\bibinfo {author} {\bibfnamefont {Z.-G.}\ \bibnamefont {Zhu}},
		\bibinfo {author} {\bibfnamefont {J.-S.}\ \bibnamefont {You}},
		\bibinfo {author} {\bibfnamefont {T.}\ \bibnamefont {Low}},
		\bibinfo {author} {\bibfnamefont {G.}\ \bibnamefont {Su}}},\
	\bibfield {title} {\bibinfo {title} {Topological nonlinear anomalous Nernst effect in strained transition metal dichalcogenides},\ }
	\href {https://doi.org/10.1103/physrevb.99.201410}
	{\bibfield {journal} {\bibinfo {journal} {Phys. Rev. B}\ }\textbf {\bibinfo {volume} {99}},\ \bibinfo {pages} {201410} (\bibinfo {year} {2019})}
	\BibitemShut {NoStop}%
	\bibitem [{\citenamefont {Zhou}\ \emph {et~al.}(2022)\citenamefont {Zhou},
		\citenamefont {Zhang}, \citenamefont {Yu}, \citenamefont {Zhu},\ and\
		\citenamefont {Su}}]{Zhou2022}%
	\BibitemOpen
	\bibfield  {author} {\bibinfo {author} {\bibfnamefont {D.-K.}\ \bibnamefont
			{Zhou}}, \bibinfo {author} {\bibfnamefont {Z.-F.}\ \bibnamefont {Zhang}},
		\bibinfo {author} {\bibfnamefont {X.-Q.}\ \bibnamefont {Yu}}, \bibinfo
		{author} {\bibfnamefont {Z.-G.}\ \bibnamefont {Zhu}},\ and\ \bibinfo {author}
		{\bibfnamefont {G.}~\bibnamefont {Su}},\ }\bibfield  {title} {\bibinfo
		{title} {Fundamental distinction between intrinsic and extrinsic nonlinear
			thermal {H}all effects},\ }\href
	{https://doi.org/10.1103/physrevb.105.l201103} {\bibfield  {journal}
		{\bibinfo  {journal} {Phys. Rev. B}\ }\textbf {\bibinfo {volume} {105}},\
		\bibinfo {pages} {l201103} (\bibinfo {year} {2022})}\BibitemShut {NoStop}%
	\bibitem [{\citenamefont {Li}\ and\ \citenamefont {Zhu}(2024)}]{Li2024}%
	\BibitemOpen
	\bibfield  {author} {\bibinfo {author} {\bibfnamefont {J.-C.}\ \bibnamefont
			{Li}}\ and\ \bibinfo {author} {\bibfnamefont {Z.-G.}\ \bibnamefont {Zhu}},\
	}\bibfield  {title} {\bibinfo {title} {Intrinsic second-order magnon thermal
			{H}all effect},\ }\href {https://doi.org/10.1088/1361-648x/ad5bb0} {\bibfield
		{journal} {\bibinfo  {journal} {J. Phys. Condens. Matter}\ }\textbf
		{\bibinfo {volume} {36}},\ \bibinfo {pages} {395802} (\bibinfo {year}
		{2024})}\BibitemShut {NoStop}%
				\bibitem [{\citenamefont {Itskov}(2025)}]{Itskov2025Tensor}%
		\BibitemOpen
		\bibfield  {author} {\bibinfo {author} {\bibfnamefont {M.}~\bibnamefont
				{Itskov}},\ }\bibfield  {title} {\bibinfo {title} {Tensor Algebra and
				Tensor Analysis for Engineers: With Applications to Continuum
				Mechanics},\ }\href {https://doi.org/10.1007/978-3-031-92357-9}
		{\bibfield  {publisher} {\bibinfo  {publisher} {Springer},\ \bibinfo
				{address} {Cham}}\ (\bibinfo {year} {2025})}\BibitemShut {NoStop}%
		\bibitem [{\citenamefont {Li}(2017)}]{Li2017Nonlinear}%
		\BibitemOpen
		\bibfield  {author} {\bibinfo {author} {\bibfnamefont {C.}~\bibnamefont
				{Li}},\ }\bibfield  {title} {\bibinfo {title} {Nonlinear Optics:
				Principles and Applications},\ }\href
		{https://doi.org/10.1007/978-981-10-1488-8} {\bibfield  {publisher}
			{\bibinfo  {publisher} {Springer},\ \bibinfo {address} {Singapore}}\
			(\bibinfo {year} {2017})}\BibitemShut {NoStop}%
		\bibitem [{\citenamefont {Du}\ \emph {et~al.}(2021)}]{Du2021a}%
		\BibitemOpen
		\bibfield  {author} {\bibinfo {author} {\bibfnamefont {Z.~Z.}\ \bibnamefont {Du}},
			\bibinfo {author} {\bibfnamefont {H.-Z.}\ \bibnamefont {Lu}},\ and\
			\bibinfo {author} {\bibfnamefont {X.~C.}\ \bibnamefont {Xie}},\ }%
		\bibfield  {title} {\bibinfo {title} {Nonlinear Hall effects},\ }%
		\href{https://doi.org/10.1038/s42254-021-00359-6}{%
			\bibfield  {journal} {\bibinfo {journal} {Nat. Rev. Phys.}\ }%
			\textbf {\bibinfo {volume} {3}},\
			\bibinfo {pages} {744--752} (\bibinfo {year} {2021})%
		}%
		\BibitemShut {NoStop}%
		\bibitem [{\citenamefont {Yang}\ \emph {et~al.}(2020)}]{Yang2020}%
		\BibitemOpen
		\bibfield  {author} {\bibinfo {author} {\bibfnamefont {N.-X.}\ \bibnamefont {Yang}},
			\bibinfo {author} {\bibfnamefont {Q.}\ \bibnamefont {Yan}},\ and\
			\bibinfo {author} {\bibfnamefont {Q.-F.}\ \bibnamefont {Sun}},\ }%
		\bibfield  {title} {\bibinfo {title} {Linear and nonlinear thermoelectric transport in a magnetic topological insulator nanoribbon with a domain wall},\ }%
		\href{https://doi.org/10.1103/physrevb.102.245412}{%
			\bibfield  {journal} {\bibinfo {journal} {Phys. Rev. B}\ }%
			\textbf {\bibinfo {volume} {102}},\
			\bibinfo {pages} {245412} (\bibinfo {year} {2020})%
		}%
		\BibitemShut {NoStop}%
		\bibitem [{\citenamefont {Costi}\ and\ \citenamefont {Zlati\'c}(2010)}]{Costi2010}%
		\BibitemOpen
		\bibfield  {author} {\bibinfo {author} {\bibfnamefont {T.~A.}\ \bibnamefont {Costi}}\ and\
			\bibinfo {author} {\bibfnamefont {V.}\ \bibnamefont {Zlati\'c}},\ }%
		\bibfield  {title} {\bibinfo {title} {Thermoelectric transport through strongly correlated quantum dots},\ }%
		\href{https://doi.org/10.1103/physrevb.81.235127}{
			\bibfield  {journal} {\bibinfo  {journal} {Phys. Rev. B}\ }%
			\textbf {\bibinfo {volume} {81}},\
			\bibinfo {pages} {235127}(\bibinfo {year} {2010})}%
		\BibitemShut {NoStop}%
	\bibitem [{\citenamefont {Nakai}\ and\ \citenamefont
		{Nagaosa}(2019)}]{Nakai2019}%
	\BibitemOpen
	\bibfield  {author} {\bibinfo {author} {\bibfnamefont {R.}~\bibnamefont
			{Nakai}}\ and\ \bibinfo {author} {\bibfnamefont {N.}~\bibnamefont
			{Nagaosa}},\ }\bibfield  {title} {\bibinfo {title} {Nonreciprocal thermal and
			thermoelectric transport of electrons in noncentrosymmetric crystals},\
	}\href {https://doi.org/10.1103/physrevb.99.115201} {\bibfield  {journal}
		{\bibinfo  {journal} {Phys. Rev. B}\ }\textbf {\bibinfo {volume} {99}},\
		\bibinfo {pages} {115201} (\bibinfo {year} {2019})}\BibitemShut {NoStop}%
	\bibitem [{\citenamefont {Yamaguchi}\ \emph
		{et~al.}(2024{\natexlab{a}})\citenamefont {Yamaguchi}, \citenamefont
		{Nakazawa},\ and\ \citenamefont {Yamakage}}]{Yamaguchi2024}%
	\BibitemOpen
	\bibfield  {author} {\bibinfo {author} {\bibfnamefont {T.}~\bibnamefont
			{Yamaguchi}}, \bibinfo {author} {\bibfnamefont {K.}~\bibnamefont
			{Nakazawa}},\ and\ \bibinfo {author} {\bibfnamefont {A.}~\bibnamefont
			{Yamakage}},\ }\bibfield  {title} {\bibinfo {title} {Microscopic theory of
			nonlinear {H}all effect induced by electric field and temperature gradient},\
	}\href {https://doi.org/10.1103/physrevb.109.205117} {\bibfield  {journal}
		{\bibinfo  {journal} {Phys. Rev. B}\ }\textbf {\bibinfo {volume} {109}},\
		\bibinfo {pages} {205117} (\bibinfo {year} {2024}{\natexlab{a}})}\BibitemShut
	{NoStop}%
	\bibitem [{\citenamefont {Yang}\ and\ \citenamefont
		{Skinner}(2025)}]{Yang2025}%
	\BibitemOpen
	\bibfield  {author} {\bibinfo {author} {\bibfnamefont {X.}~\bibnamefont
			{Yang}}\ and\ \bibinfo {author} {\bibfnamefont {B.}~\bibnamefont {Skinner}},\
	}\bibfield  {title} {\bibinfo {title} {Nonlinear thermoelectric effects as a
			means to probe quantum geometry},\ }\href@noop {} {\bibfield  {journal}
		{\bibinfo  {journal} {arXiv}\ } (\bibinfo {year} {2025})},\ \Eprint
	{https://arxiv.org/abs/2505.00086} {arXiv:2505.00086 [cond-mat.mes-hall]}
	\BibitemShut {NoStop}%
	\bibitem [{\citenamefont {Yamaguchi}\ \emph
		{et~al.}(2024{\natexlab{b}})\citenamefont {Yamaguchi}, \citenamefont
		{Nakazawa},\ and\ \citenamefont {Yamakage}}]{Yamaguchi2024a}%
	\BibitemOpen
	\bibfield  {author} {\bibinfo {author} {\bibfnamefont {T.}~\bibnamefont
			{Yamaguchi}}, \bibinfo {author} {\bibfnamefont {K.}~\bibnamefont
			{Nakazawa}},\ and\ \bibinfo {author} {\bibfnamefont {A.}~\bibnamefont
			{Yamakage}},\ }\bibfield  {title} {\bibinfo {title} {Theory of nonlinear
			{H}all effect induced by electric field and temperature gradient in 3{D}
			chiral magnetic textures},\ }\href@noop {} {\bibfield  {journal} {\bibinfo
			{journal} {arXiv}\ } (\bibinfo {year} {2024}{\natexlab{b}})},\ \Eprint
	{https://arxiv.org/abs/2410.00563} {arXiv:2410.00563 [cond-mat.mes-hall]}
	\BibitemShut {NoStop}%
	\bibitem [{\citenamefont {Gao}(2019)}]{Gao2019}%
	\BibitemOpen
	\bibfield  {author} {\bibinfo {author} {\bibfnamefont {Y.}~\bibnamefont
			{Gao}},\ }\bibfield  {title} {\bibinfo {title} {Semiclassical dynamics and
			nonlinear charge current},\ }\href
	{https://doi.org/10.1007/s11467-019-0887-2} {\bibfield  {journal} {\bibinfo
			{journal} {Front Phys}\ }\textbf {\bibinfo {volume} {14}},\ \bibinfo {pages}
		{33404} (\bibinfo {year} {2019})}\BibitemShut {NoStop}%
		\bibitem [{\citenamefont {Sinitsyn}\ \emph {et~al.}(2005)}]{Sinitsyn2005}%
		\BibitemOpen
		\bibfield  {author} {\bibinfo {author} {\bibfnamefont {N.~A.}\ \bibnamefont {Sinitsyn}},
			\bibinfo {author} {\bibfnamefont {Q.}\ \bibnamefont {Niu}},
			\bibinfo {author} {\bibfnamefont {J.}\ \bibnamefont {Sinova}},\ and\
			\bibinfo {author} {\bibfnamefont {K.}\ \bibnamefont {Nomura}},\ }%
		\bibfield  {title} {\bibinfo {title} {Disorder effects in the anomalous {H}all effect induced by {B}erry curvature},\ }%
		\href{https://doi.org/10.1103/physrevb.72.045346}{%
			\bibfield  {journal} {\bibinfo {journal} {Phys. Rev. B}\ }%
			\textbf {\bibinfo {volume} {72}},\
			\bibinfo {pages} {045346} (\bibinfo {year} {2005})%
		}%
		\BibitemShut {NoStop}%
		\bibitem [{\citenamefont {Sinitsyn}\ \emph {et~al.}(2006)}]{Sinitsyn2006}%
		\BibitemOpen
		\bibfield  {author} {\bibinfo {author} {\bibfnamefont {N.~A.}\ \bibnamefont {Sinitsyn}},
			\bibinfo {author} {\bibfnamefont {Q.}\ \bibnamefont {Niu}},\ and\
			\bibinfo {author} {\bibfnamefont {A.~H.}\ \bibnamefont {MacDonald}},\ }%
		\bibfield  {title} {\bibinfo {title} {Coordinate shift in the semiclassical {B}oltzmann equation and the anomalous {H}all effect},\ }%
		\href{https://doi.org/10.1103/physrevb.73.075318}{%
			\bibfield  {journal} {\bibinfo {journal} {Phys. Rev. B}\ }%
			\textbf {\bibinfo {volume} {73}},\
			\bibinfo {pages} {075318} (\bibinfo {year} {2006})%
		}%
		\BibitemShut {NoStop}%
	\bibitem [{\citenamefont {Kuroda}\ \emph {et~al.}(2010)\citenamefont {Kuroda},
		\citenamefont {Arita}, \citenamefont {Miyamoto}, \citenamefont {Ye},
		\citenamefont {Jiang}, \citenamefont {Kimura}, \citenamefont {Krasovskii},
		\citenamefont {Chulkov}, \citenamefont {Iwasawa}, \citenamefont {Okuda},
		\citenamefont {Shimada}, \citenamefont {Ueda}, \citenamefont {Namatame},\
		and\ \citenamefont {Taniguchi}}]{Kuroda2010}%
	\BibitemOpen
	\bibfield  {author} {\bibinfo {author} {\bibfnamefont {K.}~\bibnamefont
			{Kuroda}}, \bibinfo {author} {\bibfnamefont {M.}~\bibnamefont {Arita}},
		\bibinfo {author} {\bibfnamefont {K.}~\bibnamefont {Miyamoto}}, \bibinfo
		{author} {\bibfnamefont {M.}~\bibnamefont {Ye}}, \bibinfo {author}
		{\bibfnamefont {J.}~\bibnamefont {Jiang}}, \bibinfo {author} {\bibfnamefont
			{A.}~\bibnamefont {Kimura}}, \bibinfo {author} {\bibfnamefont {E.~E.}\
			\bibnamefont {Krasovskii}}, \bibinfo {author} {\bibfnamefont {E.~V.}\
			\bibnamefont {Chulkov}}, \bibinfo {author} {\bibfnamefont {H.}~\bibnamefont
			{Iwasawa}}, \bibinfo {author} {\bibfnamefont {T.}~\bibnamefont {Okuda}},
		\bibinfo {author} {\bibfnamefont {K.}~\bibnamefont {Shimada}}, \bibinfo
		{author} {\bibfnamefont {Y.}~\bibnamefont {Ueda}}, \bibinfo {author}
		{\bibfnamefont {H.}~\bibnamefont {Namatame}},\ and\ \bibinfo {author}
		{\bibfnamefont {M.}~\bibnamefont {Taniguchi}},\ }\bibfield  {title} {\bibinfo
		{title} {Hexagonally deformed {F}ermi surface of the 3{D} topological insulator
			{B}i$_2${S}e$_3$},\ }\href {https://doi.org/10.1103/physrevlett.105.076802}
	{\bibfield  {journal} {\bibinfo  {journal} {Phys. Rev. Lett.}\ }\textbf
		{\bibinfo {volume} {105}},\ \bibinfo {pages} {076802} (\bibinfo {year}
		{2010})}\BibitemShut {NoStop}%
	\bibitem [{\citenamefont {Fu}(2009)}]{Fu2009}%
	\BibitemOpen
	\bibfield  {author} {\bibinfo {author} {\bibfnamefont {L.}~\bibnamefont
			{Fu}},\ }\bibfield  {title} {\bibinfo {title} {Hexagonal warping effects in
			the surface states of the topological insulator {B}e$_2${T}e$_3$},\ }\href
	{https://doi.org/10.1103/physrevlett.103.266801} {\bibfield  {journal}
		{\bibinfo  {journal} {Phys. Rev. Lett.}\ }\textbf {\bibinfo {volume} {103}},\
		\bibinfo {pages} {266801} (\bibinfo {year} {2009})}\BibitemShut {NoStop}%
		\bibitem [{\citenamefont {Gao}\ \emph {et~al.}(2025)\citenamefont {Gao},
			\citenamefont {Nagaosa}, \citenamefont {Ni}\ and\ \citenamefont
			{Xu}}]{Gao2025}%
		\BibitemOpen
		\bibfield  {author} {\bibinfo {author} {\bibfnamefont {A.}\ \bibnamefont
				{Gao}}, \bibinfo {author} {\bibfnamefont {N.}\ \bibnamefont {Nagaosa}},
			\bibinfo {author} {\bibfnamefont {N.}\ \bibnamefont {Ni}},\ and\ \bibinfo
			{author} {\bibfnamefont {S.-Y.}\ \bibnamefont {Xu}},\ }\bibfield  {title}
		{\bibinfo {title} {Quantum geometry phenomena in condensed matter systems},\
		}\href {https://doi.org/10.48550/ARXIV.2508.00469} {\bibfield  {journal}
			{\bibinfo  {journal} {arXiv:2508.00469}\ } (\bibinfo {year}
			{2025})}\BibitemShut {NoStop}%
			\bibitem [{\citenamefont {Liu}\ \emph {et~al.}(2024)\citenamefont {Liu},
				\citenamefont {Qiang}, \citenamefont {Lu}\ and\ \citenamefont
				{Xie}}]{Liu2024a}%
			\BibitemOpen
			\bibfield  {author} {\bibinfo {author} {\bibfnamefont {T.}\ \bibnamefont
					{Liu}}, \bibinfo {author} {\bibfnamefont {X.-B.}\ \bibnamefont {Qiang}},
				\bibinfo {author} {\bibfnamefont {H.-Z.}\ \bibnamefont {Lu}},\ and\
				\bibinfo {author} {\bibfnamefont {X.~C.}\ \bibnamefont {Xie}},\ }\bibfield
			{title} {\bibinfo {title} {Quantum geometry in condensed matter},\ }\href
			{https://doi.org/10.1093/nsr/nwae334} {\bibfield  {journal} {\bibinfo
					{journal} {Natl. Sci. Rev.}\ }\textbf {\bibinfo {volume} {12}},\ \bibinfo
				{pages} {nwae334} (\bibinfo {year} {2024})}\BibitemShut {NoStop}%
				\bibitem [{\citenamefont {Yu}\ \emph {et~al.}(2025)\citenamefont {Yu},
					\citenamefont {Bernevig}, \citenamefont {Queiroz}, \citenamefont {Rossi},
					\citenamefont {T{\"o}rm{\"a}}\ and\ \citenamefont {Yang}}]{Yu2025}%
				\BibitemOpen
				\bibfield  {author} {\bibinfo {author} {\bibfnamefont {J.}\ \bibnamefont
						{Yu}}, \bibinfo {author} {\bibfnamefont {B.~A.}\ \bibnamefont {Bernevig}},
					\bibinfo {author} {\bibfnamefont {R.}\ \bibnamefont {Queiroz}}, \bibinfo
					{author} {\bibfnamefont {E.}\ \bibnamefont {Rossi}}, \bibinfo {author}
					{\bibfnamefont {P.}\ \bibnamefont {T{\"o}rm{\"a}}},\ and\ \bibinfo {author}
					{\bibfnamefont {B.-J.}\ \bibnamefont {Yang}},\ }\bibfield  {title} {\bibinfo
					{title} {Quantum geometry in quantum materials},\ }\href
				{https://doi.org/10.1038/s41535-025-00801-3} {\bibfield  {journal} {\bibinfo
						{journal} {npj Quantum Mater.}\ }\textbf {\bibinfo {volume} {10}},\ \bibinfo
					{pages} {1} (\bibinfo {year} {2025})}\BibitemShut {NoStop}%
	\bibitem [{\citenamefont {He}\ \emph {et~al.}(2021)\citenamefont {He},
		\citenamefont {Isobe}, \citenamefont {Zhu}, \citenamefont {Hsu},
		\citenamefont {Fu},\ and\ \citenamefont {Yang}}]{He2021}%
	\BibitemOpen
	\bibfield  {author} {\bibinfo {author} {\bibfnamefont {P.}~\bibnamefont
			{He}}, \bibinfo {author} {\bibfnamefont {H.}~\bibnamefont {Isobe}}, \bibinfo
		{author} {\bibfnamefont {D.}~\bibnamefont {Zhu}}, \bibinfo {author}
		{\bibfnamefont {C.-H.}\ \bibnamefont {Hsu}}, \bibinfo {author} {\bibfnamefont
			{L.}~\bibnamefont {Fu}},\ and\ \bibinfo {author} {\bibfnamefont
			{H.}~\bibnamefont {Yang}},\ }\bibfield  {title} {\bibinfo {title} {Quantum
			frequency doubling in the topological insulator {Bi}$_2${Se}$_3$},\ }\href
	{https://doi.org/10.1038/s41467-021-20983-1} {\bibfield  {journal} {\bibinfo
			{journal} {Nat. Commun.}\ }\textbf {\bibinfo {volume} {12}},\ \bibinfo
		{pages} {698} (\bibinfo {year} {2021})}\BibitemShut {NoStop}%
		\bibitem [{\citenamefont {Lu}\ \emph {et~al.}(2024)\citenamefont {Lu},
			\citenamefont {Zhang}, \citenamefont {Wang}, \citenamefont {Zhao},
			\citenamefont {Zhou}, \citenamefont {Gao}, \citenamefont {Chen},
			\citenamefont {Law}\ and\ \citenamefont {Loh}}]{Lu2024}%
		\BibitemOpen
		\bibfield  {author} {\bibinfo {author} {\bibfnamefont {X.~F.}\ \bibnamefont
				{Lu}}, \bibinfo {author} {\bibfnamefont {C.-P.}\ \bibnamefont {Zhang}},
			\bibinfo {author} {\bibfnamefont {N.}\ \bibnamefont {Wang}}, \bibinfo
			{author} {\bibfnamefont {D.}\ \bibnamefont {Zhao}}, \bibinfo {author}
			{\bibfnamefont {X.}\ \bibnamefont {Zhou}}, \bibinfo {author} {\bibfnamefont
				{W.}\ \bibnamefont {Gao}}, \bibinfo {author} {\bibfnamefont {X.~H.}\
				\bibnamefont {Chen}}, \bibinfo {author} {\bibfnamefont {K.~T.}\ \bibnamefont
				{Law}},\ and\ \bibinfo {author} {\bibfnamefont {K.~P.}\ \bibnamefont
				{Loh}},\ }\bibfield  {title} {\bibinfo {title} {Nonlinear transport and
				radio frequency rectification in {BiTeBr} at room temperature},\ }\href
		{https://doi.org/10.1038/s41467-023-44439-w} {\bibfield  {journal}
			{\bibinfo  {journal} {Nat. Commun.}\ }\textbf {\bibinfo {volume} {15}},\
			\bibinfo {pages} {1} (\bibinfo {year} {2024})}\BibitemShut {NoStop}%
	\bibitem [{\citenamefont {Gong}\ \emph {et~al.}(2024)\citenamefont {Gong},
		\citenamefont {Du}, \citenamefont {Sun}, \citenamefont {Lu},\ and\
		\citenamefont {Xie}}]{Gong2024}%
	\BibitemOpen
	\bibfield  {author} {\bibinfo {author} {\bibfnamefont {Z.-H.}\ \bibnamefont
			{Gong}}, \bibinfo {author} {\bibfnamefont {Z.~Z.}\ \bibnamefont {Du}},
		\bibinfo {author} {\bibfnamefont {H.-P.}\ \bibnamefont {Sun}}, \bibinfo
		{author} {\bibfnamefont {H.-Z.}\ \bibnamefont {Lu}},\ and\ \bibinfo {author}
		{\bibfnamefont {X.~C.}\ \bibnamefont {Xie}},\ }\bibfield  {title} {\bibinfo
		{title} {Nonlinear transport theory at the order of quantum metric},\
	}\href@noop {} {\bibfield  {journal} {\bibinfo  {journal} {arXiv}\ }
		(\bibinfo {year} {2024})},\ \Eprint {https://arxiv.org/abs/2410.04995}
	{arXiv:2410.04995 [cond-mat.mes-hall]} \BibitemShut {NoStop}%
	\bibitem [{\citenamefont {Xiao}\ \emph {et~al.}(2019)\citenamefont {Xiao},
		\citenamefont {Du},\ and\ \citenamefont {Niu}}]{Xiao2019}%
	\BibitemOpen
	\bibfield  {author} {\bibinfo {author} {\bibfnamefont {C.}~\bibnamefont
			{Xiao}}, \bibinfo {author} {\bibfnamefont {Z.~Z.}\ \bibnamefont {Du}},\ and\
		\bibinfo {author} {\bibfnamefont {Q.}~\bibnamefont {Niu}},\ }\bibfield
	{title} {\bibinfo {title} {Theory of nonlinear {H}all effects: Modified
			semiclassics from quantum kinetics},\ }\href
	{https://doi.org/10.1103/physrevb.100.165422} {\bibfield  {journal} {\bibinfo
			{journal} {Phys. Rev. B}\ }\textbf {\bibinfo {volume} {100}},\ \bibinfo
		{pages} {165422} (\bibinfo {year} {2019})}\BibitemShut {NoStop}%
	\bibitem [{\citenamefont {Mehraeen}(2024)}]{Mehraeen2024}%
	\BibitemOpen
	\bibfield  {author} {\bibinfo {author} {\bibfnamefont {M.}~\bibnamefont
			{Mehraeen}},\ }\bibfield  {title} {\bibinfo {title} {Quantum kinetic theory
			of quadratic responses},\ }\href
	{https://doi.org/10.1103/physrevb.110.174423} {\bibfield  {journal} {\bibinfo
			{journal} {Phys. Rev. B}\ }\textbf {\bibinfo {volume} {110}},\ \bibinfo
		{pages} {174423} (\bibinfo {year} {2024})}\BibitemShut {NoStop}%
	\bibitem [{\citenamefont {Sinitsyn}\ \emph {et~al.}(2007)\citenamefont
		{Sinitsyn}, \citenamefont {MacDonald}, \citenamefont {Jungwirth},
		\citenamefont {Dugaev},\ and\ \citenamefont {Sinova}}]{Sinitsyn2007}%
	\BibitemOpen
	\bibfield  {author} {\bibinfo {author} {\bibfnamefont {N.~A.}\ \bibnamefont
			{Sinitsyn}}, \bibinfo {author} {\bibfnamefont {A.~H.}\ \bibnamefont
			{MacDonald}}, \bibinfo {author} {\bibfnamefont {T.}~\bibnamefont
			{Jungwirth}}, \bibinfo {author} {\bibfnamefont {V.~K.}\ \bibnamefont
			{Dugaev}},\ and\ \bibinfo {author} {\bibfnamefont {J.}~\bibnamefont
			{Sinova}},\ }\bibfield  {title} {\bibinfo {title} {Anomalous hall effect in a
			two-dimensional dirac band: The link between the {K}ubo-{S}treda formula and
			the semiclassical {B}oltzmann equation approach},\ }\href
	{https://doi.org/10.1103/physrevb.75.045315} {\bibfield  {journal} {\bibinfo
			{journal} {Phys. Rev. B}\ }\textbf {\bibinfo {volume} {75}},\ \bibinfo
		{pages} {045315} (\bibinfo {year} {2007})}\BibitemShut {NoStop}%
	\bibitem [{\citenamefont {Parker}\ \emph {et~al.}(2019)\citenamefont {Parker},
		\citenamefont {Morimoto}, \citenamefont {Orenstein},\ and\ \citenamefont
		{Moore}}]{Parker2019}%
	\BibitemOpen
	\bibfield  {author} {\bibinfo {author} {\bibfnamefont {D.~E.}\ \bibnamefont
			{Parker}}, \bibinfo {author} {\bibfnamefont {T.}~\bibnamefont {Morimoto}},
		\bibinfo {author} {\bibfnamefont {J.}~\bibnamefont {Orenstein}},\ and\
		\bibinfo {author} {\bibfnamefont {J.~E.}\ \bibnamefont {Moore}},\ }\bibfield
	{title} {\bibinfo {title} {Diagrammatic approach to nonlinear optical
			response with application to {W}eyl semimetals},\ }\href
	{https://doi.org/10.1103/physrevb.99.045121} {\bibfield  {journal} {\bibinfo
			{journal} {Phys. Rev. B}\ }\textbf {\bibinfo {volume} {99}},\ \bibinfo
		{pages} {045121} (\bibinfo {year} {2019})}\BibitemShut {NoStop}%
	\bibitem [{\citenamefont {Bruus}\ and\ \citenamefont
		{Flensberg}(2004)}]{Bruus2004}%
	\BibitemOpen
	\bibfield  {author} {\bibinfo {author} {\bibfnamefont {H.}~\bibnamefont
			{Bruus}}\ and\ \bibinfo {author} {\bibfnamefont {K.}~\bibnamefont
			{Flensberg}},\ }\href@noop {} {\emph {\bibinfo {title} {Many-Body Quantum
				Theory in Condensed Matter Physics: An Introduction}}},\ edited by\ \bibinfo
	{editor} {\bibnamefont {online}}\ (\bibinfo  {publisher} {Oxford University
		Press},\ \bibinfo {year} {2004})\BibitemShut {NoStop}%
	\bibitem [{\citenamefont {Bhalla}(2021)}]{Bhalla2021}%
	\BibitemOpen
	\bibfield  {author} {\bibinfo {author} {\bibfnamefont {P.}~\bibnamefont
			{Bhalla}},\ }\bibfield  {title} {\bibinfo {title} {Intrinsic contribution to
			nonlinear thermoelectric effects in topological insulators},\ }\href
	{https://doi.org/10.1103/physrevb.103.115304} {\bibfield  {journal} {\bibinfo
			{journal} {Phys. Rev. B}\ }\textbf {\bibinfo {volume} {103}},\ \bibinfo
		{pages} {115304} (\bibinfo {year} {2021})}\BibitemShut {NoStop}%
	\bibitem [{\citenamefont {Varshney}\ \emph {et~al.}(2023)\citenamefont
		{Varshney}, \citenamefont {Das}, \citenamefont {Bhalla},\ and\ \citenamefont
		{Agarwal}}]{Varshney2023}%
	\BibitemOpen
	\bibfield  {author} {\bibinfo {author} {\bibfnamefont {H.}~\bibnamefont
			{Varshney}}, \bibinfo {author} {\bibfnamefont {K.}~\bibnamefont {Das}},
		\bibinfo {author} {\bibfnamefont {P.}~\bibnamefont {Bhalla}},\ and\ \bibinfo
		{author} {\bibfnamefont {A.}~\bibnamefont {Agarwal}},\ }\bibfield  {title}
	{\bibinfo {title} {Quantum kinetic theory of nonlinear thermal current},\
	}\href {https://doi.org/10.1103/physrevb.107.235419} {\bibfield  {journal}
		{\bibinfo  {journal} {Phys. Rev. B}\ }\textbf {\bibinfo {volume} {107}},\
		\bibinfo {pages} {235419} (\bibinfo {year} {2023})}\BibitemShut {NoStop}%
	\bibitem [{\citenamefont {Ma}\ and\ \citenamefont {Xie}(2025)}]{Ma2025b}%
	\BibitemOpen
	\bibfield  {author} {%
		\bibinfo {author} {\bibfnamefont {D.}\ \bibnamefont {Ma}},
		\bibinfo {author} {\bibfnamefont {Z.~F.}\ \bibnamefont {Zhang}},
		\bibinfo {author} {\bibfnamefont {H.}\ \bibnamefont {Jiang}},\ and\
		\bibinfo {author} {\bibfnamefont {X.~C.}\ \bibnamefont {Xie}},\ }%
	\bibfield  {title} {\bibinfo {title} {Quantum kinetic theory of the semiclassical side jump, skew scattering, and longitudinal velocity},\ }%
	\href{https://doi.org/10.1103/rgr5-yzy2}{%
		\bibfield  {journal} {\bibinfo  {journal} {Phys. Rev. B}\textbf {\bibinfo {volume}
				{112}},\ \bibinfo {pages} {045136}  }%
		(\bibinfo {year} {2025})%
	}%
	\BibitemShut {NoStop}%
	\bibitem [{\citenamefont {Wei}\ \emph {et~al.}(2022)\citenamefont {Wei},
		\citenamefont {Wang}, \citenamefont {Yu}, \citenamefont {Xu},\ and\
		\citenamefont {Wang}}]{Wei2022}%
	\BibitemOpen
	\bibfield  {author} {\bibinfo {author} {\bibfnamefont {M.}~\bibnamefont
			{Wei}}, \bibinfo {author} {\bibfnamefont {B.}~\bibnamefont {Wang}}, \bibinfo
		{author} {\bibfnamefont {Y.}~\bibnamefont {Yu}}, \bibinfo {author}
		{\bibfnamefont {F.}~\bibnamefont {Xu}},\ and\ \bibinfo {author}
		{\bibfnamefont {J.}~\bibnamefont {Wang}},\ }\bibfield  {title} {\bibinfo
		{title} {Nonlinear {H}all effect induced by internal {C}oulomb interaction
			and phase relaxation process in a four-terminal system with time-reversal
			symmetry},\ }\href {https://doi.org/10.1103/physrevb.105.115411} {\bibfield
		{journal} {\bibinfo  {journal} {Phys. Rev. B}\ }\textbf {\bibinfo {volume}
			{105}},\ \bibinfo {pages} {115411} (\bibinfo {year} {2022})}\BibitemShut
	{NoStop}%
	\bibitem [{\citenamefont {Zhang}\ \emph {et~al.}(2023)\citenamefont {Zhang},
		\citenamefont {Xu}, \citenamefont {Chen}, \citenamefont {Xing},\ and\
		\citenamefont {Wang}}]{Zhang2023}%
	\BibitemOpen
	\bibfield  {author} {\bibinfo {author} {\bibfnamefont {L.}~\bibnamefont
			{Zhang}}, \bibinfo {author} {\bibfnamefont {F.}~\bibnamefont {Xu}}, \bibinfo
		{author} {\bibfnamefont {J.}~\bibnamefont {Chen}}, \bibinfo {author}
		{\bibfnamefont {Y.}~\bibnamefont {Xing}},\ and\ \bibinfo {author}
		{\bibfnamefont {J.}~\bibnamefont {Wang}},\ }\bibfield  {title} {\bibinfo
		{title} {Quantum nonlinear ac transport theory at low frequency},\ }\href
	{https://doi.org/10.1088/1367-2630/ad05a4} {\bibfield  {journal} {\bibinfo
			{journal} {New J. Phys.}\ }\textbf {\bibinfo {volume} {25}},\ \bibinfo
		{pages} {113006} (\bibinfo {year} {2023})}\BibitemShut {NoStop}%
	\bibitem [{\citenamefont {Li}\ \emph {et~al.}(2025)\citenamefont {Li},
		\citenamefont {Wei}, \citenamefont {Xu},\ and\ \citenamefont
		{Wang}}]{Li2025}%
	\BibitemOpen
	\bibfield  {author} {\bibinfo {author} {\bibfnamefont {G.}~\bibnamefont
			{Li}}, \bibinfo {author} {\bibfnamefont {M.}~\bibnamefont {Wei}}, \bibinfo
		{author} {\bibfnamefont {F.}~\bibnamefont {Xu}},\ and\ \bibinfo {author}
		{\bibfnamefont {J.}~\bibnamefont {Wang}},\ }\bibfield  {title} {\bibinfo
		{title} {General method for calculating transport properties of disordered
			mesoscopic systems based on the nonequilibrium {G}reen's function
			formalism},\ }\href {https://doi.org/10.1103/physrevb.111.035409} {\bibfield
		{journal} {\bibinfo  {journal} {Phys. Rev. B}\ }\textbf {\bibinfo {volume}
			{111}},\ \bibinfo {pages} {035409} (\bibinfo {year} {2025})}\BibitemShut
	{NoStop}%
	\bibitem [{\citenamefont {Pasqua}\ and\ \citenamefont
		{Fabrizio}(2025)}]{Pasqua2025}%
	\BibitemOpen
	\bibfield  {author} {\bibinfo {author} {\bibfnamefont {I.}~\bibnamefont
			{Pasqua}}\ and\ \bibinfo {author} {\bibfnamefont {M.}~\bibnamefont
			{Fabrizio}},\ }\bibfield  {title} {\bibinfo {title} {Fermi-liquid corrections
			to the intrinsic anomalous {H}all conductivity of topological metals},\
	}\href {https://doi.org/10.21468/SciPostPhys.19.1.014} {\bibfield  {journal}
		{\bibinfo  {journal} {SciPost Phys.}\ }\textbf {\bibinfo {volume} {19}},\
		\bibinfo {pages} {014} (\bibinfo {year} {2025})}\BibitemShut {NoStop}%
	\bibitem [{\citenamefont {Jho}\ \emph {et~al.}(2017)\citenamefont {Jho},
		\citenamefont {Han},\ and\ \citenamefont {Kim}}]{Jho2017}%
	\BibitemOpen
	\bibfield  {author} {\bibinfo {author} {\bibfnamefont {Y.-S.}\ \bibnamefont
			{Jho}}, \bibinfo {author} {\bibfnamefont {J.-H.}\ \bibnamefont {Han}},\ and\
		\bibinfo {author} {\bibfnamefont {K.-S.}\ \bibnamefont {Kim}},\ }\bibfield
	{title} {\bibinfo {title} {Topological fermi-liquid theory for interacting
			{W}eyl metals with time reversal symmetry breaking},\ }\href
	{https://doi.org/10.1103/physrevb.95.205113} {\bibfield  {journal} {\bibinfo
			{journal} {Phys. Rev. B}\ }\textbf {\bibinfo {volume} {95}},\ \bibinfo
		{pages} {205113} (\bibinfo {year} {2017})}\BibitemShut {NoStop}%
	\bibitem [{\citenamefont {Chen}\ and\ \citenamefont {Son}(2017)}]{Chen2017}%
	\BibitemOpen
	\bibfield  {author} {\bibinfo {author} {\bibfnamefont {J.-Y.}\ \bibnamefont
			{Chen}}\ and\ \bibinfo {author} {\bibfnamefont {D.~T.}\ \bibnamefont {Son}},\
	}\bibfield  {title} {\bibinfo {title} {Berry {F}ermi liquid theory},\ }\href
	{https://doi.org/10.1016/j.aop.2016.12.017} {\bibfield  {journal} {\bibinfo
			{journal} {Ann. Phys-new. York.}\ }\textbf {\bibinfo {volume} {377}},\
		\bibinfo {pages} {345} (\bibinfo {year} {2017})}\BibitemShut {NoStop}%
	\bibitem [{\citenamefont {Wang}\ \emph {et~al.}(2025)\citenamefont {Wang},
		\citenamefont {Zhou}\ and\ \citenamefont {Yao}}]{Wang2025}%
	\BibitemOpen
	\bibfield  {author} {\bibinfo {author} {\bibfnamefont {M.}\ \bibnamefont
			{Wang}}, \bibinfo {author} {\bibfnamefont {J.}\ \bibnamefont {Zhou}},\ and\
		\bibinfo {author} {\bibfnamefont {Y.}\ \bibnamefont {Yao}},\ }\bibfield
	{title} {\bibinfo {title} {Linear and nonlinear optical responses in
			{Green}'s function formula},\ }\href
	{https://doi.org/10.48550/ARXIV.2508.07280} {\bibfield  {journal} {\bibinfo
			{journal} {arXiv:2508.07280}} (\bibinfo {year}
		{2025}).}\BibitemShut {NoStop}%
		\bibitem [{\citenamefont {Nielsen}\ and\ \citenamefont {Taylor}(1968)}]{Nielsen1968}%
		\BibitemOpen
		\bibfield  {author} {\bibinfo {author} {\bibfnamefont {P.~E.}\ \bibnamefont
				{Nielsen}}\ and\ \bibinfo {author} {\bibfnamefont {P.~L.}\ \bibnamefont
				{Taylor}},\ }\bibfield  {title} {\bibinfo {title} {Low-temperature
				thermopower of Dilute alloys},\ }\href
		{https://doi.org/10.1103/PhysRevLett.21.893} {\bibfield  {journal}
			{\bibinfo  {journal} {Phys. Rev. Lett.}\ }\textbf {\bibinfo {volume} {21}},\
			\bibinfo {pages} {893} (\bibinfo {year} {1968})}\BibitemShut {NoStop}%
		\bibitem [{\citenamefont {Nielsen}\ and\ \citenamefont {Taylor}(1974)}]{Nielsen1974}%
		\BibitemOpen
		\bibfield  {author} {\bibinfo {author} {\bibfnamefont {P.~E.}\ \bibnamefont
				{Nielsen}}\ and\ \bibinfo {author} {\bibfnamefont {P.~L.}\ \bibnamefont
				{Taylor}},\ }\bibfield  {title} {\bibinfo {title} {Theory of thermoelectric
				effects in metals and alloys},\ }\href
		{https://doi.org/10.1103/PhysRevB.10.4061} {\bibfield  {journal}
			{\bibinfo  {journal} {Phys. Rev. B}\ }\textbf {\bibinfo {volume} {10}},\
			\bibinfo {pages} {4061} (\bibinfo {year} {1974})}\BibitemShut {NoStop}%
		\bibitem [{\citenamefont {Dudenhoeffer}\ and\ \citenamefont {Bourassa}(1972)}]{Dudenhoeffer1972}%
		\BibitemOpen
		\bibfield  {author} {\bibinfo {author} {\bibfnamefont {A.~W.}\ \bibnamefont
				{Dudenhoeffer}}\ and\ \bibinfo {author} {\bibfnamefont {R.~R.}\ \bibnamefont
				{Bourassa}},\ }\bibfield  {title} {\bibinfo {title} {Nielsen-Taylor effect
				in the thermopower of Dilute Aluminum alloys},\ }\href
		{https://doi.org/10.1103/PhysRevB.5.1651} {\bibfield  {journal}
			{\bibinfo  {journal} {Phys. Rev. B}\ }\textbf {\bibinfo {volume} {5}},\
			\bibinfo {pages} {1651} (\bibinfo {year} {1972})}\BibitemShut {NoStop}%
		\bibitem [{\citenamefont {Howson}\ \emph {et~al.}(1988)\citenamefont {Howson},
			\citenamefont {Hickey},\ and\ \citenamefont {Morgan}}]{Howson1988}%
		\BibitemOpen
		\bibfield  {author} {\bibinfo {author} {\bibfnamefont {M.~A.}\ \bibnamefont
				{Howson}}, \bibinfo {author} {\bibfnamefont {B.~J.}\ \bibnamefont
				{Hickey}},\ and\ \bibinfo {author} {\bibfnamefont {G.~J.}\ \bibnamefont
				{Morgan}},\ }\bibfield  {title} {\bibinfo {title} {Quantum interference
				effects and the magnitude of the resistivity and thermopower of Ca-Al
				metallic glasses},\ }\href
		{https://doi.org/10.1103/PhysRevB.38.5267} {\bibfield  {journal}
			{\bibinfo  {journal} {Phys. Rev. B}\ }\textbf {\bibinfo {volume} {38}},\
			\bibinfo {pages} {5267} (\bibinfo {year} {1988})}\BibitemShut {NoStop}%
		\bibitem [{\citenamefont {Reizer}\ and\ \citenamefont {Sergeev}(2000)}]{Reizer2000}%
		\BibitemOpen
		\bibfield  {author} {\bibinfo {author} {\bibfnamefont {M.}\ \bibnamefont
				{Reizer}}\ and\ \bibinfo {author} {\bibfnamefont {A.}\ \bibnamefont
				{Sergeev}},\ }\bibfield  {title} {\bibinfo {title} {Weak-localization effect
				on thermomagnetic phenomena},\ }\href
		{https://doi.org/10.1103/PhysRevB.61.7340} {\bibfield  {journal}
			{\bibinfo  {journal} {Phys. Rev. B}\ }\textbf {\bibinfo {volume} {61}},\
			\bibinfo {pages} {7340} (\bibinfo {year} {2000})}\BibitemShut {NoStop}%
			\bibitem [{\citenamefont {Huang}\ \emph {et~al.}(2025)\citenamefont {Huang},
				\citenamefont {Xiao}, \citenamefont {Yang},\ and\ \citenamefont {Li}}]{Huang2025}%
			\BibitemOpen
			\bibfield  {author} {\bibinfo {author} {\bibfnamefont {Y.-X.}\ \bibnamefont
					{Huang}}, \bibinfo {author} {\bibfnamefont {C.}\ \bibnamefont {Xiao}},
				\bibinfo {author} {\bibfnamefont {S. A.}\ \bibnamefont {Yang}},\ and\
				\bibinfo {author} {\bibfnamefont {X.}\ \bibnamefont {Li}},\ }\bibfield  {title}
			{\bibinfo {title} {Scaling law and extrinsic mechanisms for time-reversal-odd second-order nonlinear transport},\ }\href
			{https://doi.org/10.1103/PhysRevB.111.155127} {\bibfield  {journal} {\bibinfo
					{journal} {Phys. Rev. B}\ }\textbf {\bibinfo {volume} {111}},\ \bibinfo
				{pages} {155127} (\bibinfo {year} {2025})}\BibitemShut {NoStop}%
			\bibitem [{\citenamefont {Guo}\ \emph {et~al.}(2026)\citenamefont {Guo},
				\citenamefont {Liu}, \citenamefont {Xiao},\ and\ \citenamefont {Yuan}}]{Guo2026}%
			\BibitemOpen
			\bibfield  {author} {\bibinfo {author} {\bibfnamefont {R.-D.}\ \bibnamefont
					{Guo}}, \bibinfo {author} {\bibfnamefont {Y.}\ \bibnamefont {Liu}},
				\bibinfo {author} {\bibfnamefont {C.}\ \bibnamefont {Xiao}},\ and\
				\bibinfo {author} {\bibfnamefont {Z.}\ \bibnamefont {Yuan}},\ }\bibfield  {title}
			{\bibinfo {title} {Disorder induced time-reversal-odd nonlinear spin and orbital Hall effects},\ }\href
			{https://doi.org/10.48550/arXiv.2604.20592} {\bibfield  {journal} {\bibinfo
					{journal} {arXiv:2604.20592}\ } (\bibinfo {year} {2026})}\BibitemShut {NoStop}%
\end{thebibliography}

%

\end{document}